\documentclass[preprint,11pt,authoryear]{elsarticle}
\usepackage{geometry}
\usepackage{mathpazo}
\usepackage{fancyhdr}
\usepackage{amssymb}
\usepackage{graphicx} 
\usepackage{amsthm}
\usepackage{amsmath}
\usepackage{lineno}
\usepackage{hyperref}
\usepackage{tabularx}
\usepackage{url}
\usepackage{makecell}
\setcitestyle{numbers,square}
\usepackage{bm}
\usepackage{amsmath}
\usepackage{isomath}
\usepackage{amsfonts}
\usepackage[dvipsnames]{xcolor}
\usepackage{float}
\usepackage{xr}
\usepackage{framed}
\usepackage{makecell}
\usepackage{arydshln}
\usepackage{booktabs}
\usepackage{fullpage}
\usepackage{enumitem}
\usepackage{subcaption}

\usepackage[linesnumbered,boxed,ruled,vlined,longend]{algorithm2e}
\renewcommand{\vec}[1]{\boldsymbol{\mathrm{#1}}}

\renewcommand{\vec}[1]{\boldsymbol{\mathbfit{#1}}}

\definecolor{darkgreen}{rgb}{0.8,.4,0.4}
\newcommand{\edit}[1]{{\color{black}{#1}}}
\journal{arXiv.org}
\begin{document}
\begin{frontmatter}
\title{\textbf{\Large Scalable analysis of stop-and-go waves: \\ Representation, measurements and insights}}
\author[inst1,inst2]{Junyi Ji}
\ead{junyi.ji@vanderbilt.edu}
\author[inst2]{Derek Gloudemans}
\author[inst5]{Yanbing Wang}
\author[inst2]{Gergely Zach\'ar}
\author[inst2]{William Barbour}
\author[inst1,inst2,inst3]{Jonathan Sprinkle}
\author[inst6]{Benedetto Piccoli}
\author[inst1,inst2,inst3]{Daniel B. Work}

\affiliation[inst1]{organization={Department of Civil and Environmental Engineering, Vanderbilt University},country={United States}}
\affiliation[inst2]{organization={Institute for Software Integrated Systems, Vanderbilt University},country={United States}}
\affiliation[inst3]{organization={Department of Computer Science, Vanderbilt University},country={United States}}
\affiliation[inst5]{organization={School of Sustainable Engineering and the Built Environment, Arizona State University},country={United States}}
\affiliation[inst6]{organization={Department of Mathematical Sciences, Rutgers University–Camden},country={United States}}

\begin{abstract}
Analyzing stop-and-go waves at the scale of miles and hours of data is an emerging challenge in traffic research. 
The past 5 years have seen an explosion in the availability of large-scale traffic data containing traffic waves and complex congestion patterns, making existing approaches unsuitable for repeatable and scalable analysis of traffic waves in these data. This paper makes a first step towards addressing this challenge by introducing an automatic and scalable stop-and-go wave identification method capable of capturing wave generation, propagation, dissipation, as well as bifurcation and merging, which have previously been observed only very rarely. Using a concise and simple critical-speed based definition of a stop-and-go wave, the proposed method identifies all wave boundaries that encompass spatio-temporal points where vehicle speed is below a chosen critical speed. The method is built upon a graph representation of the spatio-temporal points associated with stop-and-go waves, specifically wave front (start) points and wave tail (end) points, and approaches the solution as a graph component identification problem. It enables the measurement of wave properties at scale. The method is implemented in Python and demonstrated on a large-scale dataset, I-24 MOTION INCEPTION. Our results show insights on the complexity of traffic waves. Traffic waves can bifurcate and merge at a scale that has never been observed or described before. The clustering analysis of all the identified wave components reveals the different topological structures of traffic waves. We explored that the wave merge or bifurcation points can be explained by spatial features. The gallery of all the identified wave topologies is demonstrated at \url{https://trafficwaves.github.io/}.
\end{abstract}

\end{frontmatter}

\newpage
\section{Introduction}
\subsection{Motivation and challenges}
Stop-and-go waves have garnered considerable attention from the fields of mathematics, physics, and traffic engineering, and remains a persistent and complex challenge. Since their first reporting in the Holland and Lincoln Tunnel experiments \cite{edie1960effect,edie1961,edie1961car,edie1967generation}, understanding stop-and-go waves has become one of the focal points of traffic science. Empirical data allows us to observe and analyze stop-go-wave patterns; though many legacy datasets \cite{edie1967generation,treiterer1974hysteresis,NGSIM} enable these efforts, the field of traffic science continues to suffer from a paucity of comprehensive datasets \cite{li2020trajectory} encompassing a broader spectrum of traffic wave phenomena, which are essential for deriving more profound insights through empirical analysis.

In recent years, advancements in computer vision technologies have facilitated the expansion of large-scale traffic trajectory datasets, such as pNEUMA drone data \cite{barmpounakis2020new}, Zen Traffic Data \citep{ztd2018}, DLR highway traffic (DLR-HT) data \cite{schicktanz2025dlr} and I-24 MOTION data \cite{gloudemans202324}, which can provide million or even billions of trajectory points. The field of trajectory data analysis has entered a new era of massive data \cite{barmpounakis2020new} and brings up new research opportunities \cite{li2020trajectory}. Scaling up the analysis to handle massive datasets has been both a clear research need and a significant challenge. In the context of stop-and-go wave analysis, it is feasible to manually label each wave in the NGSIM dataset and handle each case individually, as only a limited number of waves are observed. However, when the number of waves scales up to hundreds or even thousands, the necessity for an automated method to segment wave components becomes critical — a task that, to the best of our knowledge, has not been accomplished before. This motivates us to develop a method and release a tool designed to analyze large-scale datasets with massive trajectory data, thereby accelerating the traffic flow research and lowering the barrier for conducting the analyses.

\begin{figure}[t]
    \centering
    \includegraphics[width=\linewidth]{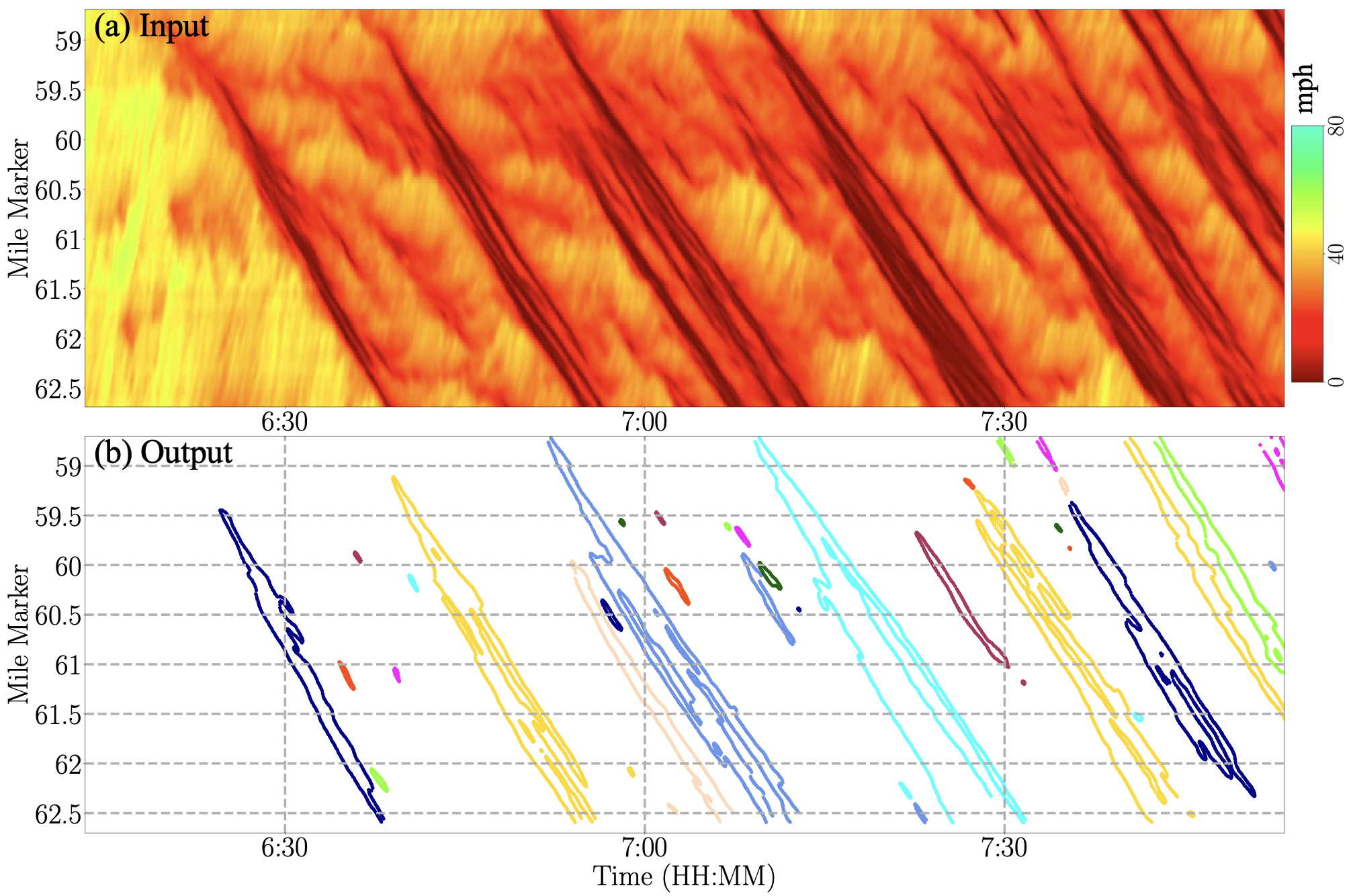}
    \caption{\textbf{Illustration of the problem solved in this paper}: (a) a lane-level macroscopic speed field on the space-time diagram with x-axis is time and y-axis is the mile marker as the model input (example on I-24 MOTION testbed lane 1, dated from November 22, 2022); (b) the automatically identified components of stop-and-go waves as the model output, with each component being independent of the others. The demonstration here highlights the complexity and richness of wave dynamics.}
    \label{fig:overview}
\end{figure}

Moreover, large-scale data collection also reveals a broader range of complex and rich phenomena associated with these waves. The oscillatory dynamics of stop-and-go waves are intricately complex, involving stages of precursor, growth, stable, and decay \citep{chen2014periodicity}. Empirically, the wave propagation speed is consistent \citep{treiber2000congested} and displays a concave growth pattern \citep{tian2016empirical}. Wave patterns, including characteristics such as wave speed and wave duration, are documented and summarized in the literature \citep{edie1967generation, treiterer1974hysteresis, laval2010mechanism}. It has been observed that waves can split \citep{yeo2009understanding}, leading to bifurcation and merging \cite{jin2015understanding} behaviors within wave dynamics. However, these phenomena have not been thoroughly explored, primarily due to the limited observations from empirical data, and existing methods are not designed to effectively capture or analyze bifurcation and merging, particularly in terms of the wave topology. A more detailed review of the characteristics observed in both experimental and empirical data is provided in Section \ref{sec:lit}.  As shown in Figure \ref{fig:overview}a, multiple waves propagate diagonally, showcasing the complexity and richness of dynamic behaviors such as wave bifurcation and merging, which form phenomena resembling a ``river delta" in the space-time diagram. This growing complexity underscores the need for scalable methods to capture the topology of waves, which could significantly accelerate research into the nature of traffic waves. To date, no existing approaches are capable of unifying all observed phenomena to provide a comprehensive understanding of the dynamics of stop-and-go waves, which motivates us to develop a method capable of analyzing these complex phenomena.

\subsection{Research questions and contributions}
\label{sec:problem_state}
Building upon the aforementioned research motivations and challenges, the aim of this paper is to develop a tool to enable stop-and-go wave analysis for massive trajectory data. We address the problem illustrated in Figure \ref{fig:overview}: Given the space-time diagram in Figure \ref{fig:overview}a and a critical speed, (i) can we develop a method to automatically identify all stop-and-go wave boundaries, ensuring that the sets within these boundaries, which fall below the critical speed, are independent of each other as illustrated in Figure \ref{fig:overview}b? (ii) can we report their related wave properties? (iii) can the developed tools provide new insights into the dynamics of stop-and-go waves?

Specifically, it is crucial to identify key spatio-temporal features for stop-and-go wave analysis, including: (i)  the temporal and spatial coordinates marking the start and end of the waves, (ii) the paths of wave propagation, (iii) wave speed and duration, and (iv) the processes of wave bifurcation and merging. To address the research questions, this paper introduces a graph-based method that automatically identifies stop-and-go wave fronts and tails in both space and time, as well as their topology. This approach enables a comprehensive life-cycle analysis of stop-and-go wave characteristics and dynamics at scale.
The contributions of this paper are outlined as follows: 

\begin{enumerate}[label=(\roman*),noitemsep]
    \item We propose a graph representation of stop-and-go waves, capable of capturing the complex traffic wave topological phenomena, including wave generation, propagation, dissipation, bifurcation, and merging.
    \item We develop an automatic approach for identifying and measuring stop-and-go wave boundaries and topologies at scale with implementation to a large-scale dataset.
    \item Our results show insights on the complexity of traffic waves. Traffic waves can bifurcate and merge at a scale that has never been observed or described before. The clustering analysis of all the identified wave components reveals the different topological structures of traffic waves. We explored that the wave merge or bifurcation points can be explained by spatial features. The gallery of all the identified wave topologies is demonstrated at \url{https://trafficwaves.github.io/}.
\end{enumerate}

This paper is organized as follows. Section \ref{sec:lit} presents a literature review on the stop-and-go wave analysis characteristics and its identification techniques. Section \ref{sec:method} outlines the method for modeling stop-and-go waves as a graph, including the use of connected components to represent these complex phenomena. Section \ref{sec:experiment} presents the data utilized in this study, along with the parameters and hyperparameters employed. Section \ref{sec:result} showcases the identified results. \edit{Section \ref{sec:dis} discusses the statistics of traffic wave measurements and explore the insights that graph topological analysis offers to traffic modeling.}  Section \ref{sec:conclusion} summarize the findings and future work.

\section{Literature review}
\label{sec:lit}

\subsection{Stop-and-go wave and its characteristics}
\label{sec:lit1}
Stop-and-go waves, also called traffic oscillations \cite{ahn2005formation,li2010measurement,zheng2011applications}, or wide-moving jammed waves \cite{helbing1999phase,helbing2009theoretical,schonhof2009criticism,treiber2010three,kerner1999physics,kerner2002empirical,kerner2008theory,kerner2013criticism} or shock waves \cite{lu2007freeway,izadpanah2009automatic,yang2023data}, have different terminologies in different research communities \cite{orosz2010traffic}. In this paper, stop-and-go waves are referred to the repeated cycles of deceleration and acceleration engaged by vehicles \cite{laval2010mechanism,zheng2011freeway,li2014stop}. At the very beginning, inductive loop detectors were the main data sources used by researchers to analyze stop-and-go traffic \cite{cassidy1995methodology,mauch2002freeway,coifman2002estimating,bertini2004empirical}. Later, the well-known NGSIM dataset \cite{NGSIM} provided access to trajectory data, enabling the community to investigate wave phenomena on both microscopic and macroscopic scales. This made a significant contribution to the empirical analysis of stop-and-go waves, which was followed by numerous other trajectory data collection efforts worldwide \cite{wan2020spatiotemporal,ztd2018,Talebpour2024,gloudemans202324}. 

Another significant push in understanding stop-and-go waves comes from field experiments, which enable researchers to more effectively discern the causes of wave generation and propagation. \cite{sugiyama2008traffic} conducted a single-lane ring-road experiment in Japan, revealing for the first time that stop-and-go waves can occur due to instabilities in driving behavior, even in the absence of any bottlenecks. \cite{stern2018dissipation} later replicated the experiment under the same conditions and used one automated vehicle to dissipate the traffic waves. A research team in China \cite{jiang2015some,jin2015understanding,zheng2021experimental,zheng2022empirical} also conducted multiple comprehensive field experiments with multiple-vehicle platoon on open road and closed track in different locations, showing wave features in lightly congested traffic \cite{zheng2021experimental}, hyper-congested traffic \cite{jin2015understanding} and the growing pattern of traffic oscillations \cite{zheng2022empirical}. In recent years, there have been numerous studies on field testing automated cruise control vehicles \cite{gunter2020commercially,zhao2020field,makridis2021openacc,li2021car,shi2021empirical,li2022fundamental,yang2024microsimacc,lapardhaja2024unlocking} to gather further insights into how these vehicles impact and control stop-and-go waves.

\begin{table}[ht]
\centering
\caption{\textbf{Summary of stop-and-go waves experimental and empirical trajectory data:} For the experimental data, the size is  described by the size of the platoon, whereas for empirical data, it is characterized by the coverage in time and space. Note that the Treiterer’73 and TGSIM datasets were captured by helicopter and only include a selection of vehicles within the overall traffic flow. Dataset in \textbf{bold} is used in this work.
}
\label{tab:wave_data}
{\scriptsize
\begin{tabular}{lllll}
\toprule
\textbf{Dataset}  & \textbf{Size} & \textbf{Setting} & \textbf{Site}  & \textbf{Sensor} \\ \hline
\multicolumn{5}{c}{\textbf{Experimental dataset}}\\ \hline
Sugiyama'08 \cite{sugiyama2008traffic}
&  22 vehicles
& Ring track
& Japan    
& Camera\\
Nagoya Dome \cite{tadaki2013phase}
&  10 - 40 vehicles
& Ring track
& Japan    
& Camera\\
Jiang'14 \cite{jiang2014traffic}
& 25 vehicles     
& Closed track
& China          
& GPS\\
Jiang'17 \cite{jiang2018experimental}
& 51 vehicles
& Closed track
& China            
& GPS\\
Jiang'18 \cite{huang2018experimental}
& 11 vehicles
& Open freeway
& China            
& GPS\\
Arizona'18~\cite{stern2018dissipation}
& 19 - 22 vehicles     
& Ring track
& United States             
& Camera\\
Arizona'20~\cite{gunter2020commercially}
& 8 vehicles     
& Closed track
& United States             
& GPS\\
CATS'20~\cite{yao2020study}
& 12 vehicles     
& Open track
& China             
& GPS\\
OpenACC N.1~\cite{makridis2021openacc}
& 3 vehicles     
& Off-peak freeway
& Italy       
& GPS\\
OpenACC N.2~\cite{makridis2021openacc}
& 5 vehicles     
& Off-peak freeway
& Italy       
& GPS\\
OpenACC N.3~\cite{makridis2021openacc}
& 5 vehicles     
& Closed track
& Sweden       
& GPS\\
OpenACC N.4~\cite{makridis2021openacc}
& 11 vehicles     
& Closed track
& Hungary   
& GPS\\
MA~\cite{li2021car}
& 3 vehicles     
& Closed track
& United States  
& GPS\\
GA~\cite{li2022fundamental}
& 2 vehicles     
& Closed track
& United States  
& GPS\\
Jiang'21 \cite{zheng2021experimental}
& 40 vehicles
& Ring track
& China            
& GPS\\
\hline
\multicolumn{5}{c}{\textbf{Empirical dataset}}\\ \hline
Treiterer'73 \cite{myers1973experimental,treiterer1974hysteresis}
&  209 vehicles
& Open freeway
& United States 
& Camera\\
Coifman'97 \cite{coifman1997time}
&  0.15hr $\times$ 0.10km
& Open freeway
& United States 
& Camera\\
NGSIM \cite{NGSIM}
&  0.75hr $\times$ 0.64km
& Open freeway
& United States 
& Camera\\
UTE \cite{wan2020spatiotemporal}
& 10 $\times$ 0.15hr $\times$ 0.4km
& Open freeway
& China 
& Camera\\
TGSIM \cite{Talebpour2024}
&  2hr $\times$ 1.3km
& Open freeway
& United States 
& Camera\\
HIGH-SIM \cite{shi2021video}
&  2hr $\times$ 2.44km
& Open freeway
& United States 
& Camera\\
ZTD \cite{seo2020evaluation}
&  5hr $\times$ 2km
& Open freeway
& Japan 
& Camera\\
\edit{MiTra~\cite{chaudhari2025mitra}}
&  2.1hr $\times$ 0.9km
& Open freeway
& Germany
& Camera\\
\textbf{I-24 MOTION} \cite{gloudemans202324}
& \textbf{47hr $\times$ 6.75km}
& \textbf{Open freeway}
&\textbf{United States}
&\textbf{Camera }\\ 
\bottomrule
\end{tabular}}
\end{table} 

 Table \ref{tab:wave_data} summarizes historical experimentation on stop and go waves by setting and data source. The empirical dataset used in this paper stands out due to its extensive scale, covering 47 hours over a 6.75 km stretch on an open freeway in the United States. \textbf{This scale unlocks the potential for analyzing stop-and-go waves at a greater level of detail and comprehensiveness.} 
 
From both the empirical and the experimental data, the stop-and-go waves typically have the life-cycles including generation, propagation, merge and bifurcation, and dissipation \citep{treiber2010three,chen2014periodicity,li2024jam,suh2016empirical}. Below is a summary of the patterns reported in the literature:
\begin{enumerate}[label=(\roman*),noitemsep]
\item \textbf{Generation.} Stop-and-go waves can be generated from various causes, including endogenous reasons like instability \cite{sugiyama2008traffic,laval2010mechanism}, driving behaviour \cite{yeo2009understanding,laval2011hysteresis} and self-organized criticality \cite{laval2023self} and exogenous reasons including fixed bottleneck, moving bottleneck \cite{newell1998moving,gazis1992moving,munoz2002moving}, lane-changing maneuvers \cite{laval2005linking,ahn2007freeway} and other disturbances or perturbations to the system \cite{cooper2009investigation}.
Meta-stability \cite{nakayama2009metastability,tadaki2013phase} is also discussed in relation to wave generation, particularly in terms of how a temporary homogeneous flow with high velocity can occur just before the waves are generated. However, the collection of empirical data is still insufficient to statistically reason through the generation of waves.

\item \textbf{Propagation.} Wave propagation is the most frequently discussed pattern and is the most intuitive to experience during daily driving \cite{trafficwaves}. 
One characteristic of these waves is their propagation against the traffic flow at a relatively constant speed of 10-20 km/hr, with variations depending on location \cite{myers1973experimental,kerner1996experimental,treiber2000congested,zielke2008empirical}. 
The upstream and downstream fronts of congested traffic are discussed separately in \cite{treiber2010three,li2010measurement,zheng2011applications}, and both warrant further investigation. At the microscopic level, the wave speed may vary over time depending on driver behavior \cite{yeo2009understanding}. Another characteristic observed in experimental and empirical data is that the growth of wave propagation exhibits a concave pattern, as measured by the standard deviation of individual vehicle speeds \cite{tian2016empirical,jiang2018experimental,zheng2022empirical,zheng2023comparison}. Other measurements for wave propagation stage include duration and period \cite{zheng2011applications,li2014stop}.

\item \textbf{Merge and bifurcation.} Only a few pieces of literature discuss these phenomena in detail. \cite{jin2015understanding} reported that the structure of hyper-congested traffic may involve small jams merging into larger ones slowly, with larger jams occasionally breaking into smaller ones. The studies by \cite{nishi2013theory,li2024jam} considered the effect of wave-absorbing behavior in traffic control, which may lead to additional wave phenomena like wave bifurcation, this has also been observed in simulations \cite{vishnoi2023cav,li2024jam}. These patterns are often obscured within the internal dynamics of congested traffic, where empirical aggregated traffic measurements tend to average out these phenomena. 

\item \textbf{Dissipation.} Dissipation is a frequently discussed topic in the field of control systems, examining whether stop-and-go waves can be alleviated using advanced technologies such as Connected and Automated Vehicles (CAVs) \cite{stern2018dissipation, lee2025traffic, goatin2024dissipation} or Intelligent Traffic Systems (ITS) \cite{hegyi2005model, hegyi2008specialist, chen2014variable}. These studies investigate the potential of these technologies to smooth traffic flow and minimize the incidence of such waves.
However, from an empirical perspective, the pattern and dynamics surrounding the dissipation of these waves remain insufficiently explored. Analyses often overlook aspects such as the opposite side of the concave pattern observed during the growth phase \cite{tian2016empirical}, due to a lack of observations.
\end{enumerate}

In summary, there is consensus that stop-and-go waves can be generated with or without explicit disturbances, and once initiated, they propagate at a relatively consistent speed opposite to the direction of traffic flow. However, the more intricate patterns within these waves require further exploration and detailed empirical data observations.

\subsection{Stop-and-go waves identification techniques}
As summarized in Section \ref{sec:lit1}, wave generation and propagation are the most frequently discussed patterns in the literature. Consequently, most identification methods are predicated on these phenomena. Prior to the availability of trajectory data, the majority of these methods were developed using fixed sensor data as input \cite{edie1967generation,kerner2004recognition,zielke2008empirical,li2010measurement,raghavan2023identifying}, as summarized in Table \ref{tab:methods}. These models often assume that waves propagate at a constant speed, which constrains the understanding of the dynamics of stop-and-go waves. Since the release of NGSIM data, there has been a surge in the development of methods to analyze vehicle trajectory data. These methods \cite{lu2007freeway,yeo2009understanding,zheng2011applications,zheng2011freeway,li2012prediction,li2014stop, pu2025online} have expanded the capabilities of researchers to study traffic dynamics in greater detail. In addition to detailed trajectory data, recent research has also explored the use of sparse connected vehicle trajectories to identify stop-and-go waves \cite{sakhare2023methodology,zhang2024time,pu2025online}. This approach leverages limited but strategically connected data points from vehicles, offering a promising method for understanding and managing traffic dynamics with less comprehensive datasets.

\begin{table}[H]
  \centering
  \caption{\textbf{Summary of stop-and-go waves identification methods}: sorted by input data type (fixed sensor data or trajectory data). \textbf{Topology} indicates whether the method considers a graph-like topology for wave characterization, \textbf{Code} indicates whether code for reproducing the method is open-sourced, and \textbf{Phenomena} indicates which of the identified wave phenomena from Section \ref{sec:lit1} are considered by the method (G for generation, P for propagation, M for merge, B for bifurcation, and D for dissipation.} 
  \label{tab:methods}
  \renewcommand{\arraystretch}{1} 
  {\scriptsize
  \begin{tabularx}{0.65\textwidth}{@{}lccl@{}} 
    \toprule
    \textbf{Method} & \textbf{Topology} &\textbf{Code} &\textbf{Phenomena} \\
    \midrule
    \textbf{Fixed sensor data} \\ \hline
    \makecell[l]{Speed contour line~\citep{edie1967generation}}& No & No  & G,~P \\
    \makecell[l]{ASDA~\citep{kerner2004recognition}} & No & Yes  & G,~P,~D \\
    \makecell[l]{Cross-correlation~\citep{zielke2008empirical}}& No & No  & P \\
    \makecell[l]{Fourier Transform~\citep{li2010measurement}}& No & No  & P \\
    \makecell[l]{Clustering~\citep{nguyen2019feature}} & No & No  & G,~P,~D \\
    \midrule
    \textbf{Trajectory data} \\ \hline
    \makecell[l]{Speed contour line~\citep{myers1973experimental}} & No & No  & G,~P \\
    \makecell[l]{Clustering~\citep{lu2007freeway,yang2023data,tgsimwave}}& No & No  & P \\
    \makecell[l]{Wavelet Transform~\citep{zheng2011applications,zheng2011freeway}}& No & No  & G,~P \\
    \makecell[l]{Trajectory decomposition~\citep{li2012prediction,li2014stop}}& No & No  & G,~P \\
    \makecell[l]{Dynamic time warping~\citep{zheng2023parsimonious}}& No & No  & P \\
    \midrule
    Ours &Yes & Yes &G,~P,~M,~B,~D \\
    \bottomrule
  \end{tabularx}}
\end{table}

On the identification method, the central challenge lies in identifying the areas of interest across both spatial and temporal dimensions. Table \ref{tab:methods} summarizes the method used to identify the stop-and-go waves. Using speed contour lines \cite{edie1967generation,myers1973experimental} is the most straightforward approach for analyzing stop-and-go waves; however, it is inadequate for distinguishing the individual structures of each wave. Time-series data analysis \cite{zielke2008empirical,zheng2023parsimonious} within the \textit{time domain} for each spatial or vehicle observation is another method to measure stop-and-go waves. Another mainstream method is to transform the data to the \textit{frequency domain}, which allows for the separation of low-frequency and high-frequency signals \cite{li2010measurement,zheng2011applications,zheng2011freeway}, where the Wavelet Transform is viewed as the best of practice. More recent studies~\citep{nguyen2024interpretable} leverage graph-based representations to provide insights into congestion patterns. Moreover, the aforementioned methods primarily focus on wave generation and propagation, often neglecting the phenomena of wave merging and bifurcation. \cite{yeo2009understanding} provided a detailed behavioral analysis using asymmetric traffic theory, illustrating the wave propagation path across both space and time. This analysis revealed that the path of acceleration curves may split, potentially generating a secondary wave. This highlights the need for methods that not only detect and track waves but also \textit{stitch} them together over time to support the full life-cycle wave analysis.

\subsection{Emerging large-scale trajectory data}
As stated by \cite{li2020trajectory}, \textit{more trajectory data are needed}. In recent years, more trajectory data collection efforts comes from drones \cite{krajewski2018highd,barmpounakis2020new,wan2020spatiotemporal,ma2022magic,zheng2024citysim,Talebpour2024}, helicopter \cite{Talebpour2024}, and roadside infrastructures \cite{ztd2018,zhang2022design,shi2021video,gloudemans202324}. As more data is generated, the need for tools to analyze this data and enhance research reproducibility within the community becomes more evident \cite{zheng2021reasons}. Additionally, with the continuous collection of data, the size of the datasets has become significantly larger than previous ones, adding more challenges for researchers to address \cite{zheng2015trajectory}. Moreover, these datasets may uncover previously unobserved scenarios, such as the effects of automated control vehicles \cite{Talebpour2024}, traffic smoothing strategies \cite{lee2025traffic,nice2024middle}, electric vehicles \cite{lapardhaja2024unlocking}, and traffic incidents \cite{wang2024surrogate}. \textit{While generating more data is a crucial first step, the development of more analytical tools is equally paramount.} The majority of existing tools do not open-source their analysis software (shown in Table \ref{tab:methods}), which potentially limits subsequent analyses on different datasets. This paper aims to fill this gap by exploring the analysis of massive trajectory data to make stop-and-go wave analysis at scale possible.
\section{Terminology}
We define the terminology for key spatio-temporal features of stop-and-go waves as follows:
\begin{figure}
    \centering
    \includegraphics[width=\linewidth]{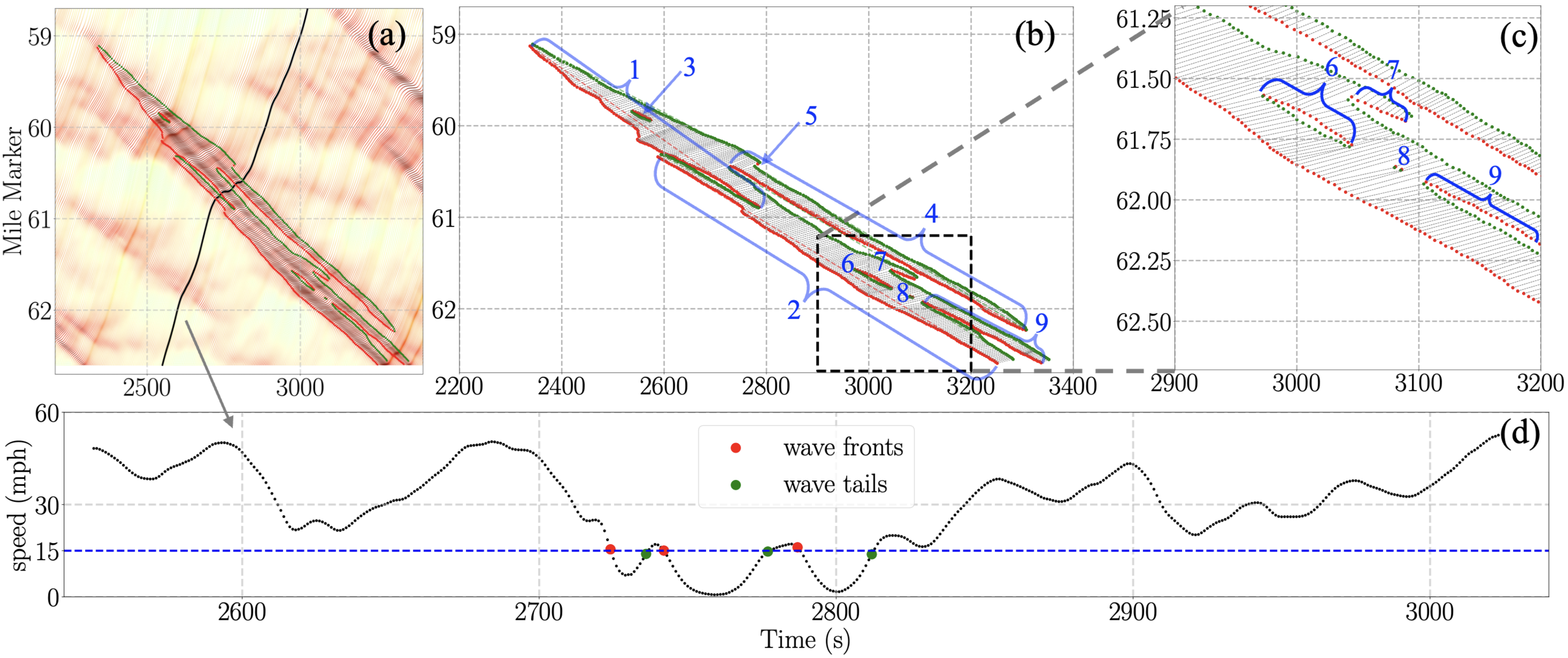}
    \caption{\textbf{Terminology in this paper:} (a) a wave boundary is illustrated on the space-time diagram, where red dots indicate wave fronts and green dots represent wave tails; (b) the wave boundary composed of 9 wave front paths and 9 wave tail paths is shown, with individual wave front paths manually labeled for clarity; (c) a zoomed-in view of wave front paths 6 to 9 is provided for closer inspection; (d) the wave fronts and tails along a vehicle trajectory in a speed-time diagram is demonstrated, highlighting the critical timestamps when the vehicle trajectory encounters the wave fronts and tails.}
    \label{fig:terminology}
\end{figure}
\begin{enumerate}[label=(\roman*),noitemsep]
    \item \textbf{Wave boundary:} defined as the boundary encompassing spatio-temporal points where the speed falls below the critical threshold $v_c$. As demonstrated in Figure~\ref{fig:terminology}a, it depicts a single wave boundary.
    \item \textbf{Wave front points:} defined as the spatio-temporal points where the speed of a vehicle decreases down to the critical speed $v_c$ from the high speed range. As demonstrated in Figure~\ref{fig:terminology}d, the red dots are the wave front points.
    \item \textbf{Wave tail points:} defined as the spatio-temporal points where the speed increases up to critical speed $v_c$ from the low speed range. As demonstrated in Figure~\ref{fig:terminology}d, the green dots are the wave tail points.
    \item \textbf{Wave front:} defined as the continuous trajectories formed by connecting adjacent wave fronts as they propagate in the space-time diagram. Figure~\ref{fig:terminology}c manually labels each wave front within the demonstrated wave boundary in the space-time diagram.
    \item \textbf{Wave tail:} defined as the continuous trajectories formed by connecting adjacent wave tails as they propagate in the space-time diagram.
\end{enumerate}

\textbf{Note:} (i) Multiple wave boundaries may exist within a single space-time diagram, with each boundary being distinct and separate from the others. (ii) A single wave boundary can consist of multiple wave fronts and wave tails as illustrated in Figure~\ref{fig:terminology}b and Figure~\ref{fig:terminology}c.

Given these definitions, multiple methods exist which would produce roughly the same wave boundaries, fronts and tails. In the next section, we present one such method to identify the above wave features. Our claim is not that this is the most computationally efficient or best method; rather, it is a simple and fairly intuitive approach, and a first attempt at providing an automatic and scalable method for wave characterization in response to a flood of emerging large-scale traffic data.

\section{Methodology}
\label{sec:method}
The problem addressed in this paper is stated as follows: Given an arbitrary critical speed $v_c$, identify the wave fronts, wave tails, and wave boundaries as defined above. In this paper, we propose a method for identifying stop-and-go waves using trajectory data. The stop-and-go waves are then modeled as a graph, incorporating the definitions of nodes, edges, and connected components. Table \ref{tab:variables} provides a comprehensive list of the variables and parameters used throughout the paper.

\begin{table}[ht]
\caption{\textbf{Variables and parameters defined and utilized in the paper}: the parameter column indicates whether it is a parameter in the paper.}
\centering
{\scriptsize
\begin{tabular}{clcc}
\toprule
\textbf{Notation} & \textbf{Description} & \textbf{Unit}& \textbf{Parameter?} \\ \hline
$N$ &  the number of vehicle trajectories & -  & -\\
$T$ & the set of all $N$ trajectories & - & - \\
$v_c$    &  the critical speed threshold  & mph  & Yes \\
$\vec{\tau}_i$    &  vector representing the trajectory of vehicle $i$  & -  & -\\
$\vec{\tau}_i^c$    &  vector representing the stationary component trajectory of vehicle $i$  & -  & -\\
$\vec{\tau}_i^{osc}$    &  vector representing the oscillation component trajectory of vehicle $i$  & -  & -\\
$M_o$         & the start point of the dataset extents & miles & - \\
\hline
$\mathcal{G}$ &  the stop-and-go graph & -  & -\\
$\mathcal{V}$ &  the nodes in the stop-and-go graph & -  & -\\
$\mathcal{E}$ &  the edges in the stop-and-go graph & -  & -\\
$\mathcal{D}$&  wave fronts set for all vehicles & -  & -\\
$\mathcal{A}$&  wave tails set for all vehicles & -  & -\\
$d_i$ &  wave fronts set for vehicle $i$ & -  & -\\
$a_i$ &  wave tails set for vehicle $i$ & -  & -\\
$K_i$ &  the number of wave fronts and tails for vehicle $i$ & -  & -\\
$\mathcal{E_\text{inner}}$ &  the inner edges in the stop-and-go graph & -  & -\\
$\mathcal{E_\text{cross}}$ &  the cross edges in the stop-and-go graph & -  & -\\
$\mathcal{S(\cdot)}$ & the search function for the cross edges for a given node  & -  & Yes\\
$\mathcal{G}_d$ &  the wave fronts graph & -  & -\\
$\mathcal{G}_a$ &  the wave tails graph & -  & -\\
$N_c$ & the number of connected components for graph $\mathcal{G}$  & -  & -\\
$\mathcal{C}^d$ & the components for the wave fronts graph $\mathcal{G}_d$  & -  & -\\
$\mathcal{C}^a$ & the components for the wave tails graph $\mathcal{G}_a$  & -  & -\\
\bottomrule
\end{tabular}
}
\label{tab:variables}
\end{table}

\subsection{Preliminaries}
Long-range raw vehicle trajectory data as from \cite{gloudemans202324} is imperfect due to the challenges of tracking and re-identification in dense traffic conditions \cite{gloudemans2024so,wang2024automatic}. In essence, it is difficult to track vehicles perfectly across hundreds of cameras and through occlusions \cite{coifman2024partial}, so trajectory datasets over large spatial ranges often contain fragmented trajectories. These issues do not significantly impact local traffic measurements, such as traffic speeds, which can still be obtained with extremely high fidelity. We prepare raw data for the methods described next by utilizing the method from~\cite{ji2024virtual}, which we describe in more detail in~\ref{app:virtual}. The resulting virtual trajectories have no passing or lane change maneuvers. This allows efficient search between adjacent trajectories, but the methods described next could also be applied to raw trajectory data with lane-change maneuvers with slight modifications.

\subsection{Identifying the stop-and-go waves: the trajectory perspective}
\label{sec:def1}
\begin{figure}[H]
    \centering    \includegraphics[width=\textwidth]{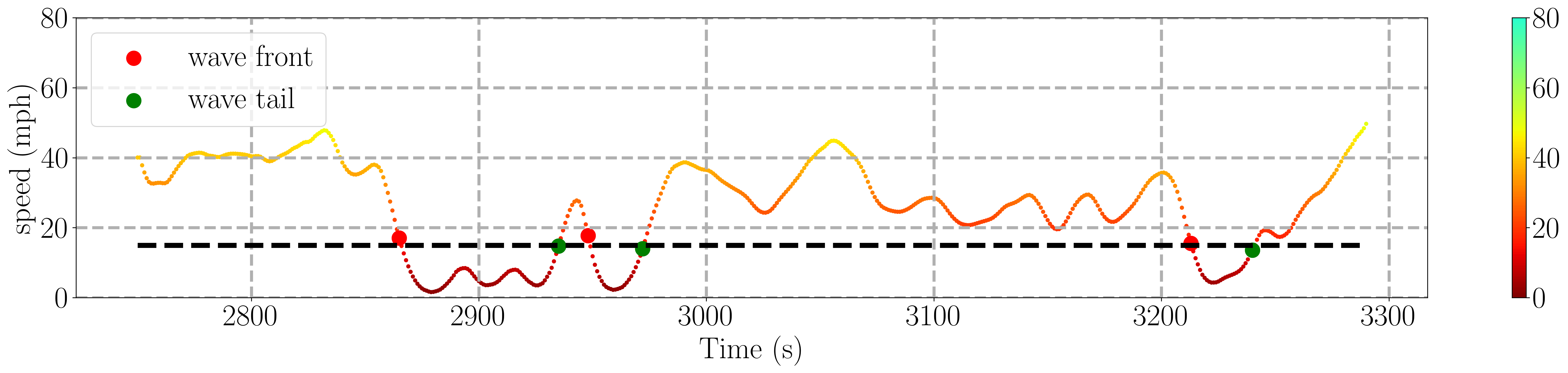}
    \caption{\textbf{Demonstration of the  stop-and-go waves:} The speed of a single trajectory on the left-most lane generated from the I-24 MOTION INCEPTION dataset (dated November 22, 2022) is plotted over time. Red and green dots indicate the wave fronts and tails, respectively. The time axis references the time in seconds from 6 AM. This analysis captures 3 distinct stop-and-go waves over a 4-mile section for a critical speed $v_c$ = 15 mph (represented by black dashed line) Each pair of front and tail dots defines the boundary of a wave for further analysis.}
    \label{fig:wave_demo}
\end{figure}

From the perspective of a vehicle trajectory, stop-and-go waves can be intuitively understood as a sequence of deceleration and acceleration events \cite{yeo2009understanding,zheng2011applications}. This pattern is discernible from the speed time-series profile of each individual vehicle (see Figure \ref{fig:wave_demo}). By analyzing these profiles, we can observe the fluctuations in speed that characterize stop-and-go waves, providing a clear visualization of how vehicles repeatedly slow down and speed up within the traffic flow. 

\edit{
The identification process involves decomposing an empirical trajectory into two components~\cite{li2014stop}, followed by detecting local maxima that represent wave fronts and minima that represent wave tails within the oscillation component. A flowchart with visual illustration of this process is provided in Fig.~\ref{fig:flowchart}.

\begin{figure}
    \centering
    \includegraphics[width=\linewidth]{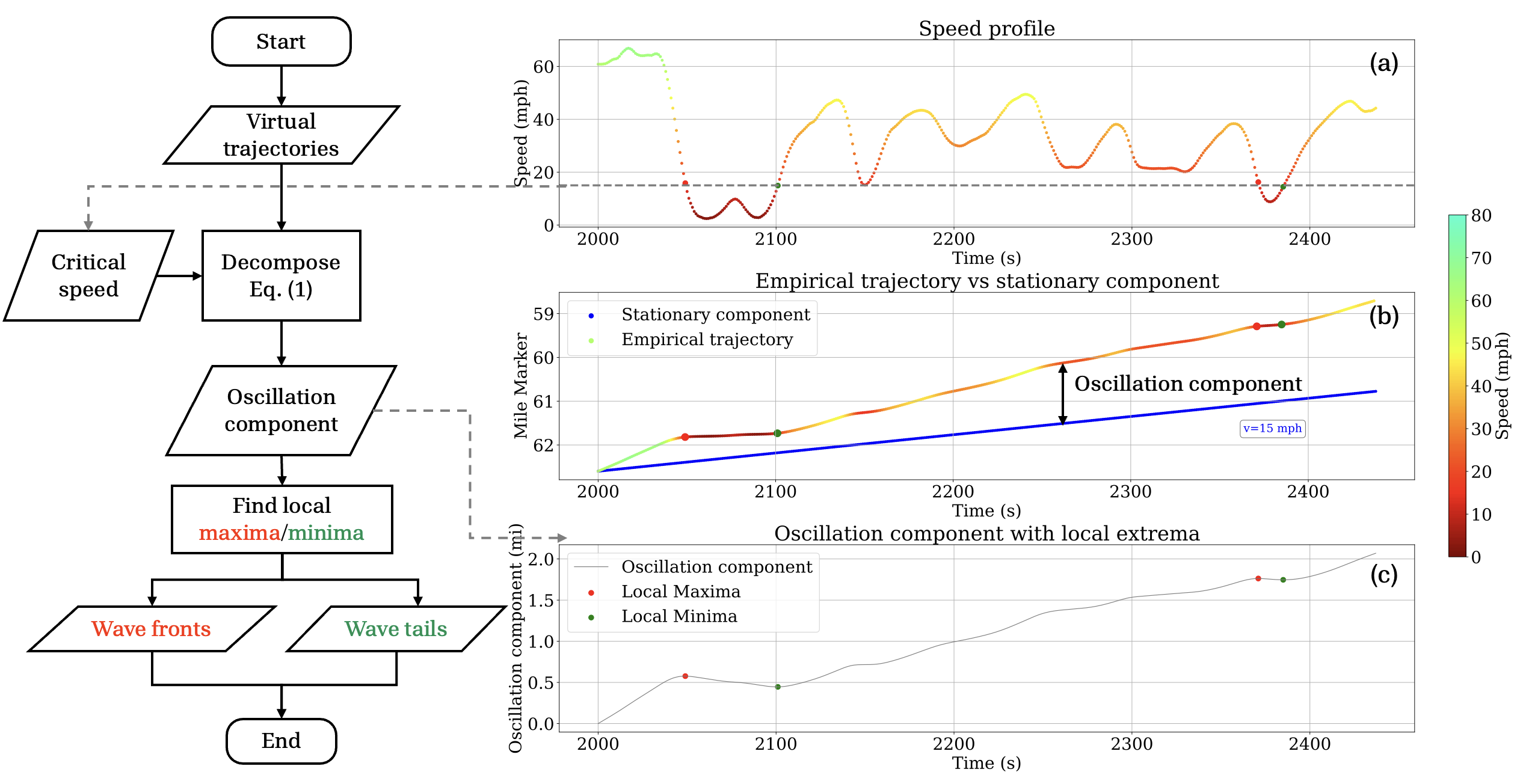}
    \caption{\textbf{Flowchart (left) and visual demonstration (right) of wave identification for each vehicle trajectory:} (a) speed profile with critical-speed threshold; (b) empirical vs. stationary component trajectory with vertical offset defining the oscillation component; (c) oscillation component time series with detected extrema. }
    \label{fig:flowchart}
\end{figure}
}

Let $\mathcal{T}$ be the set of all trajectories. Note that the preprocessing in \ref{app:virtual} ensures that trajectories are ordered by increasing time, and no trajectories enter or exit the considered lane. Let $\vec{\tau}_i \in \mathcal{T}$ represent the trajectory of vehicle $i$ as a vector of time-space points.  Throughout this section, we will use $\tau_i(t)$ to represent a continuous, functional representation of the trajectory of vehicle $i$ from which the points in $\vec{\tau}_i$ are sampled. To detect the critical events from the trajectories, each trajectory is decomposed into two components as follows:

\begin{align}
    \tau_i(t) = \tau_i^{c}(t) + \tau_i^{osc}(t),  
\end{align}
where $\tau_i^{c}(t)$ represents the stationary component  and $\tau_i^{osc}(t)$ denotes the oscillation component, as introduced and described by \cite{li2014stop}. Here we define the stationary component $\tau_i^{c}(c)$ as uniform motion at a constant speed $v_c$: 

\begin{equation}
    \tau_i^{c}(t) = M_o + v_c \cdot t,
\end{equation}

with the initial position of the vehicle $i$ set to $M_o$, the start of the dataset's observation window. Differentiating equation \ref{eq:decomp} with respect to time yields:

\begin{equation}
    \dot{\tau_i}(t) = v_c + \dot{\tau}_i^{\text{osc}}(t),
    \label{eq:decomp}
\end{equation}

By definition, a wave front or tail occurs when trajectory velocity $\dot{\tau}_i(t) = v_c$, or when $\dot{\tau}_i^{\text{osc}}(t) = 0$. Thus, wave front and tail points correspond to a local maximum or minimum of $\tau_i^{osc}(t)$, respectively.

\edit{
\begin{itemize}
\item \textbf{Wave front points}. The wave front is associated with a local maximum of $\tau_i^{\text{osc}}(t)$, denoted by $t^{\text{front}} = t_{\text{local}}^{\text{max}}$.
\item \textbf{Wave tail points}. The wave tail is associated with a local minimum of $\tau_i^{\text{osc}}(t)$, denoted by $t^{\text{tail}} = t_{\text{local}}^{\text{min}}$.
\end{itemize}
In our implementation, the local maxima and minima of $\tau_i^{\text{osc}}(t)$ are identified using peak-finding algorithms \texttt{scipy.signal.find\_{peaks}}~\cite{2020SciPy-NMeth}.
}  We find all local maxima and minima in $\vec{\tau}_i^{osc}$, yielding a set $d_i$ of wave front and a set $a_i$ of wave tail points for trajectory $i$.
We keep only those wave fronts that are followed by a wave tail and those wave tails that are preceded by a wave front, excluding any wave fronts or tails that occur at the boundaries of the trajectories, as they do not contribute to a complete stop-and-go cycle. This ensures that there are the same number $K_i$ of wave front points and wave tail points for trajectory $i$. This process is repeated for all trajectories in $\mathcal{T}$, yielding an overall set of wave front and tail points $\mathcal{D} = \{d_0,d_1, ... d_i\}$ and $\mathcal{A} = \{a_0,a_1, ... a_i\}$.

\subsection{Representing stop-and-go waves as a graph}
Identified wave front and tail points correspond to acceleration and deceleration events experienced by individual vehicles. To be useful for many traffic analyses, these events must be associated \textit{across} vehicle trajectories. To accomplish this, we model stop-and-go waves as an undirected graph with the detected wave tails and wave fronts describe in Section \ref{sec:def1} as graph nodes. The construction of this graph is detailed next.
\subsubsection{Nodes}
Each node in the graph represents a critical event detected from the vehicle trajectories, in this case is the wave front or the wave tail. Nodes are defined in two sets based on whether they represent wave fronts or wave tails.

\begin{itemize}
    \item \textbf{Wave fronts set} $\mathcal{D}$. Let $\mathcal{D} = \{ d_{i} \}_{i=1}^{N}$, and $d_i = \{d_{i,k_i}\}_{k_i=1}^{K_i}$ where $i$ is the index of the $i$-th trajectory, and $k_i$ is the index of the $k$-th wave the vehicle $i$ passed. The coordinates representing the spatial and temporal information, as well as the index information for the node are defined as  $d_{i,k_i} = (t^\text{front}_{i,k_i}, \tau_i(t^\text{front}_{i,k_i}), i) = (t^\text{front}_{i,k_i}, s^\text{front}_{i,k_i}, i)$.  
    \item \textbf{Wave tails set} $\mathcal{A}$. Similarly, $\mathcal{A} = \{ a_{i} \}_{i=1}^{N}$, and $a_i = \{a_{i,k_i}\}_{k_i=1}^{K_i}$. The coordinates representing the spatial and temporal information, as well as the index information for the node are defined as  $a_{i,k_i} = (t^\text{tail}_{i,k_i}, \tau_i(t^\text{tail}_{i,k_i}), i)  = (t^\text{tail}_{i,k_i}, s^\text{tail}_{i,k_i}, i)$.
\end{itemize}
Note that there is a one-to-one correspondence between the elements of the two sets $\mathcal{D}$ and $\mathcal{A}$, which satisfies the following property:
\begin{equation}
    |{d_i}| = |a_i| = K_i      \: \: \:\:\:    \forall i \in \{0...N\}.
\end{equation}


\subsubsection{Edges}

\begin{figure}
    \centering
    \includegraphics[width=\textwidth]{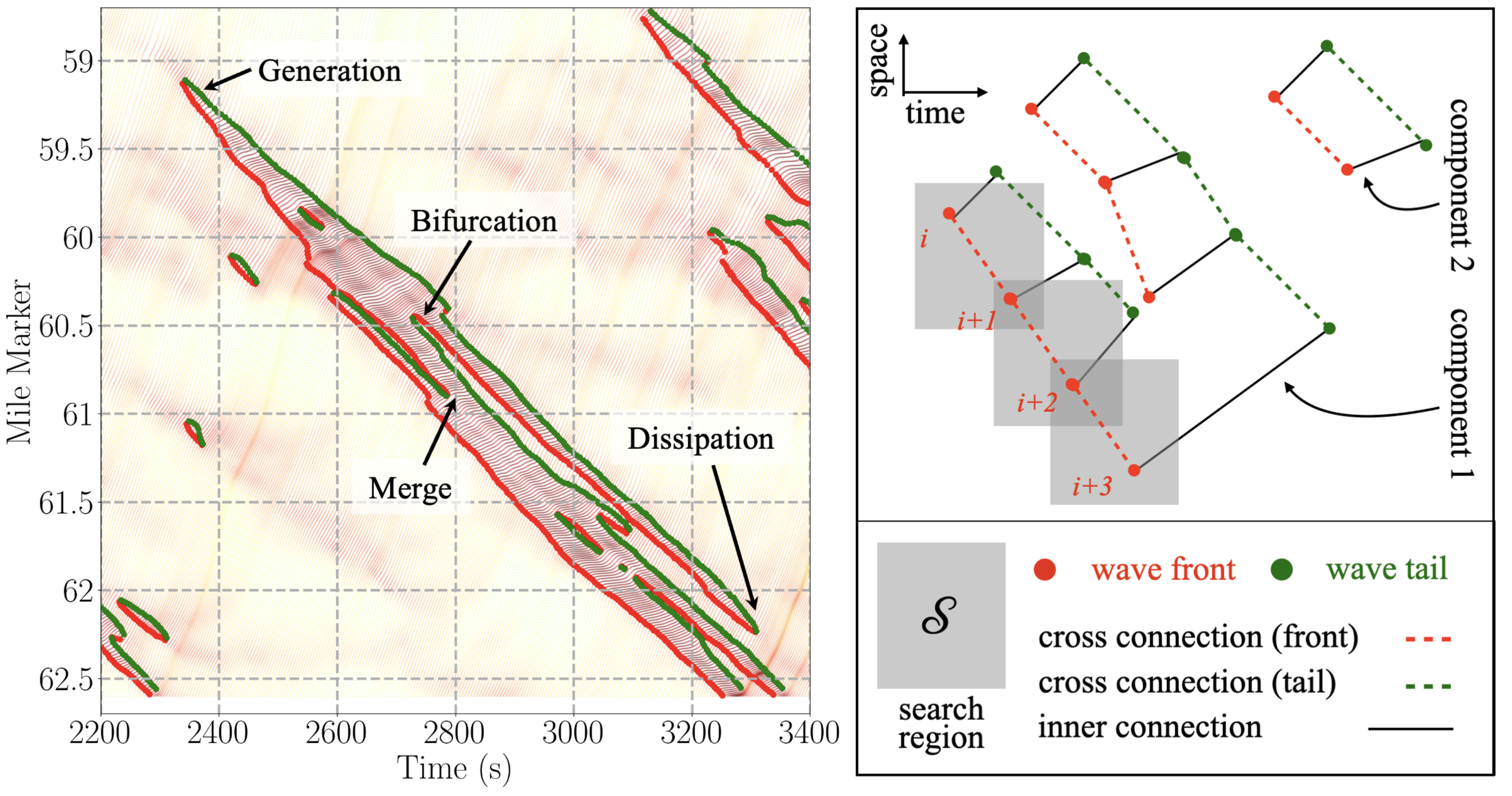}
    \caption{\textbf{Wave identification method capturing complex wave behaviors:} (a) complex phenomena observed visualized by the space-time diagram generated from trajectories (same data as in Figure \ref{fig:wave_demo}); (b) example demonstrating $N_c^d = 3$ connected components for the wave fronts $\mathcal{G}_d$, $N_c^a = 3$ connected components for the wave tails $\mathcal{G}_a$, and $N_c = 2$ connected components for the graph $\mathcal{G}$. }
\label{fig:wave_case} 
\end{figure}

Next, we describe the process for adding edges to the stop-and-go wave graph such that when finished, all connected wave fronts and wave tails correspond to a single distinct traffic wave. Two types of edges are considered to connect nodes in the stop-and-go-wave graph. i). \textbf{Inner edges} connect wave fronts and tails for the same trajectory, based on the intuition that each wave front and tail are added to the graph as a pair and by definition correspond to the same stop and go wave. ii.) \textbf{Cross edges} connect either wave fronts or wave tails for nearby trajectories, based on the idea two decelerations or accelerations experienced by two trajectories in close spatio-temporal proximity are caused by the same stop-and-go wave. This framework facilitates the analysis of both individual trajectory patterns and the interactions between different trajectories  \citep{yeo2009understanding,laval2010mechanism}.
\begin{itemize}
\item  \textbf{Inner edges}: For each stop-and-go cycle $k_i$ in each trajectory $i$, the wave front node $d_{i,k_i}$ is connected to its corresponding (directly subsequent) wave tail node $a_{i,k_i}$. The connection is illustrated by the black line connecting red and green dots in Figure \ref{fig:wave_case}b. The set of inner edges $\mathcal{E}_{\text{inner}}$ is defined as follows: 
\begin{equation}
\mathcal{E}_{\text{inner}} = \bigcup_{i=1}^{N}\bigcup_{k_i=1}^{K_i} \{ \{d_{i,k_i}, a_{i,k_i}\} \}.
\label{eq:inner_edges}
\end{equation}

\item \textbf{Cross edges} $\mathcal{E}_{\text{cross}}$. The cross edges encode the insights from Newell's car-following model \cite{newell2002simplified}, where adjacent trajectories are influenced by leading vehicles. To define the cross edges, we connect the wave fronts and wave tails separately, resulting in two distinct sets of edges: $\mathcal{E}_{\text{cross}}^\mathcal{D}$ and $\mathcal{E}_{\text{cross}}^\mathcal{A}$. The total set of cross edges is then the union of these two sets:
\begin{equation}
\mathcal{E}_{\text{cross}} = \mathcal{E}_{\text{cross}}^{\mathcal{D}} \cup \mathcal{E}_{\text{cross}}^{\mathcal{A}},
\end{equation}
where the connection for wave front is illustrated by the red dashed lines connecting adjacent red dots in \ref{fig:wave_case}b, green for wave tails. \textbf{Succinctly, for each wave front point, we search in a narrow spatio-temporal box in time and in space around that point for other wave front points identified on the next upstream trajectory. If any are found, we add an edge to $\mathcal{E}_{\text{cross}}^{\mathcal{D}}$.} If there are multiple points found, it will connect to the one closest in time. The same is then done for each wave tail point. Mathematically, we define these sets of edges connecting wave fronts ($\mathcal{E}_{\text{cross}}^\mathcal{D}$) and tails ($\mathcal{E}_{\text{cross}}^\mathcal{A}$) as:
\begin{align}
\mathcal{E}_{\text{cross}}^\mathcal{D} = & \bigcup_{i=1}^{N-1} \bigcup_{k_i=1}^{K_i} \left\{ \left\{ d_{i,k_i}, \arg\min_{d_{j,m_{j}} \in \mathcal{S}(d_{i,k_i})} \left| t_{i,k_i}^{\text{front}} - t_{j,m_{j}}^{\text{front}} \right| \right\} \mid \mathcal{S}(d_{i,k_i}) \neq \emptyset , j=i+1\right\}, \label{eq:edge-cross}\\
\mathcal{E}_{\text{cross}}^\mathcal{A} =    &   \bigcup_{i=1}^{N-1} \bigcup_{k_i=1}^{K_i} \left\{ \left\{ a_{i,k_i}, \arg\min_{a_{j,m_{j}} \in \mathcal{S}(a_{i,k_i})} \left| t_{i,k_i}^{\text{tail}} - t_{j,m_{j}}^{\text{tail}} \right| \right\} \mid \mathcal{S}(a_{i,k_i}) \neq \emptyset, j=i+1 \right\},
\end{align}
where the $\mathcal{S}(d_{i,k_i})$ is a search function that is input a wave front point $d_{i,k_i}$ for trajectory $i$, and returns a set of wave fronts for trajectory $i+1$ within a spatio-temporal neighborhood. If the set is empty, then no connection is established. The same logic applies to wave tails edge connection $\mathcal{E}_{\text{cross}}^\mathcal{A}$. The neighborhood defined in this paper is a rectangular region in both space and time, as illustrated in Figure \ref{fig:wave_case}b. The search function within this neighborhood is crucial for filtering out non-physical wave propagation speeds. (Note that to modify this search for trajectory data with lane change maneuvers, the set of all wave front points within the spatio-temporal neighborhood should be considered (not just those from trajectory $i+1$.)
\end{itemize}

\subsubsection{Graphs}
With all the nodes and edges defined, the stop-and-go graph can be defined as $\mathcal{G} = (\mathcal{V}, \mathcal{E})$ where $\mathcal{V} = \mathcal{D} \cup \mathcal{A}$.  Other than the stop-and-go graph $\mathcal{G}$, the wave front graph $\mathcal{G}_d$ and wave tail graph $\mathcal{G}_a$ can also be defined as follows:
\begin{align}
    \mathcal{G}_d & = (\mathcal{D}, \mathcal{E}_{\text{cross}}^{\mathcal{D}}),\\
    \mathcal{G}_a & = (\mathcal{A}, \mathcal{E}_{\text{cross}}^{\mathcal{A}}).
\end{align}

\subsubsection{Components}
\label{sec:components}

Next we consider the independently connected components \cite{siek2001boost} present in $\mathcal{G}_d$, $\mathcal{G}_a$, and $\mathcal{G}$. For the graph $\mathcal{G}_d$ and $\mathcal{G}_a$, the number of connected components are denoted as $N_c^d$ and $N_c^a$ respectively. Figure \ref{fig:wave_case}b presents a simplified example, demonstrating the connected components for $\mathcal{G}_d$, $\mathcal{G}_a$ and $\mathcal{G}$ that do not share any nodes or edges. In this example, there are $N_c^d = 3$ connected components for $\mathcal{G}_d$, $N_c^a = 3$ connected components for $\mathcal{G}_a$, and $N_c = 2$ connected components for $\mathcal{G}$. To generalize the physical meaning of the components in each graph, we can describe them as follows:
\begin{enumerate}[label=(\roman*),noitemsep]
    \item \textbf{Components in $\mathcal{G}_d$}: Each component $C_m^d$ in the wave front graph corresponds to the paths of wave fronts that move across vehicles, interpreted as the trajectories of these wave fronts. From $C_m^d$, one can coherently define the wave front propagation time ${W_T}^d_m$, distance ${W_D}^d_m$ and average speed ${W_S}^d_m$, which are defined as follows:
    \begin{align}
        {W_T}^d_m &= \max(t_c) - \min(t_c) \quad \text{for all } c \in C_m^d, \\
        {W_D}^d_m &= \max(s_c) - \min(s_c) \quad \text{for all } c \in C_m^d, \label{eq:wave_distance} \\
        {W_S}^d_m &= \frac{{W_D}^d_m}{{W_T}^d_m}, \label{eq:wave_speed}\\
        {R^2}^d_m & = RS(C_m^d),
    \end{align}
    where the propagation average speed is estimated by a linear regression given a set of nodes in $C_m^d$. The functions $RS(\cdot)$ utilize the spatial and temporal information from the nodes in $C_m^d$ to perform a linear regression, yielding the $R^2$ value from the data points.
    
    \item \textbf{Components in $\mathcal{G}_a$}: Each component $C_m^a$ in the wave front graph represents the tracks of wave fronts that propagate across vehicles, which can be interpreted as the trajectory of these wave fronts. Similar definitions for the wave tail propagation time ${W_T}^a_m$, distance ${W_D}^a_m$ and average speed ${W_S}^a_m$ can be formulated in a manner analogous to those for the wave front.   
\end{enumerate}

To identify the components in the graphs $\mathcal{G}_d$ and $\mathcal{G}_a$, the Breadth-First Search (BFS) algorithm \cite{beamer2013direction} is applied (see \ref{app:bfs}). The process of identifying components in $\mathcal{G}_d$ serves as an example.

\begin{enumerate}[label=(\roman*),noitemsep,start=3]
\item \textbf{Components in $\mathcal{G}$}: The components in graph $\mathcal{G}$ can be understood as individual stop-and-go wave boundaries that evolve independently. These wave components do not interact with other components, meaning they do not experience merging or bifurcation events with other wave elements. However, within each component, wave fronts and tails may merge and bifurcate as shown in Figure \ref{fig:wave_case}a. These components are similarly identified via breadth-first graph traversal. See \ref{app:bfs} for full algorithm details.
\end{enumerate}

\section{Data}
\label{sec:experiment}
I-24 MOTION is a traffic instrument for the freeway traffic observation \cite{zachar2023visualization,gloudemans202324,wang2024automatic}, which is designed for the continuous freeway traffic data collection and analysis. It is expected to generate 200 million vehicle miles traveled (VMT) in trajectory data annually \cite{wang2024automatic}, which poses significant challenges for data analysis. In this paper, the I-24 MOTION INCEPTION data released by \cite{gloudemans202324} is used for the stop-and-go wave analysis with the proposed method in Section \ref{sec:method}. Table \ref{tab:data} summarizes the data used in this paper, including the corresponding week day of the date, vehicle miles traveled (VMT) and the mean speed (MS). The method input consists of virtual trajectories generated from the mean speed field, designed to address data quality issues \cite{gloudemans2024so} in individual trajectories and enable scalable measurements since traffic waves are the phenomena in between microscopic and macroscopic. Further details are provided in \ref{app:virtual}.
\begin{table}[H]
\caption{\textbf{Dates of data analyzed in this paper}: the table summarizes the date, day, the number of trajectories, showing the scale of the trajectory data
}
\centering
{\scriptsize
\begin{tabular}{clcc}
\toprule
\textbf{Date} & \textbf{Day}  & \textbf{VMT (miles)}  & \textbf{MS (mph)}\\ \hline
2022-11-22   &  Tuesday & 75041 & 36.55\\
2022-11-28    & Monday  &74775 & 36.82\\
2022-11-29    & Tuesday&66134  & 27.33  \\
2022-11-30    & Wednesday&65124 & 27.38 \\
2022-12-01   & Thursday& 68587& 29.60 \\
2022-12-02    & Friday& 71515& 38.49 \\
\bottomrule
\end{tabular}
}
\label{tab:data}
\end{table}

The hyper-parameter in this paper is the critical speed $v_c$ and is examined at various values: $\{1, 5, 10, 15, 20, 25, 30, 35, 40\}$ mph. The search region $\mathcal{S}$ as illustrated in Figure \ref{fig:wave_case}, is defined as a rectangular area centered around a given spatio-temporal feature, extending from -0.02 miles to 0.05 miles spatially and from -5 seconds to 15 seconds temporally. In this context, the negative direction in space refers to the opposite direction of traffic flow, while negative values in time indicate past moments relative to the spatio-temporal feature. The 15-second section is chosen based on the 5-second interval for sending virtual vehicles to the field, as outlined in \ref{app:virtual}. The selected size should neither be excessively large nor too small; the criterion is to ensure adequate coverage of the next vehicle’s adjacent critical spatio-temporal point (either the front or tail). The design of looking backward in time and space is to account for specific cases where waves are stationary \cite{schonhof2007empirical} or when drivers engage in preventive driving, resulting in braking points occurring even earlier than those of the preceding vehicle.

\section{Results: Wave boundaries identification}
\label{sec:result}
\subsection{Demonstration}
\begin{figure}[H]
    \begin{subfigure}[b]{0.49\linewidth}
        \centering
        \includegraphics[width=\linewidth]{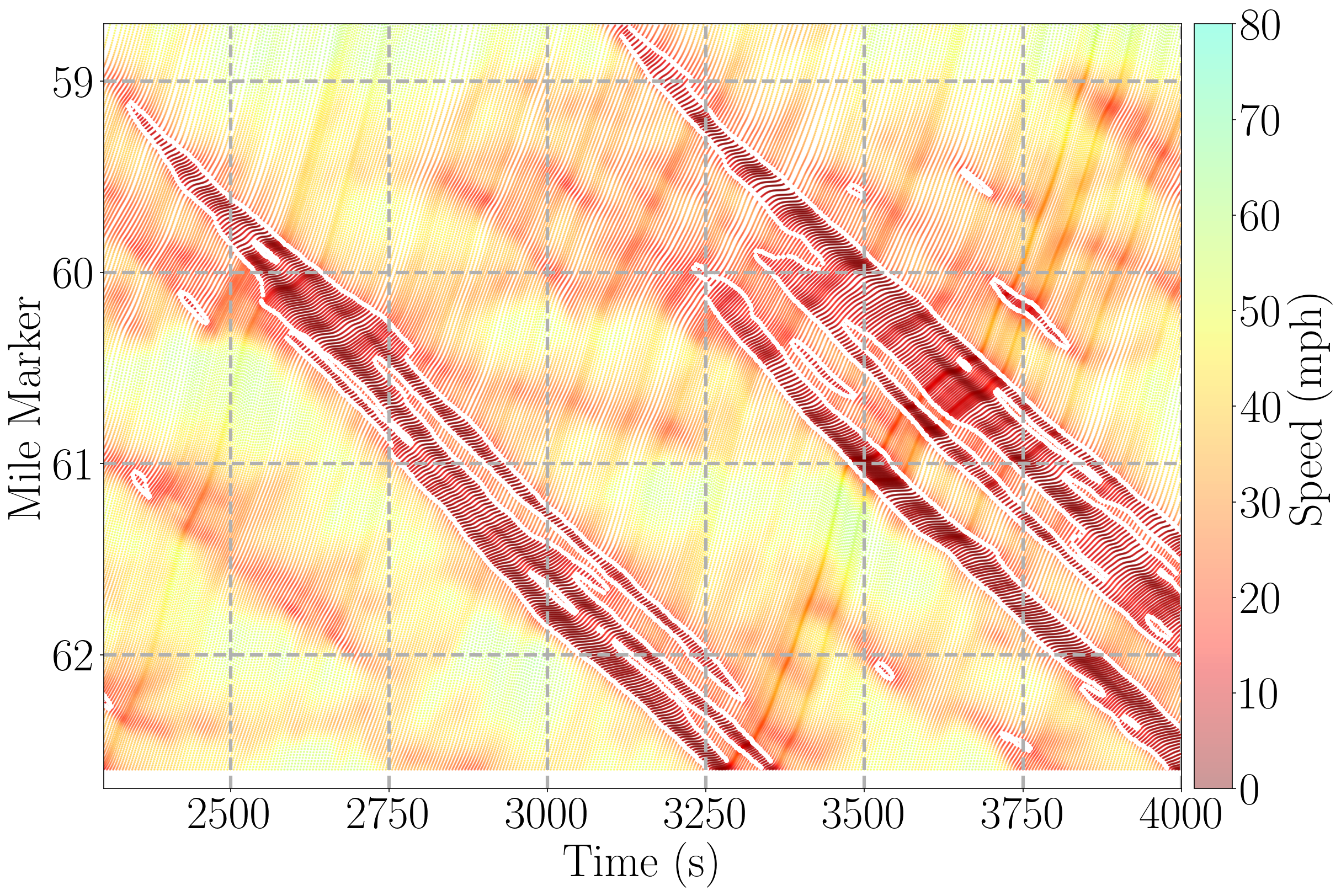}
        \caption{Original space-time diagram}
        \label{fig:demo_white}
    \end{subfigure}
    \hfill
   \begin{subfigure}[b]{0.49\linewidth}
        \centering
        \includegraphics[width=\linewidth]{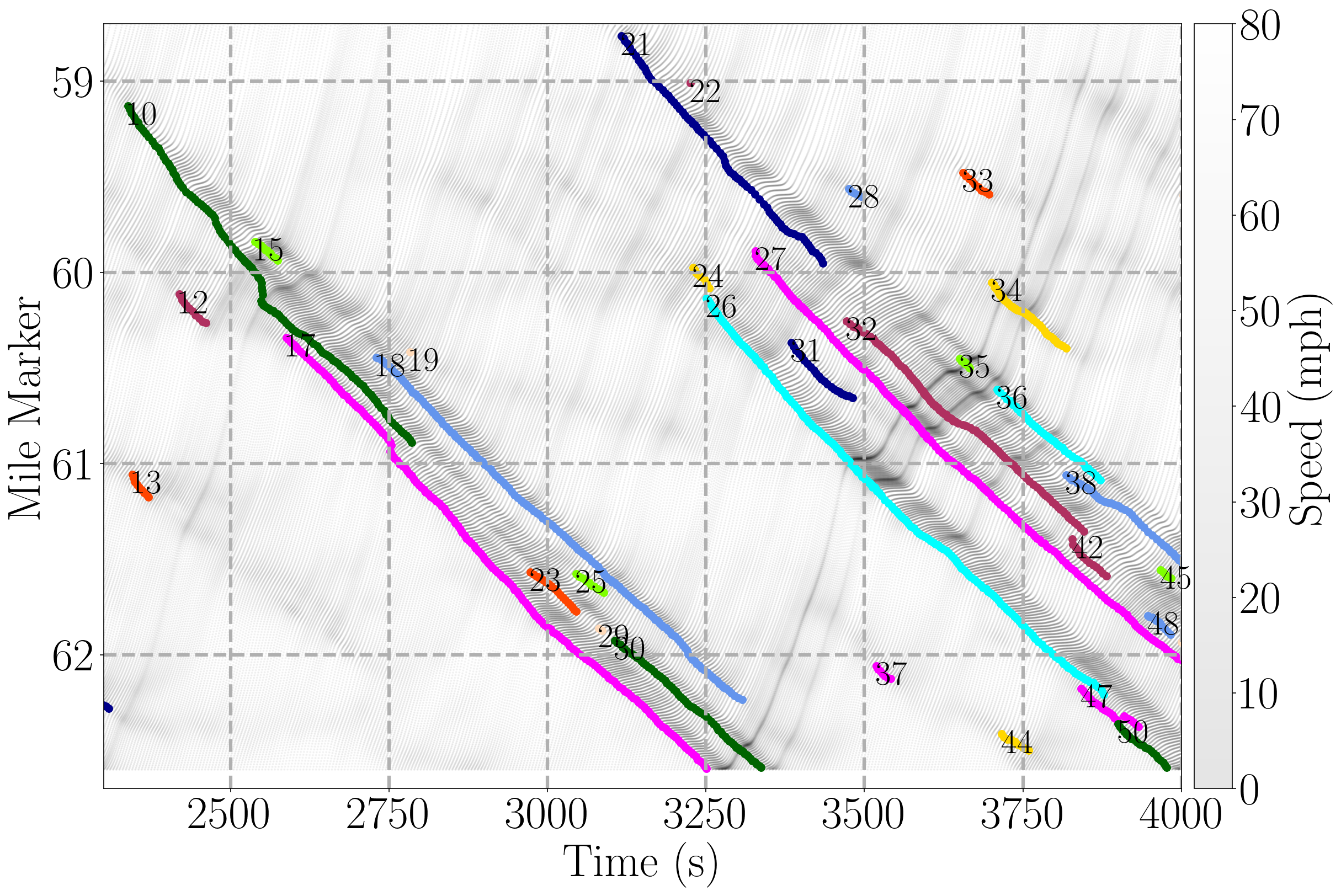}
        \caption{Identified fronts}
        \label{fig:demo_component}
    \end{subfigure}
    \\
    \begin{subfigure}[b]{0.49\linewidth}
        \centering
        \includegraphics[width=\linewidth]{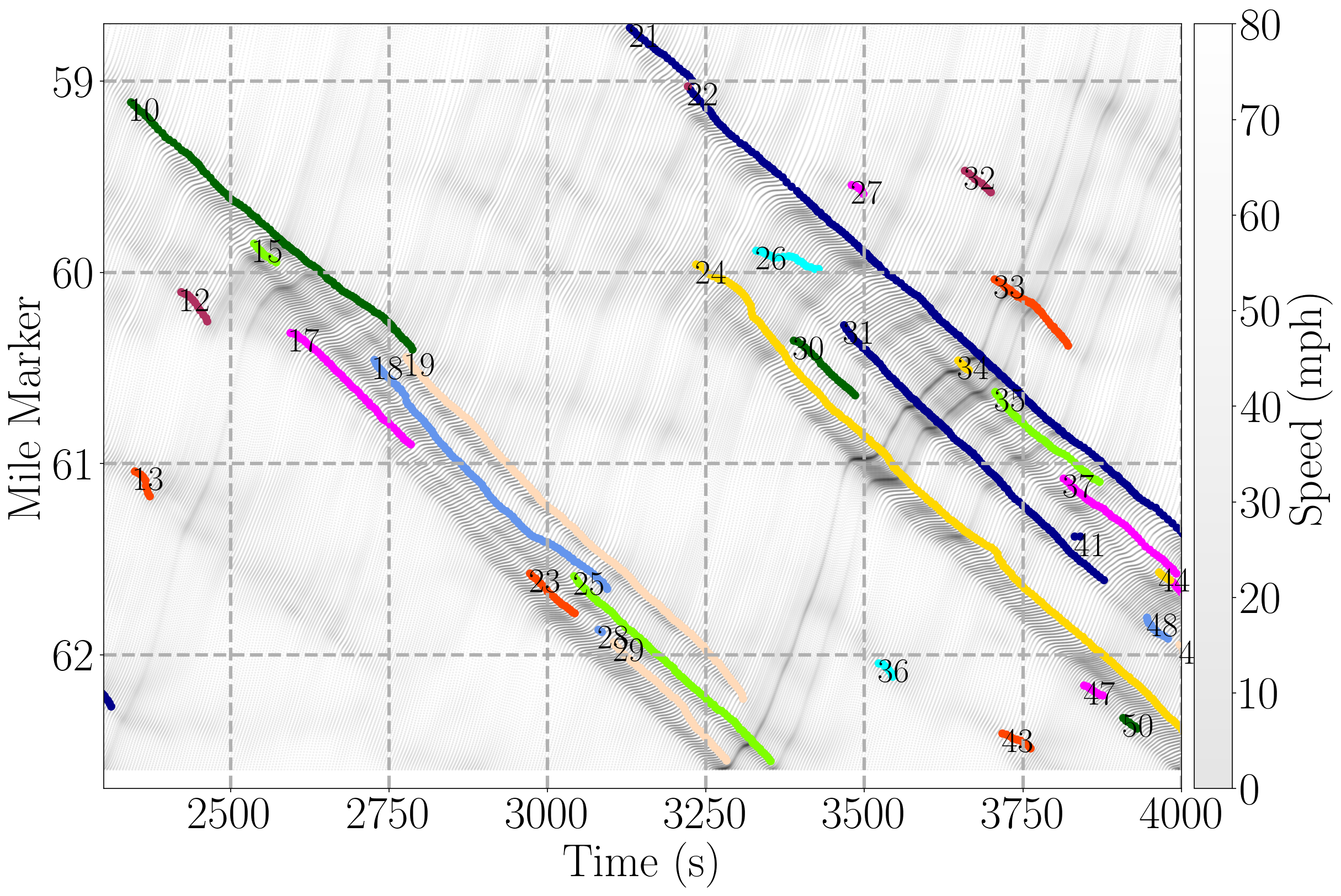}
        \caption{Identified tails}
        \label{fig:demo_front}
    \end{subfigure}
    \hfill
   \begin{subfigure}[b]{0.49\linewidth}
        \centering
        \includegraphics[width=\linewidth]{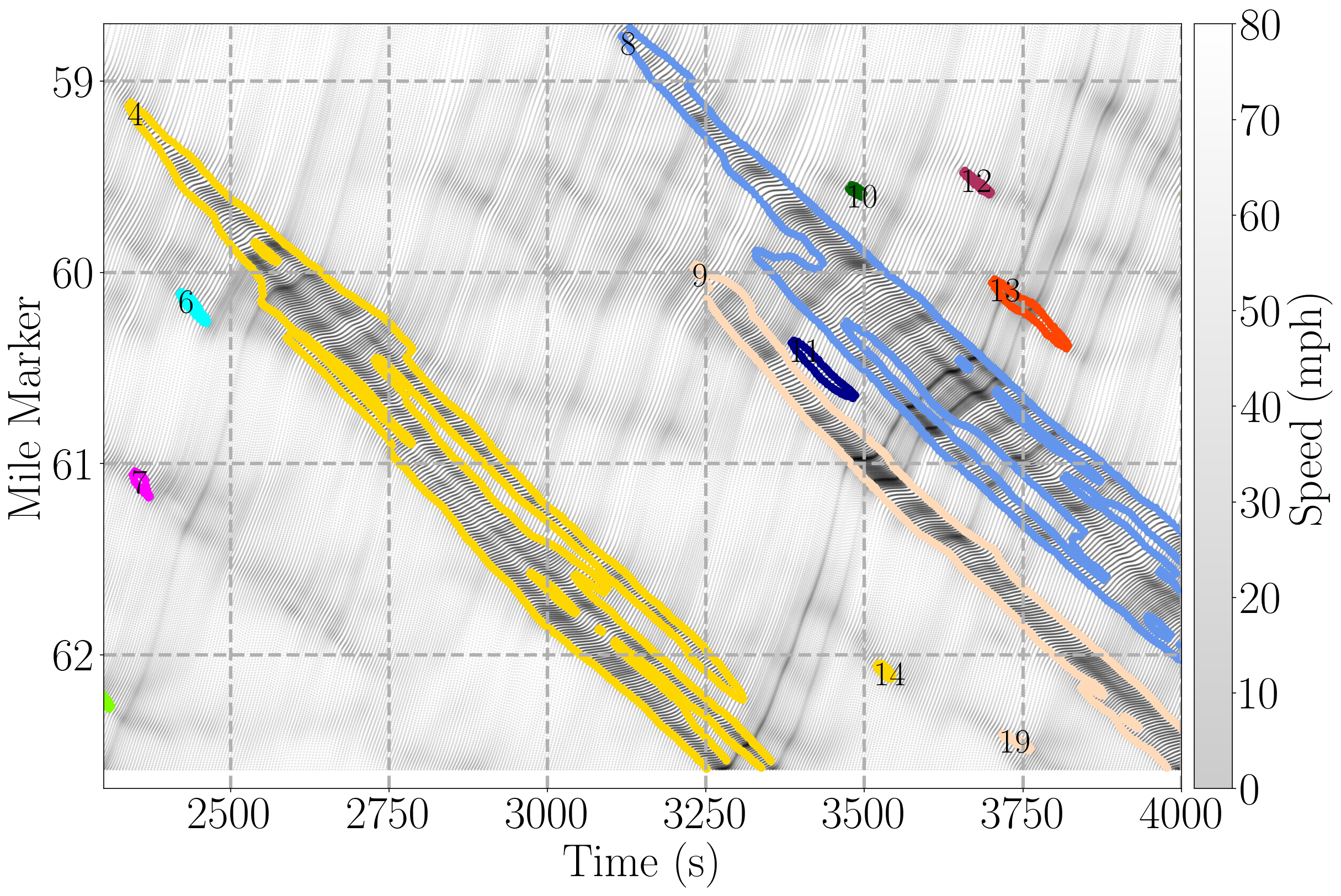}
        \caption{Identified boundaries}
        \label{fig:demo_tail}
    \end{subfigure}
    \caption{\textbf{Demonstration of a mini-scale analysis for the critical speed 15mph}: (a) the raw space-time diagram from the input data with while lines overlapped shown the boundary of the 15mph contour; (b) the wave fronts identified by our method, with 33 unique fronts identified; (c) the wave tails identified by our method, with 32 unique tails identified; (d) the wave boundaries identified by our method, with 12 unique boundaries identified.}
    \label{fig:demo_demo}
\end{figure}

To enhance understanding of the results, we provided a mini-scale analysis as an example demonstration in Figure~\ref{fig:demo_demo}. It demonstrates a mini-scale analysis on November 22, 2022 for a critical speed of 15 mph, showcasing the raw space-time diagram in Figure~\ref{fig:demo_white}, the identified wave fronts Figure~\ref{fig:demo_front}, identified wave tails Figure~\ref{fig:demo_tail}, and the resulting wave boundaries Figure~\ref{fig:demo_component} using the method described, revealing the dynamics and interactions of wave fronts and tails across the space-time diagram. As observed in the demonstration, some wave fronts and tails travel long distances while others travel short distances, with their speeds also varying from one another. Additionally, one single wave boundary can be composed of multiple wave fronts and tails.
\subsection{Wave fronts and tails identification}
\begin{figure}[H]
    \centering
    \begin{subfigure}[b]{0.9\linewidth}
        \centering
        \includegraphics[width=\linewidth]{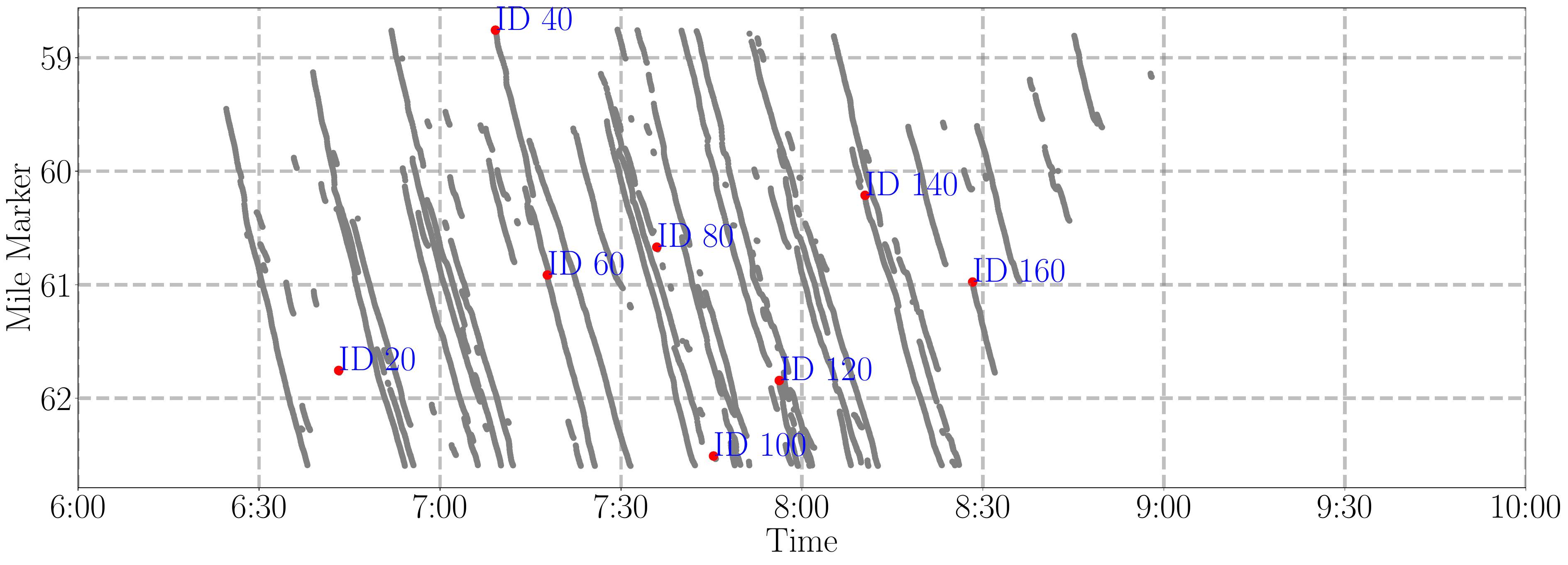}
        \caption{The identified wave fronts for lane 1 on critical speed 15 mph, 170 fronts in total are identified.}
        \label{fig:front_15}
    \end{subfigure}
    \\
   \begin{subfigure}[b]{0.9\linewidth}
        \centering
        \includegraphics[width=\linewidth]{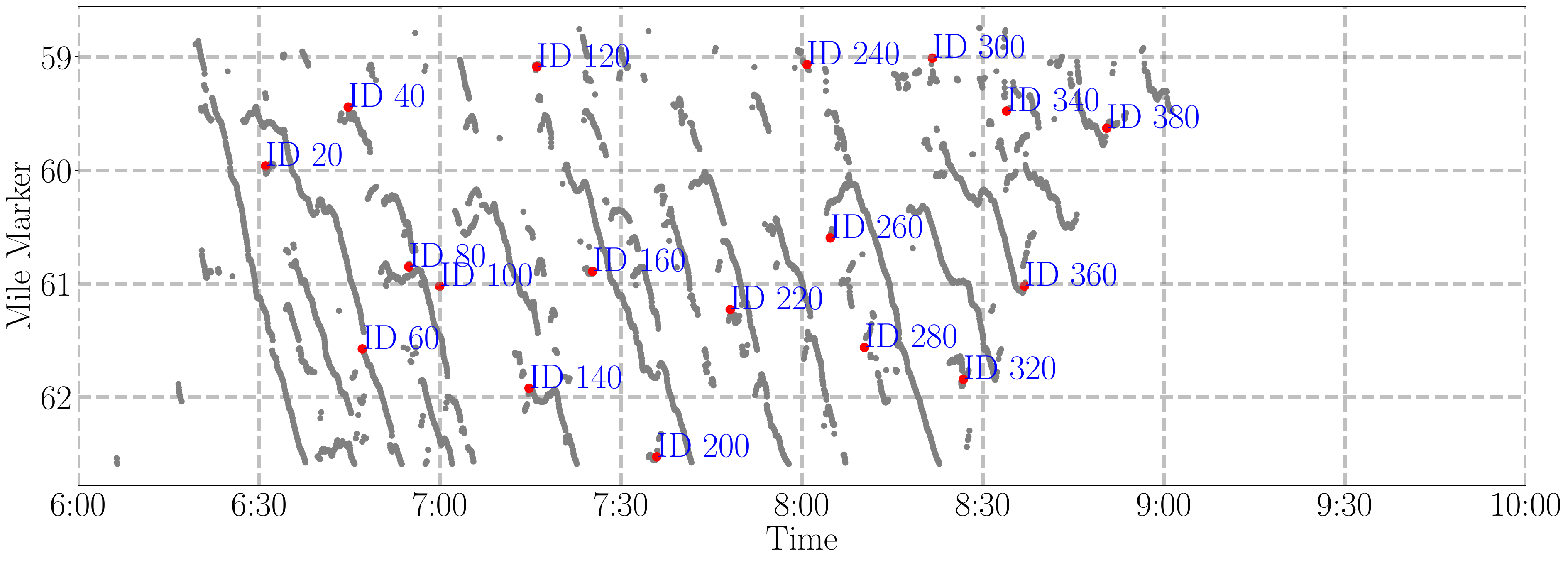}
        \caption{The identified wave fronts for lane 1 on critical speed 40 mph, 398 fronts in total are identified.}
        \label{fig:front_40}
    \end{subfigure}
    \caption{\textbf{Demonstration of the number of the identified wave fronts on lane 1 November 22, 2022}: The IDs are labeled every 20 units to illustrate the number of wave fronts.}
    \label{fig:fronts}
\end{figure}

\begin{figure}[H]
    \centering
    \begin{subfigure}[b]{0.48\linewidth}
        \centering
        \includegraphics[width=\linewidth]{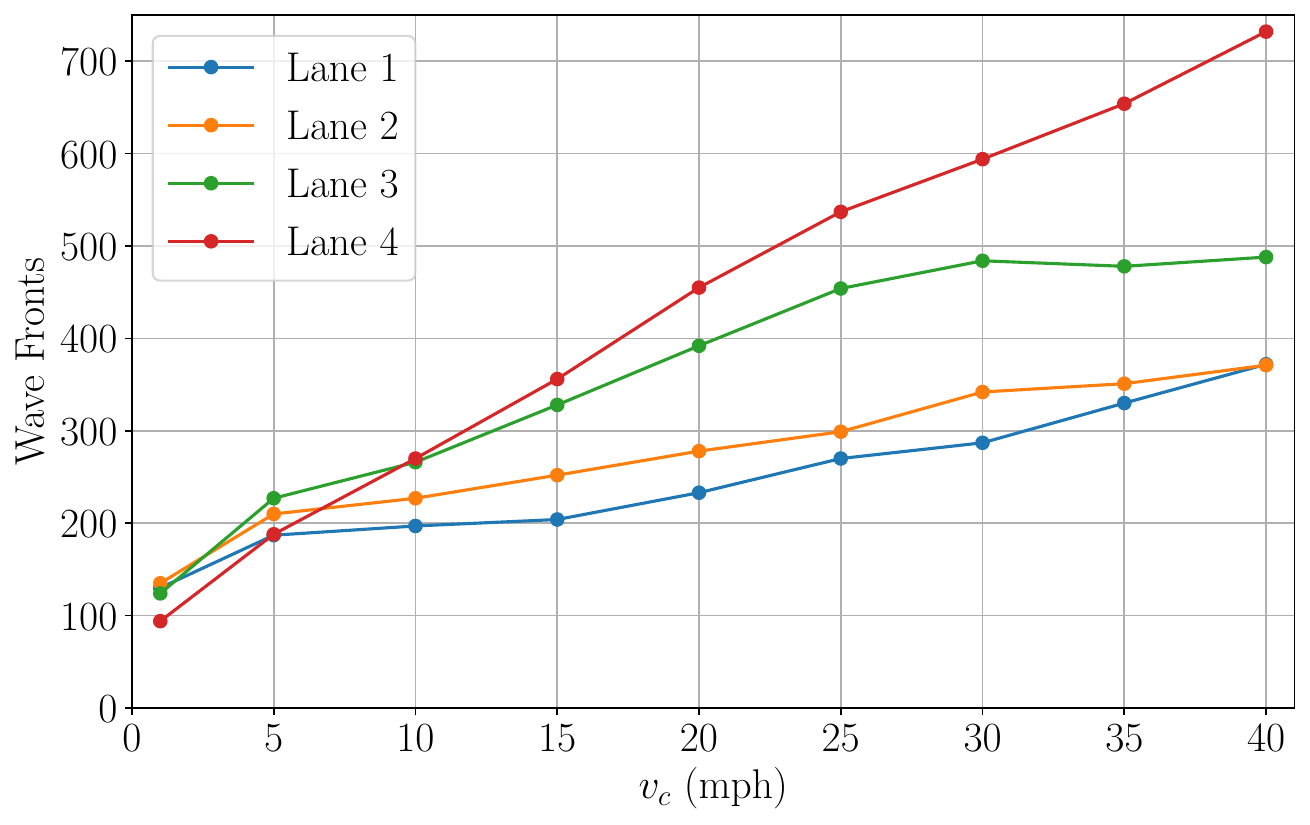}
        \caption{Number of wave fronts by different critical speed (averaged on various days)}
        \label{fig:overall_front}
    \end{subfigure}
    \hfill
    \begin{subfigure}[b]{0.48\linewidth}
        \centering
        \includegraphics[width=\linewidth]{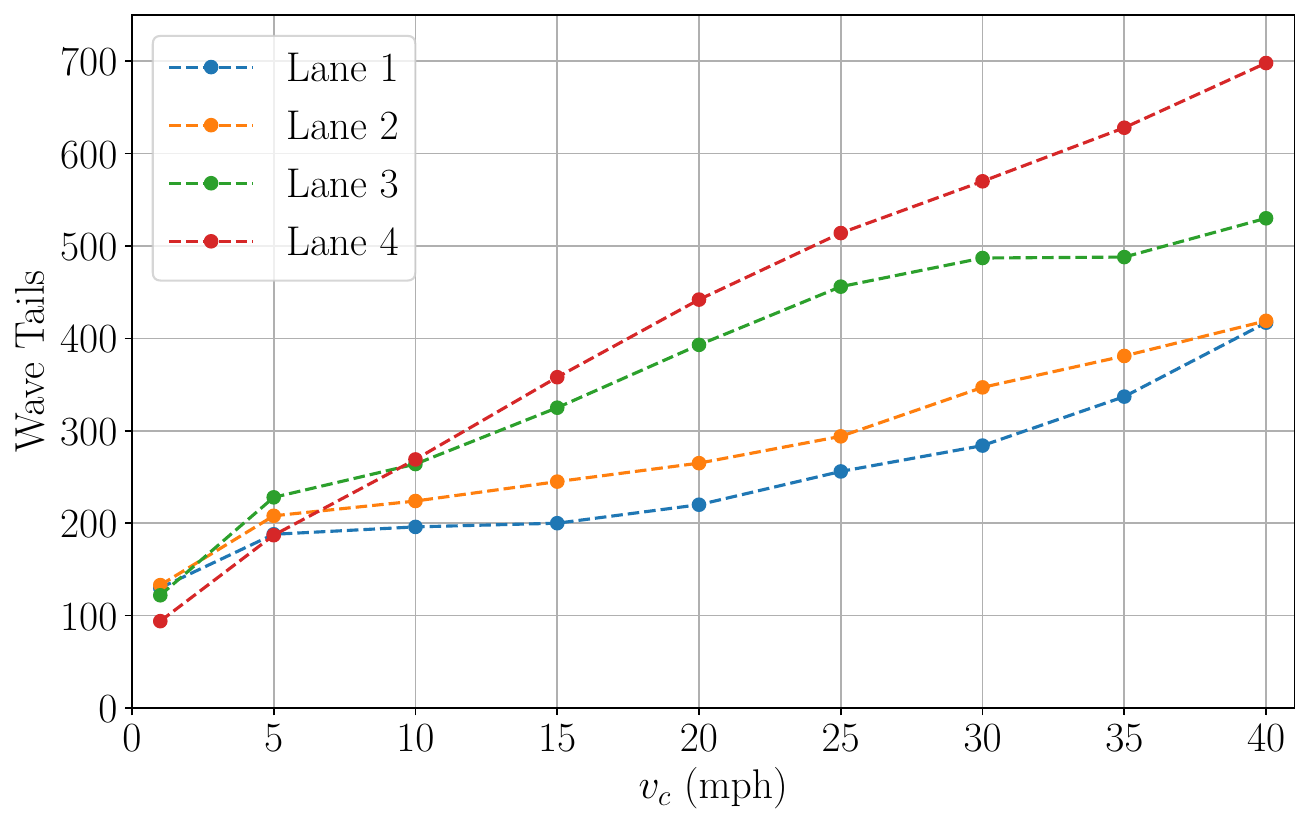} 
        \caption{Number of wave tails by different critical speed (averaged on various days)}
        \label{fig:overall_tail}
    \end{subfigure}\\
   \begin{subfigure}[b]{0.48\linewidth}
        \centering
        \includegraphics[width=\linewidth]{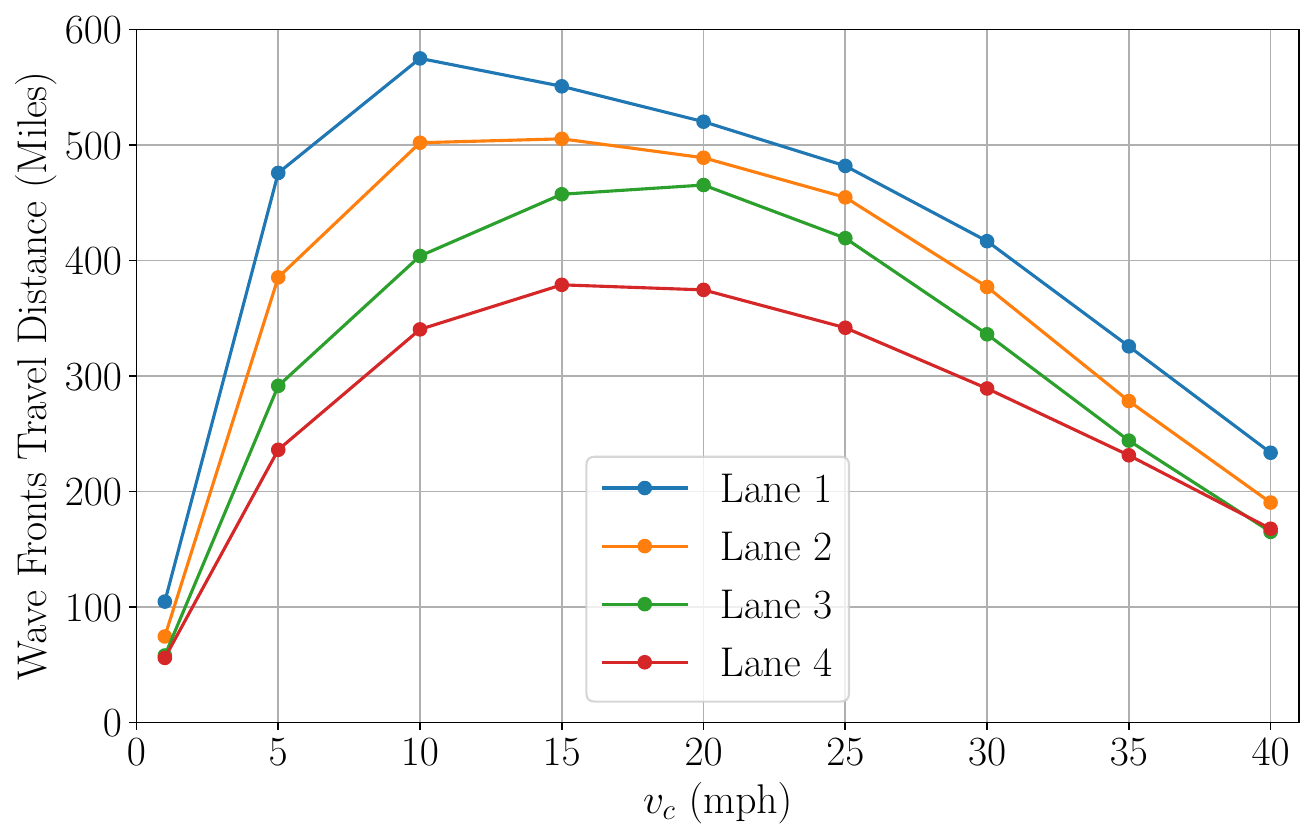}
        \caption{Total travel distance of wave fronts by different critical speed (accumulated by various days)}
        \label{fig:dis_front}
    \end{subfigure}
    \hfill
    \begin{subfigure}[b]{0.48\linewidth}
        \centering
        \includegraphics[width=\linewidth]{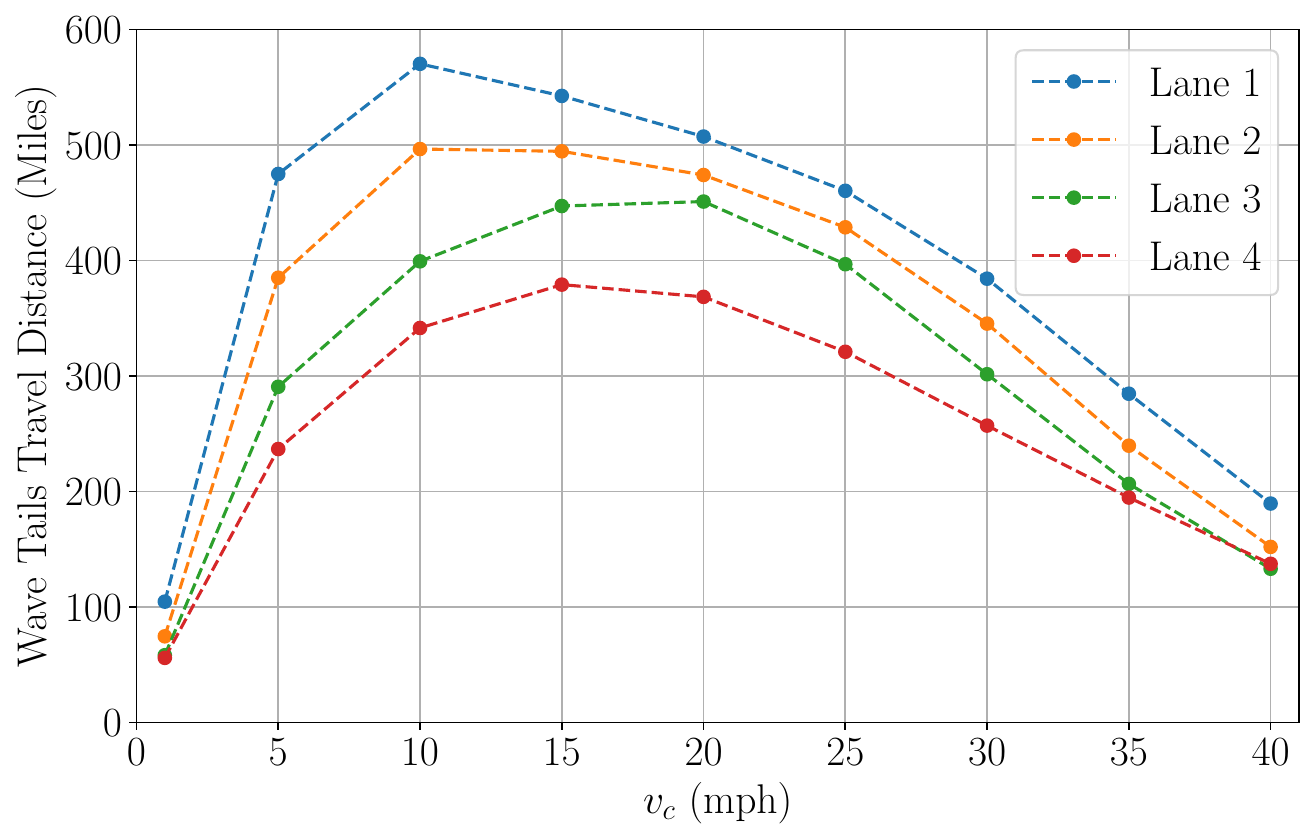} 
        \caption{Total travel distance of wave tails by different critical speed (accumulated by various days)}
        \label{fig:dis_tail}
    \end{subfigure}
    \caption{\textbf{Summary of the identified wave fronts and tails}: the average number of wave fronts and tails by different critical speed across all lanes}
    \label{fig:overall}
\end{figure}

For each $v_c$ and each lane a set of wave fronts and tails are generated. Figure \ref{fig:overall_front} and Figure \ref{fig:overall_tail} shows the average number of wave fronts and tails identified per day separately. As shown in Figure~\ref{fig:front_15}, a total of 170 wave fronts were identified for lane 1 at a critical speed of 15 mph on November 22, 2022. A higher number of fronts were identified at a different critical speed of 40 mph, with 398 wave fronts identified, as shown in Figure~\ref{fig:front_40}.

For each $v_c$ and each lane, a set of wave fronts and tails is generated. Figures \ref{fig:overall_front} and \ref{fig:overall_tail} illustrate the average number of wave fronts and tails identified per day, respectively. As illustrated in the figures, the number of wave fronts and tails increases as the critical speed $v_c$ rises. The number of wave fronts and tails tends to be higher in lane 4, followed by lane 3, lane 2, and lane 1, for speeds ranging from 10 to 40 mph. However, as shown in Figures \ref{fig:dis_front} and \ref{fig:dis_tail}, the traveling distance of the fronts and tails follows the inverse order, indicating that wave propagation is more prolonged in the inner lanes compared to the outer lanes,  which may be explained by the boundary effects of on-ramps and off-ramps. 

Figures \ref{fig:dis_front} and \ref{fig:dis_tail} demonstrate that the travel distances of wave fronts and tails exhibit nonlinear variations with the critical speed $v_c$  across four lanes. The travel distance reaches its maximum around a critical speed of 15-20 mph for all lanes, indicating that this speed range is most critical for wave propagation. Lane 1 consistently has the longest wave front travel distance, while lane 4 shows the shortest across all speed levels. These findings suggest that there is a critical speed range for maximizing wave propagation distance, with notable differences in wave dynamics between lanes, particularly at lower and higher speeds. This emphasizes the sensitivity of wave behavior to both lane-specific factors and the chosen critical speed.

\subsection{Wave components identification}
\label{sec:wave_topo}
\begin{figure}[H]
    \begin{subfigure}[b]{\linewidth}
        \centering
        \includegraphics[width=\linewidth]{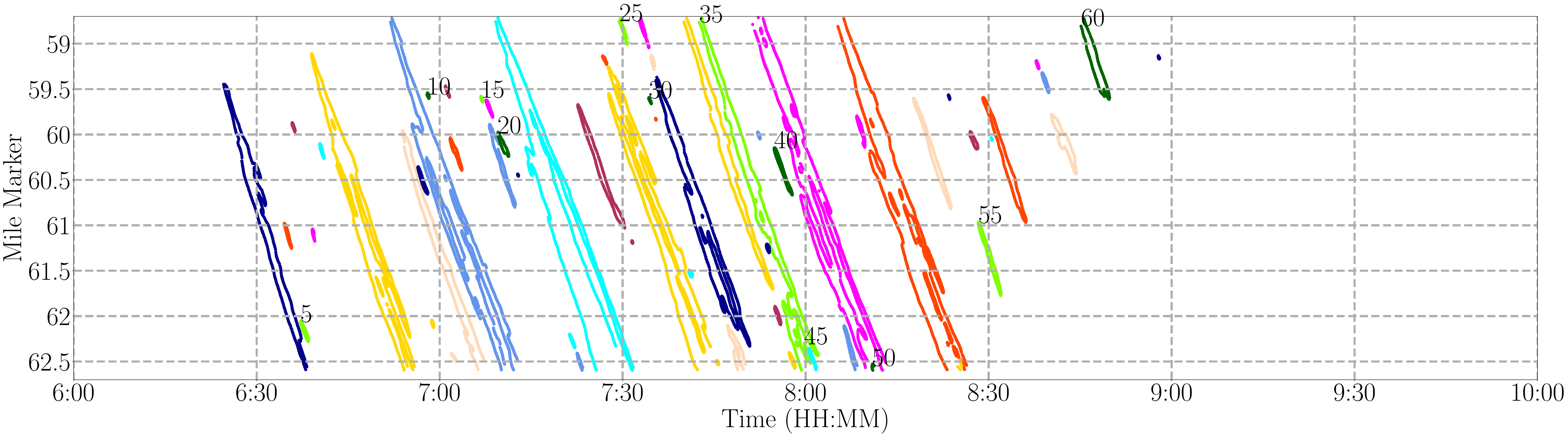}
        \caption{Wave boundaries on lane 1, 61 boundaries are identified}
        \label{fig:component_1}
    \end{subfigure}
    \\
   \begin{subfigure}[b]{\linewidth}
        \centering
        \includegraphics[width=\linewidth]{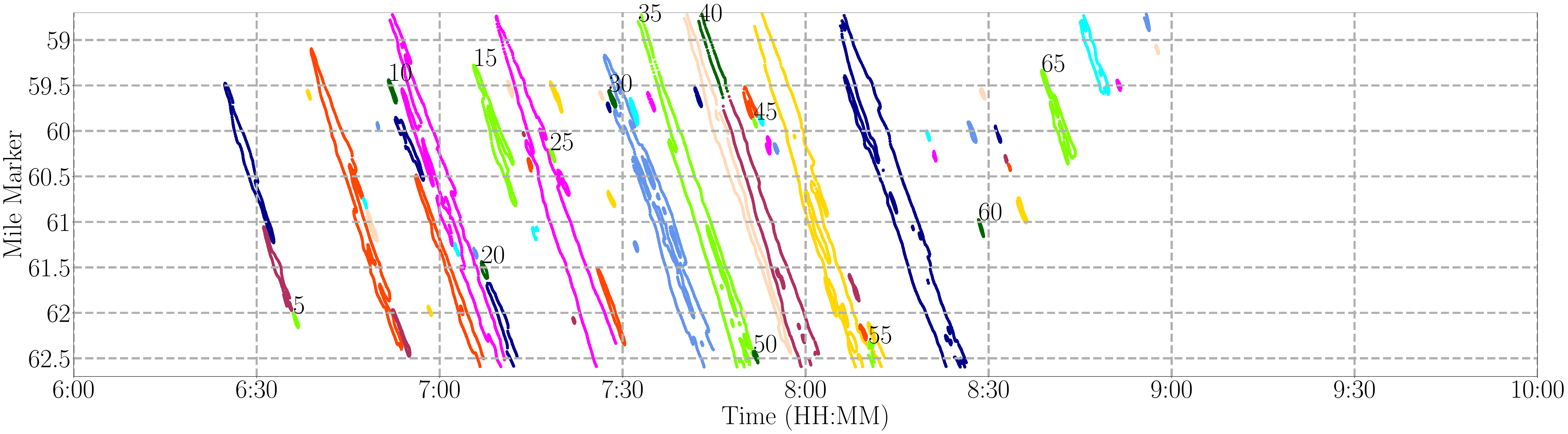}
        \caption{Wave boundaries on lane 2, 69 boundaries are identified}
        \label{fig:component_2}
    \end{subfigure}
    \\
   \begin{subfigure}[b]{\linewidth}
        \centering
        \includegraphics[width=\linewidth]{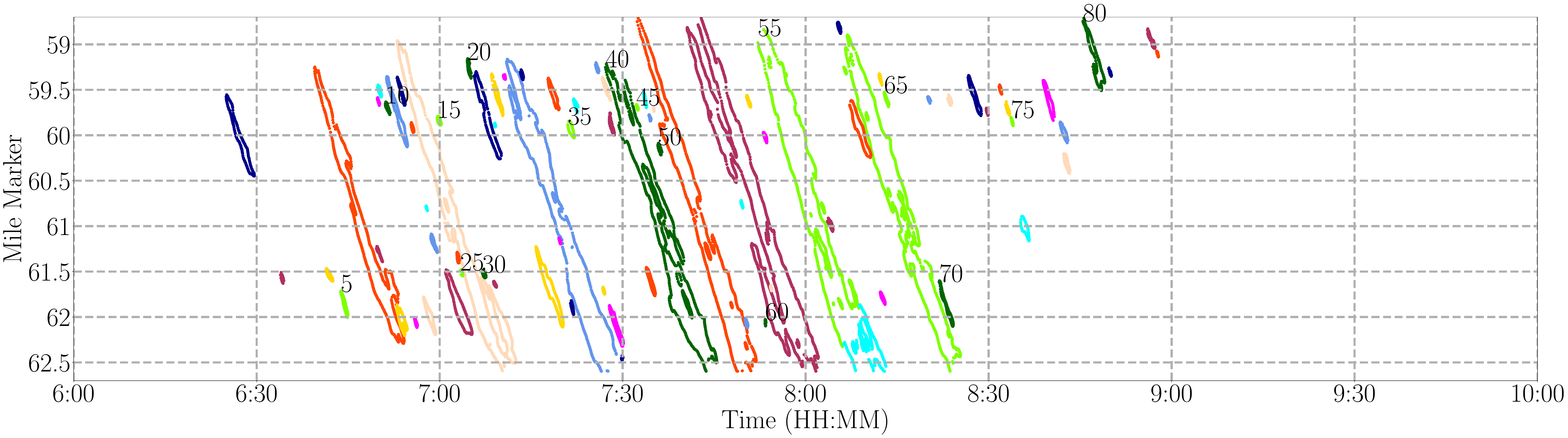}
        \caption{Wave boundaries on lane 3, 83 boundaries are identified}
        \label{fig:component_3}
    \end{subfigure}
    \\ 
   \begin{subfigure}[b]{\linewidth}
        \centering
        \includegraphics[width=\linewidth]{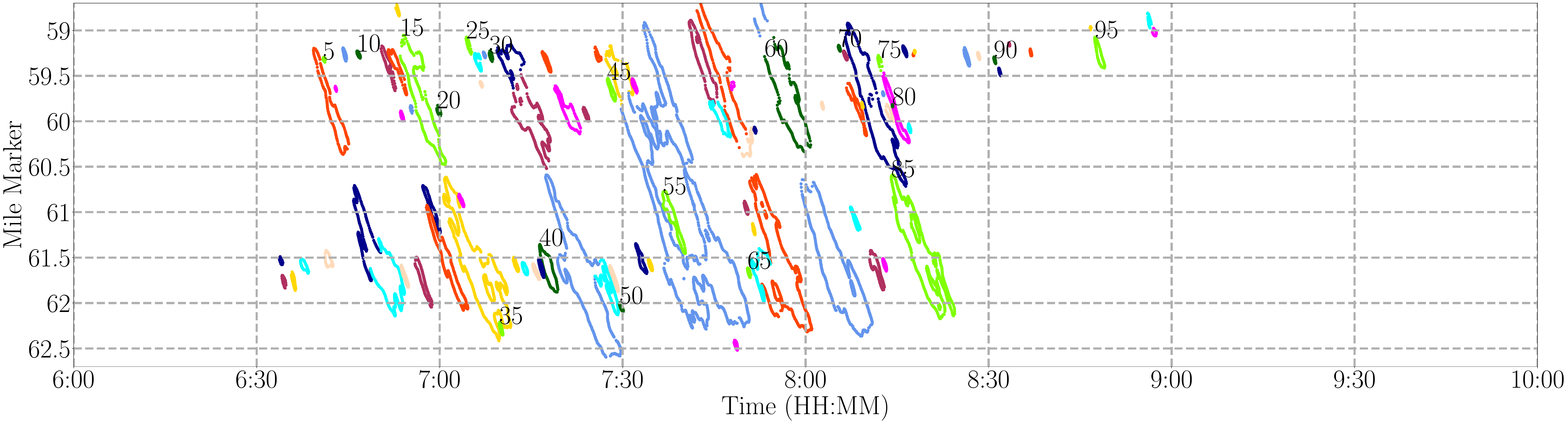}
        \caption{Wave boundaries on lane 4, 97 boundaries are identified}
        \label{fig:component_4}
    \end{subfigure}
    \caption{\textbf{Demonstration of the number of the identified wave boundaries on November 22, 2022}: The IDs are labeled every 5 units to illustrate the number of wave boundaries. Each boundary is assigned a unique color.}
    \label{fig:wave_component12}
\end{figure}

Figure \ref{fig:wave_component12} illustrates the wave component identification results for lane 1 and lane 2 under a critical speed of 15 mph using data on November 22, 2022. In this analysis, 61 wave components were identified for lane 1, while 69 components were detected for lane 2.
Table \ref{tab:compnents} summarizes the number of identified wave components for each lane, grouped by date and critical speed $v_c$ (in mph). The rows for each date show the results for lanes 1 to 4, with the mean speed (MS) of the day indicated next to each date.

The table shows that the number of identified wave components varies across lanes and critical speeds  $v_c$, with notable patterns emerging in relation to the mean speed (MS) of each day. As $v_c$ increases from 1 mph to 15 mph, the number of components generally rises, peaking around 15 mph before declining as $v_c$ continues to increase. Lane 1 consistently has more wave components than other lanes, particularly at moderate critical speeds (10-20 mph), while the differences between lanes are more distinct at lower and higher critical speeds. The mean speed (MS) of the day appears to influence these patterns, with days having lower MS (e.g., November 29, 2022) showing higher component counts across lanes, while days with higher MS (e.g., December 2, 2022) exhibit fewer components overall. This suggests an inverse relationship between mean speed and the number of identified components, indicating that lower traffic speeds lead to more frequent stop-and-go waves, captured as individual components.

\begin{table}[H]
\centering
\caption{\textbf{Summary of the number of the identified components:} summarized by date and the critical speed}
{\scriptsize
\begin{tabular}{lccccccccc}
\toprule
$v_c$ (mph) & 1 & 5 & 10 & 15 & 20 & 25 & 30 & 35 & 40 \\
\hline
\multicolumn{10}{l}{\textbf{2022-11-22 (MS = 36.55 mph)}}\\ 
\hline
Lane 1 & 66 & 84 & 63 & 61 & 73 & 71 & 55 & 35 & 35 \\
Lane 2 & 55 & 84 & 66 & 69 & 69 & 65 & 52 & 29 & 46 \\
Lane 3 & 42 & 75 & 55 & 83 & 75 & 87 & 65 & 42 & 56 \\
Lane 4 & 29 & 65 & 83 & 97 & 109 & 83 & 70 & 75 & 73 \\
\hline
\multicolumn{10}{l}{\textbf{2022-11-28 (MS = 36.82 mph)}}\\ 
\hline
Lane 1 & 63 & 67 & 57 & 70 & 60 & 66 & 50 & 40 & 35 \\
Lane 2 & 49 & 58 & 64 & 69 & 65 & 59 & 42 & 32 & 32 \\
Lane 3 & 28 & 50 & 70 & 75 & 80 & 72 & 60 & 44 & 55 \\
Lane 4 & 21 & 43 & 74 & 98 & 93 & 96 & 83 & 85 & 79 \\
\hline
\multicolumn{10}{l}{\textbf{2022-11-29 (MS = 27.33 mph)}}\\ 
\hline
Lane 1 & 90 & 160 & 122 & 105 & 98 & 60 & 39 & 29 & 31 \\
Lane 2 & 94 & 148 & 127 & 112 & 103 & 48 & 46 & 41 & 35 \\
Lane 3 & 75 & 141 & 133 & 130 & 92 & 59 & 74 & 45 & 45 \\
Lane 4 & 68 & 140 & 159 & 166 & 102 & 80 & 63 & 69 & 81 \\
\hline
\multicolumn{10}{l}{\textbf{2022-11-30 (MS = 27.38 mph)}}\\
\hline
Lane 1 & 103 & 165 & 101 & 87 & 86 & 86 & 69 & 46 & 50 \\
Lane 2 & 88 & 144 & 110 & 101 & 92 & 86 & 48 & 47 & 64 \\
Lane 3 & 87 & 134 & 129 & 121 & 118 & 87 & 69 & 65 & 71 \\
Lane 4 & 71 & 113 & 140 & 137 & 108 & 114 & 106 & 109 & 90 \\
\hline
\multicolumn{10}{l}{\textbf{2022-12-01 (MS = 29.60 mph)}}\\
\hline
Lane 1 & 76 & 142 & 99 & 84 & 72 & 59 & 49 & 53 & 47 \\
Lane 2 & 76 & 131 & 92 & 93 & 81 & 71 & 58 & 41 & 50 \\
Lane 3 & 69 & 116 & 99 & 103 & 96 & 87 & 50 & 49 & 70 \\
Lane 4 & 68 & 97 & 117 & 126 & 103 & 105 & 81 & 72 & 95 \\
\hline
\multicolumn{10}{l}{\textbf{2022-12-02 (MS = 38.49 mph)}}\\
\hline
Lane 1 & 61 & 84 & 43 & 36 & 43 & 62 & 44 & 38 & 41 \\
Lane 2 & 30 & 87 & 50 & 51 & 58 & 59 & 38 & 32 & 41 \\
Lane 3 & 25 & 69 & 52 & 50 & 68 & 59 & 46 & 45 & 64 \\
Lane 4 & 22 & 56 & 60 & 75 & 77 & 73 & 50 & 58 & 76 \\
\bottomrule
\end{tabular}
}
\label{tab:compnents}
\end{table}

\section{Results: Statistics of traffic waves dynamics}
\label{sec:dis}

\subsection{The linearity of wave fronts and tails}
To assess the \textit{linearity} for each wave (i.e. the extent to which a single wave travels at a consistent speed), linear regression is performed on the wave front points or wave tail points for each wave and tail, and the $R^2$ goodness of fit metric is used as a measure of wave linearity. Table \ref{tab:linearity} summarizes the percentage of wave fronts and tails with an $R^2$ value exceeding 0.9. The results indicate that wave fronts and tails tend to exhibit linear travel behavior under critical speeds of 10-20 mph, with variations depending on the lane. Beyond this range, the linearity decreases, suggesting more complex wave dynamics at higher critical speeds. 

\begin{table}[t]
\caption{\textbf{Percentage of the $R^2 >0.9$ for the identified wave fronts and tails:} grouped by lane and by critical speed.}
\centering
{\scriptsize
\begin{tabular}{lccccccccc}
\toprule
$v_c$ (mph) & 1      & 5      & 10     & 15     & 20     & 25     & 30     & 35     & 40     \\
\hline
\multicolumn{10}{l}{\textbf{Wave fronts}}                                                             \\ 
\hline
Lane 1      & 98.2\% & 99.3\% & 99.5\% & 98.5\% & 90.2\% & 79.8\% & 68.3\% & 56.5\% & 44.1\% \\
Lane 2      & 97.7\% & 98.7\% & 96.1\% & 93.4\% & 88.9\% & 77.8\% & 64.0\% & 53.1\% & 45.3\% \\
Lane 3      & 93.2\% & 93.9\% & 92.2\% & 88.0\% & 81.4\% & 67.6\% & 57.1\% & 48.1\% & 37.2\% \\
Lane 4      & 90.9\% & 96.8\% & 92.4\% & 84.9\% & 68.8\% & 53.1\% & 40.3\% & 34.6\% & 28.1\% \\
\hline
\multicolumn{10}{l}{\textbf{Wave tails}}                          \\
\hline
Lane 1      & 99.1\% & 99.4\% & 99.3\% & 97.5\% & 89.7\% & 74.1\% & 61.2\% & 41.3\% & 22.4\% \\
Lane 2      & 97.2\% & 98.6\% & 97.5\% & 94.0\% & 86.3\% & 74.9\% & 53.8\% & 36.5\% & 16.3\% \\
Lane 3      & 94.3\% & 93.6\% & 93.6\% & 89.6\% & 81.7\% & 63.6\% & 45.8\% & 27.9\% & 14.8\% \\
Lane 4      & 92.2\% & 95.0\% & 93.2\% & 85.4\% & 65.2\% & 48.9\% & 31.1\% & 19.1\% & 11.5\% \\
\bottomrule
\end{tabular}
}
\label{tab:linearity}
\end{table}

\begin{itemize}[noitemsep]
    \item \textbf{Wave Fronts:} the percentage of wave fronts with $R^2 > 0.9$ remains high at lower critical speeds (1-15 mph) across all lanes. Lane 1 consistently shows the highest linearity, with over 90\% of wave fronts maintaining $R^2 > 0.9$ up to 20 mph. However, linearity decreases significantly as the critical speed increases beyond 20 mph, with sharp drops observed, particularly in Lanes 3 and 4.

    \item \textbf{Wave Tails:} similarly, wave tails exhibit strong linearity at lower critical speeds (1-15 mph) across all lanes. Lane 1 shows the highest percentage of linear wave tails, maintaining over 97\% linearity up to 15 mph. As the critical speed increases beyond 20 mph, the percentage of linear wave tails drops across all lanes, with the most significant reductions occurring in Lanes 3 and 4.
\end{itemize}

\subsection{Differences in traveling speeds between wave fronts and tails}
\begin{figure}[H]
    \centering
    \includegraphics[width=0.65\linewidth]{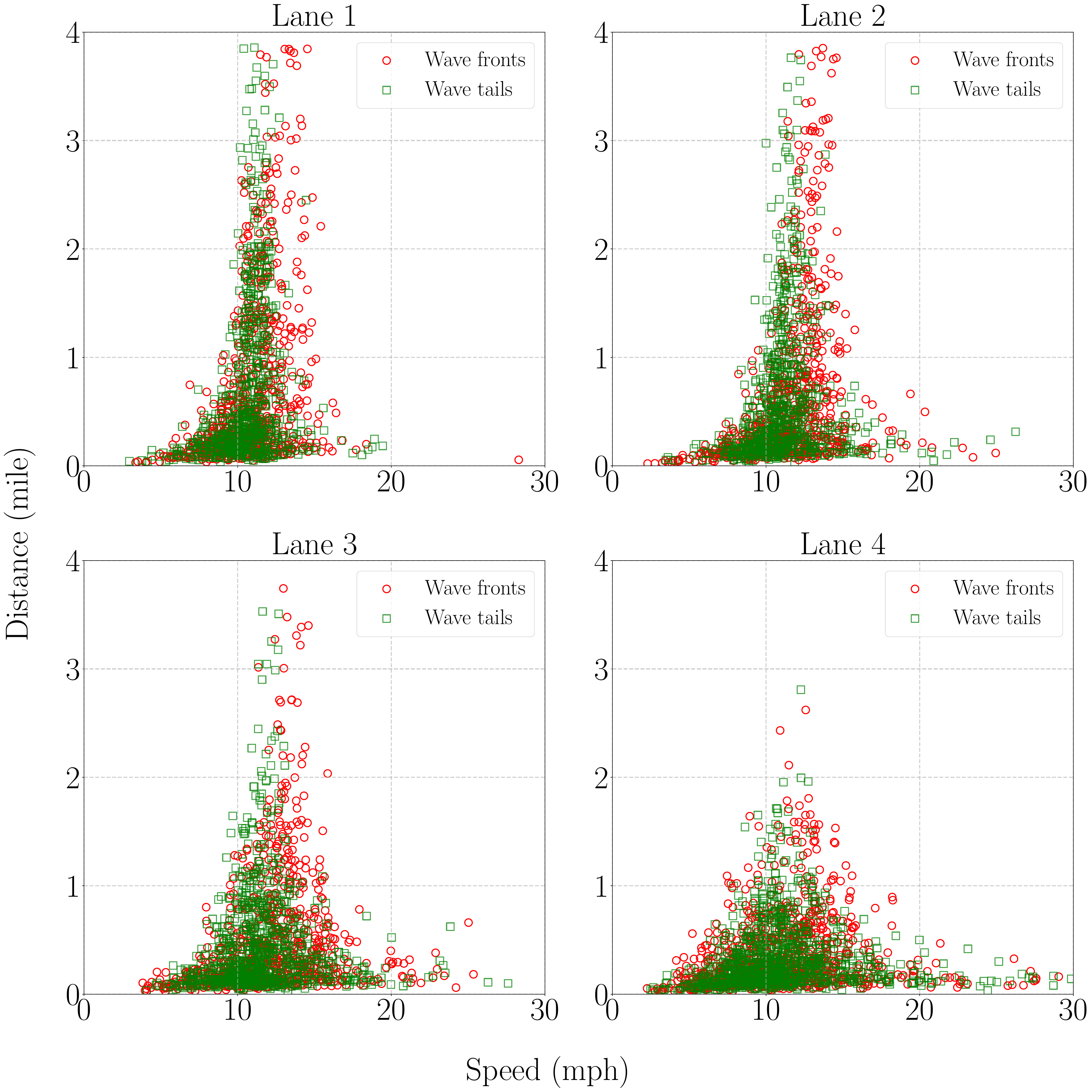}
    \caption{\textbf{The relationship between the wave travel speed and wave travel distance}  (example at a critical speed of 15 mph): the x-axis represents the wave travel speed, while the y-axis represents the wave travel distance. Red dots indicate wave fronts, and green squares represent wave tails.}
    \label{fig:wave_front_speed}
\end{figure}
Based on the definitions in Equations \eqref{eq:wave_distance} and \eqref{eq:wave_speed}, each sample (i.e., wave front or tail) can be represented in a scatter plot illustrating the relationship between wave travel speed and travel distance, as shown in Figure~\ref{fig:wave_front_speed}. \textbf{Note} that the speed shown here are absolute values; in reality, these speeds represent the traveling waves in the opposite direction of traffic flow. Figure~\ref{fig:wave_front_speed} illustrates the centrality of wave speed for both the fronts and tails as the wave travel distance increases at the critical speed 15 mph. The wave travel speed for distances below 0.5 miles is highly random across all lanes. Additional plots for different critical speeds can be found in \ref{sec:appendix3}.
\begin{table}[H]
\caption{\textbf{Wave travel speed averaged over various fronts and tails: }grouped by different lanes and different critical speeds with wave distance less than 0.5 miles filtered out. The difference indicates the variation in average speed between wave fronts and tails, where "+" signifies that the front is faster than the tail, and "-" indicates the opposite.}
\centering
{\scriptsize
\begin{tabular}{lccccccccc}
\toprule
$v_c$ (mph) & 1      & 5      & 10     & 15     & 20     & 25     & 30     & 35     & 40     \\
\hline
\multicolumn{10}{l}{\textbf{Lane 1}}\\
\hline
Wave fronts & 10.900 & 11.164 & 11.467 & 11.699 & 11.945 & 11.702 & 11.345 & 10.713 & 10.255 \\
Wave tails & 10.813 & 11.057 & 11.153 & 11.166 & 10.837 & 10.220 & 9.710 & 9.678 & 9.414 \\
Difference & +0.087 & +0.107 & +0.314 & +0.533 & +1.108 & +1.482 & +1.625 & +1.025 & +0.841 \\
\hline
\multicolumn{10}{l}{\textbf{Lane 2}}\\
\hline
Wave fronts & 11.235 & 11.946 & 12.261 & 12.435 & 12.186 & 11.709 & 10.880 & 10.321 & 10.454 \\
Wave tails & 11.364 & 11.792 & 11.756 & 11.389 & 10.942 & 10.276 & 9.888 & 9.635 & 8.876 \\
Difference & -0.129 & +0.154 & +0.505 & +1.046 & +1.244 & +1.433 & +0.992 & +0.686 & +1.578\\
\hline
\multicolumn{10}{l}{\textbf{Lane 3}}\\
\hline
Wave fronts & 13.697 & 13.333 & 13.045 & 12.830 & 12.265 & 11.589 & 11.059 & 11.174 & 10.843 \\
Wave tails & 13.063 & 13.032 & 12.224 & 11.700 & 10.989 & 10.271 & 10.330 & 9.374 & 11.424 \\
Difference & +0.634& +0.301& +0.821& +1.130& +1.276& +1.318& +0.729& +1.800& -0.581 \\
\hline
\multicolumn{10}{l}{\textbf{Lane 4}}\\
\hline
Wave fronts & 12.001 & 12.215 & 12.006 & 11.756 & 11.590 & 10.834 & 10.708 & 10.494 & 10.619 \\
Wave tails & 11.900 & 12.024 & 11.531 & 11.053 & 10.297 & 9.413 & 8.541 & 10.037 & - \\
Difference& +0.101& +0.191& +0.475& +0.703& +1.293& +1.421& +2.167& +0.457& -\\
\bottomrule
\end{tabular}
}
\label{tab:speed}
\end{table}
We further filter out wave fronts and tails with travel distances less than 0.5 miles and calculate the average wave travel speed based on the remaining fronts and tails. Table \ref{tab:speed} shows the average wave travel speeds for both wave fronts and tails, grouped by different lanes and critical speeds $v_c$ (in mph). \textbf{A key observation is that wave fronts and tails propagate at different speeds across all lanes}, with wave fronts in the analyzed dataset generally traveling faster than wave tails. This difference in speed indicates the expansion of the waves.

\subsection{Wave bifurcation and merge}
\begin{figure}[H]
    \centering
    \begin{subfigure}[b]{0.32\textwidth}
        \centering
        \includegraphics[width=\linewidth]{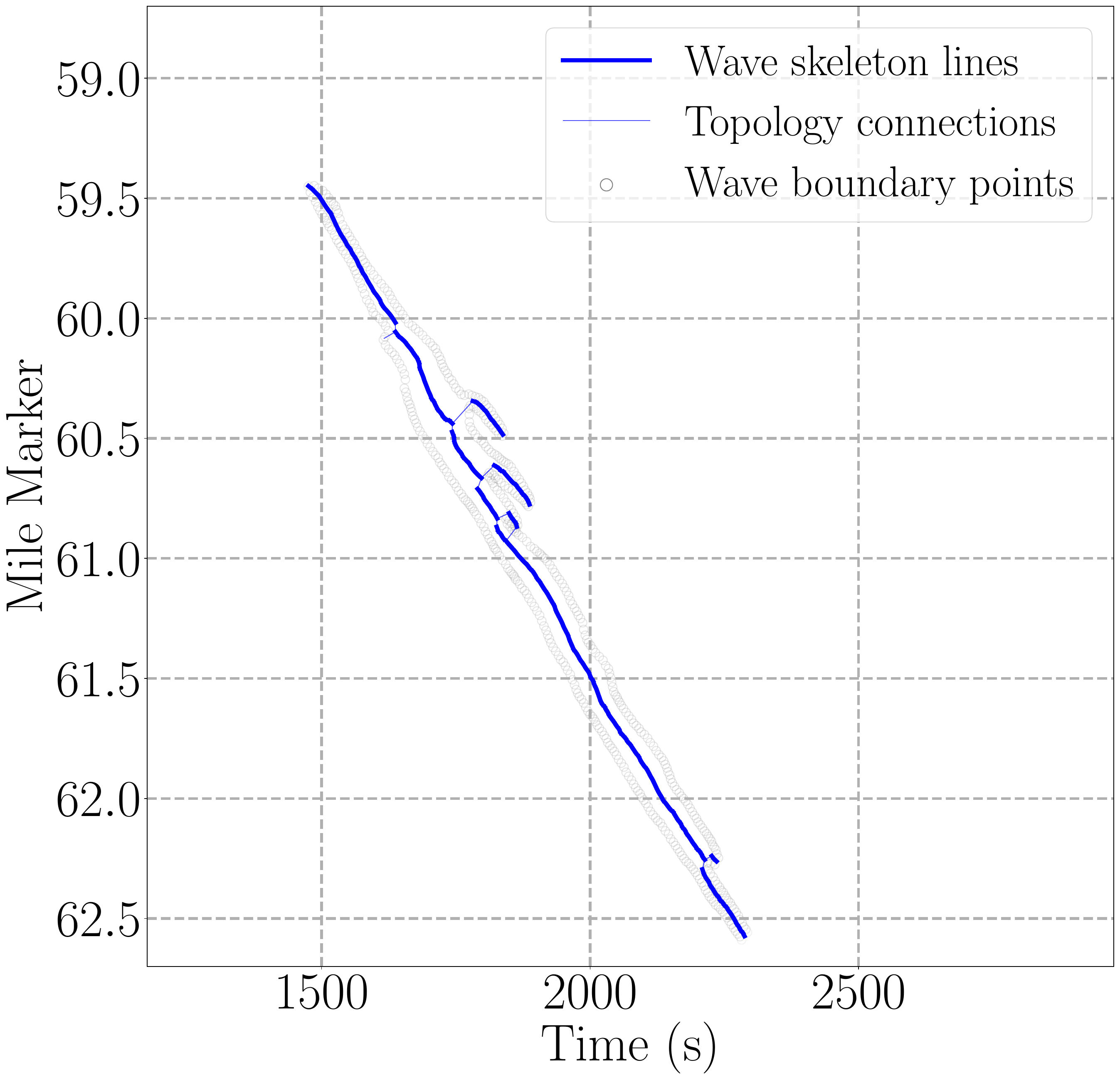}
    \end{subfigure}
    \hfill
    \begin{subfigure}[b]{0.32\textwidth}
        \centering
        \includegraphics[width=\linewidth]{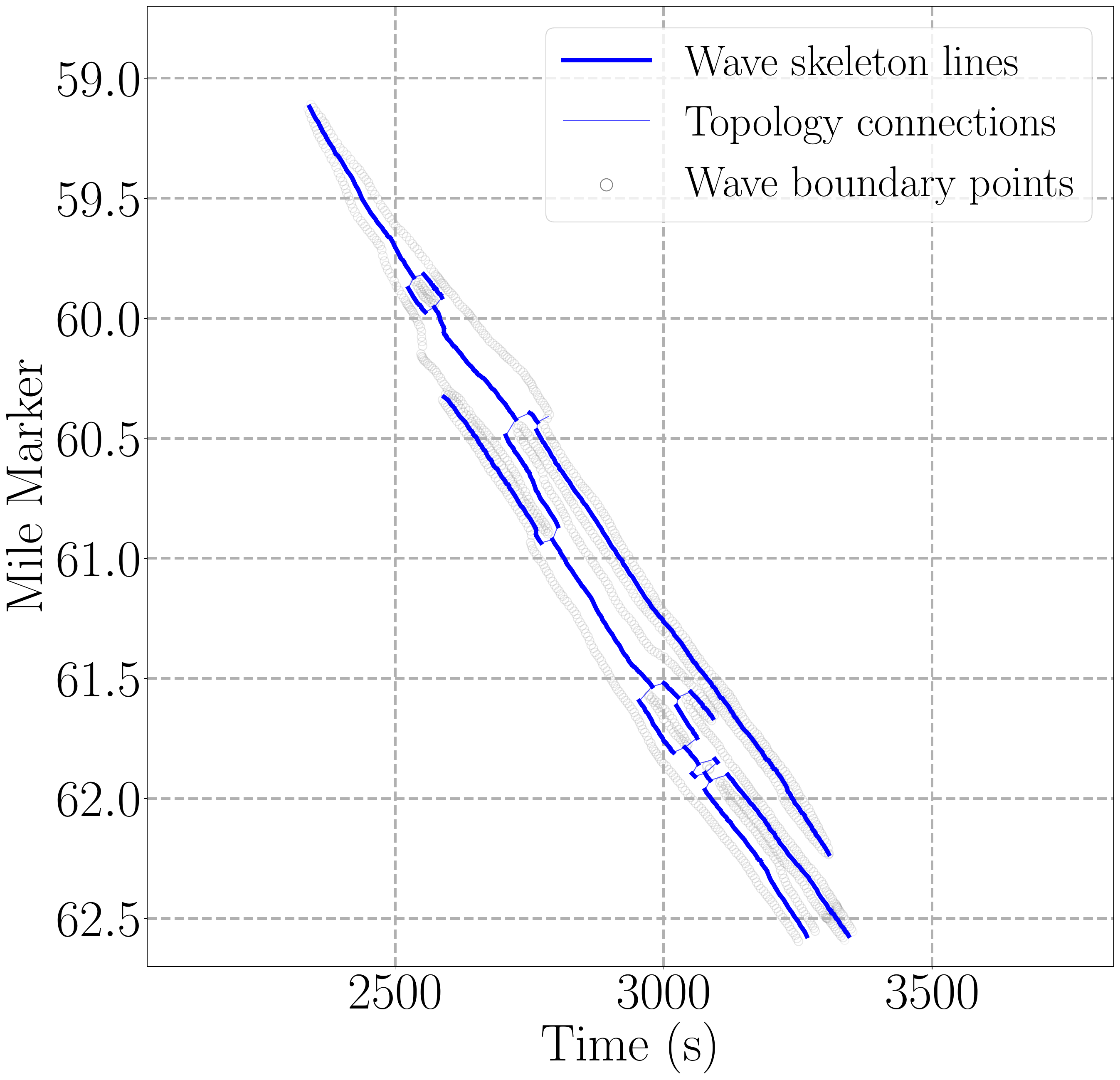}
    \end{subfigure}
    \hfill
    \begin{subfigure}[b]{0.32\textwidth}
        \centering
        \includegraphics[width=\linewidth]{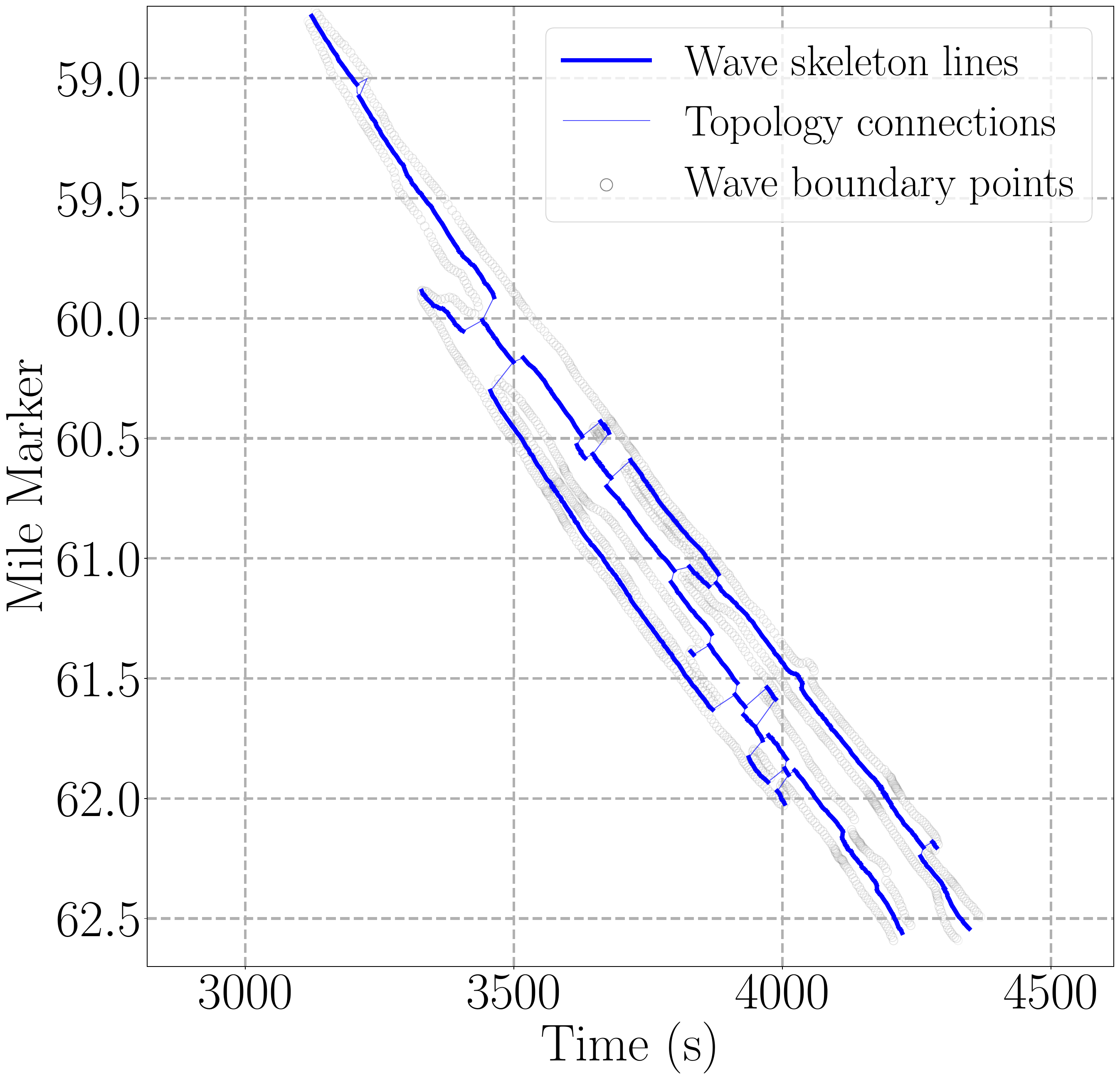}
    \end{subfigure}
    \vskip\baselineskip
    \begin{subfigure}[b]{0.32\textwidth}
        \centering
        \includegraphics[width=\linewidth]{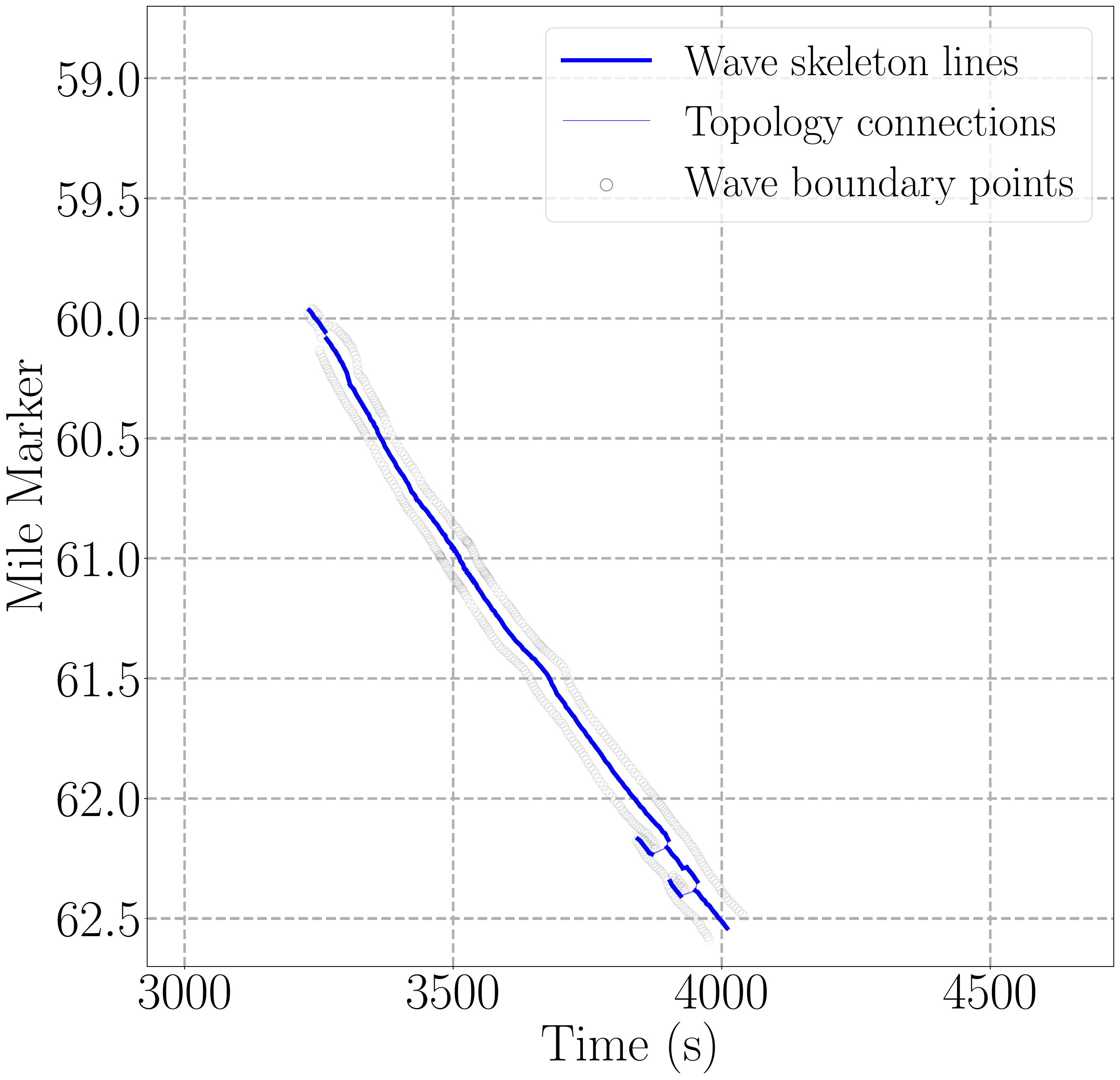}
    \end{subfigure}
    \hfill
    \begin{subfigure}[b]{0.32\textwidth}
        \centering
        \includegraphics[width=\linewidth]{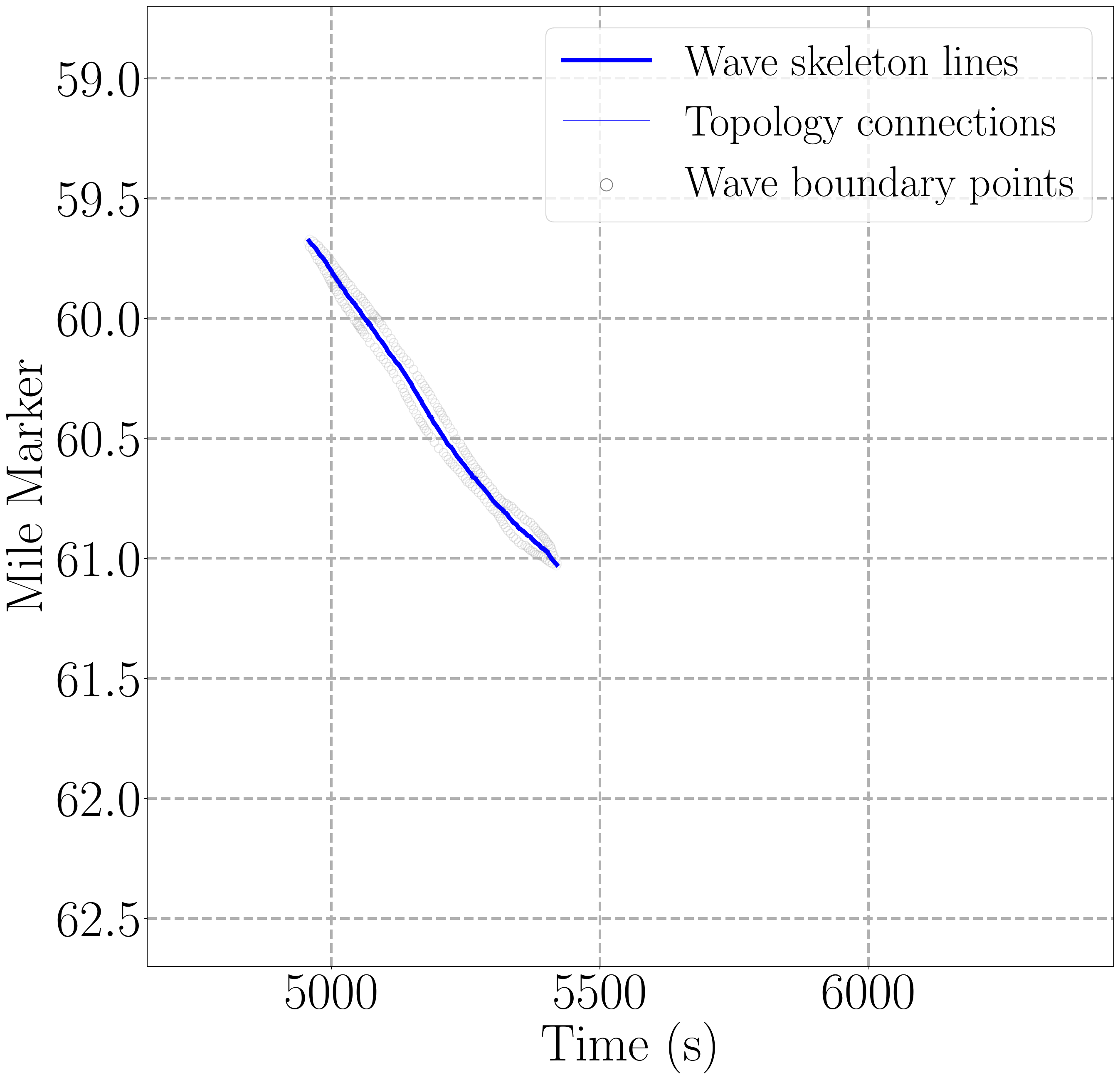}
    \end{subfigure}
    \hfill
    \begin{subfigure}[b]{0.32\textwidth}
        \centering
        \includegraphics[width=\linewidth]{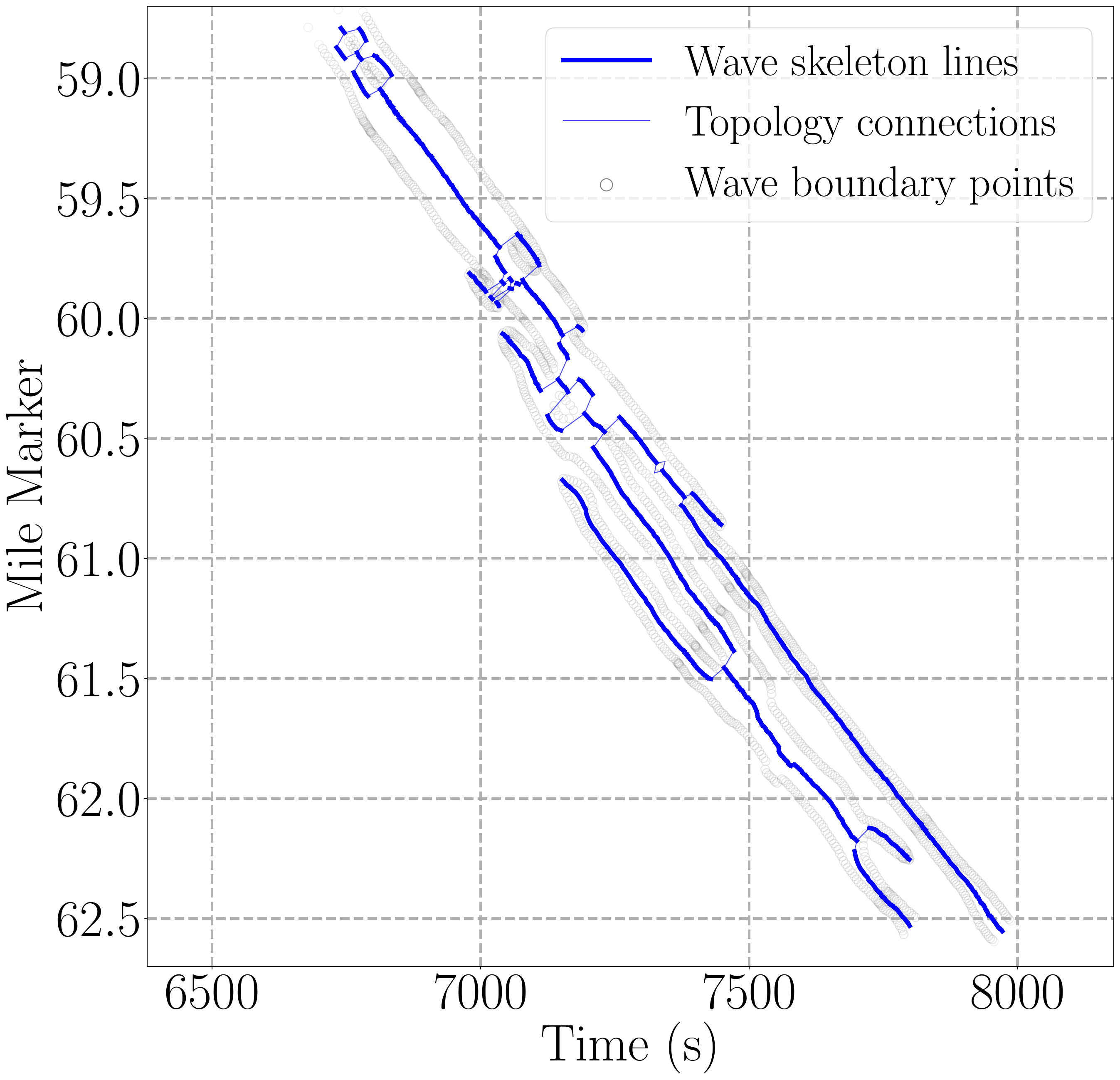}
    \end{subfigure}
    \caption{\textbf{Demonstration of wave boundary topologies:} thick blue lines represent the ``trough'' of the wave (where the middle point of the paired wave front and tail). Thin blue ``tie lines'' indicate topology connections, i.e.   a wave branching or merging. The gray outline is the overall wave boundary identified using the method from Section \ref{sec:components}.}
    \label{fig:skeleton}
\end{figure}

Lastly, Figure~\ref{fig:skeleton} shows the wave topology ``skeleton'' of several example wave boundaries, where each branch of a wave is represented by a single line. These examples clearly highlight the complexity of some traffic waves, in which the wave merges and bifurcates several times as it propagates over several miles. Additionally, some wave boundaries have no branching at all, further suggesting the variability in complexity of different waves. We further discuss the clusters of the identified wave topologies in the next section.

\subsection{Clustering analysis on the wave topology}
\begin{figure}[H]
    \centering
    \begin{subfigure}{0.48\linewidth}
        \centering
        \includegraphics[width=\linewidth]{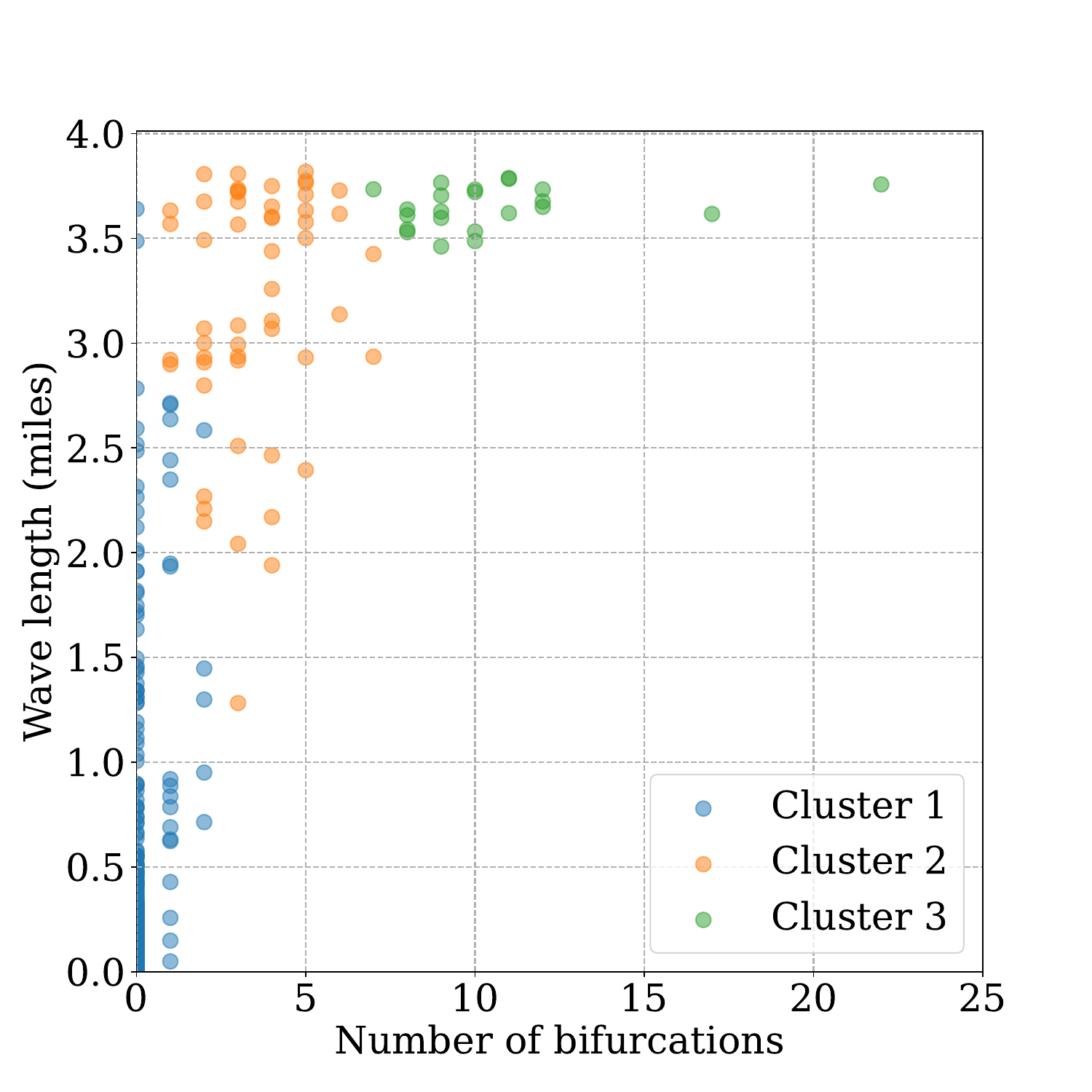}
        \caption{Bifurcation}
        \label{fig:clustering_bifurcation}
    \end{subfigure}
    \hfill
    \begin{subfigure}{0.48\linewidth}
        \centering
        \includegraphics[width=\linewidth]{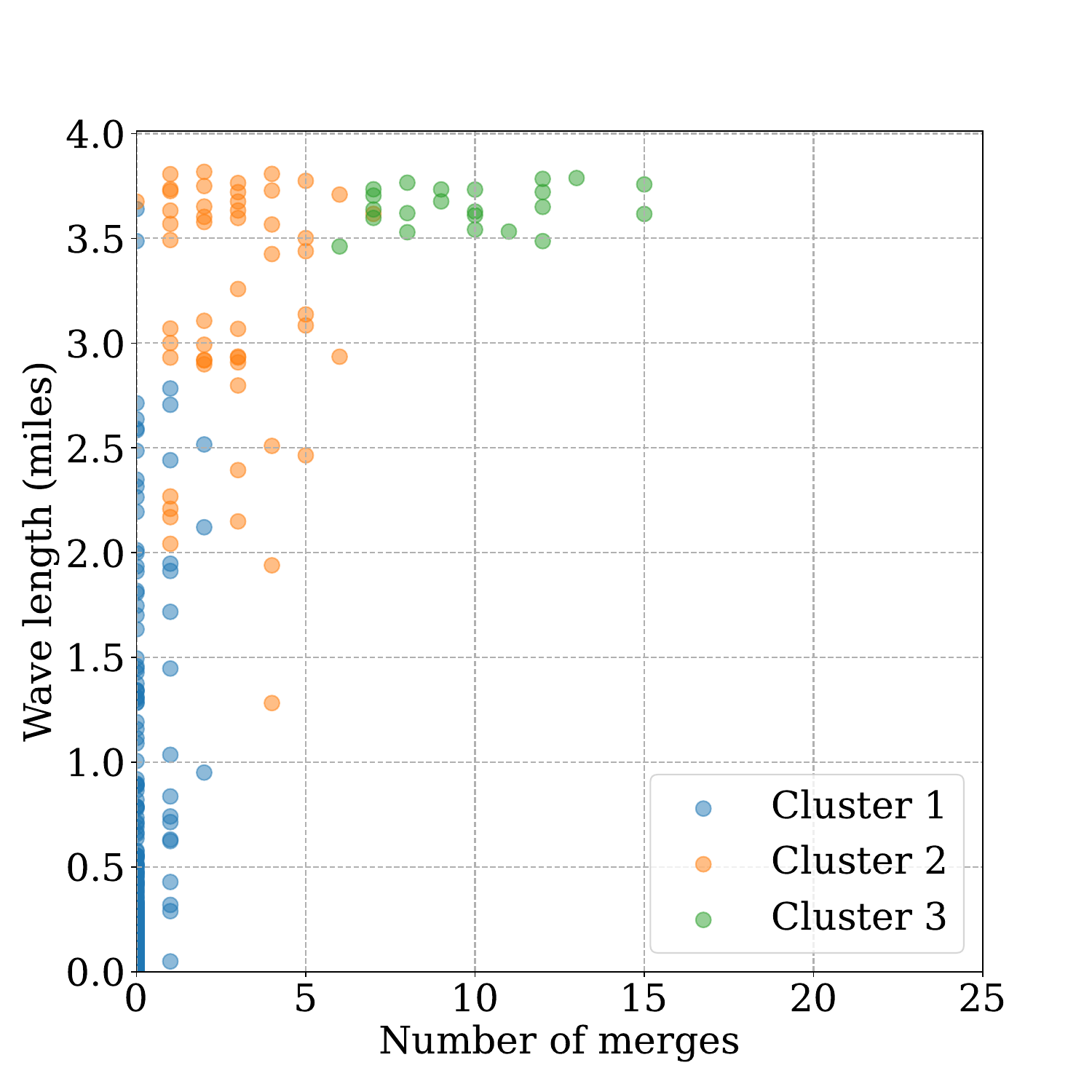}
        \caption{Merge}
        \label{fig:clustering_merge}
    \end{subfigure}
    \caption{\edit{\textbf{Cluster visualization for lane 1 with the critical speed set at 15 mph}: The x-axis represents the number of bifurcations/merges of the wave components, and the y-axis represents the wave length of the wave component.}}
    \label{fig:clustering}
\end{figure}

Clustering analysis can help better understand the topological structure of the waves. 
We use a simple clustering method, KMeans clustering via scikit-learn \cite{pedregosa2011scikit}, and visualize the results on the diagram shown in Figure~\ref{fig:clustering}. Figure~\ref{fig:clustering} is a demonstration of lane 1 on the critical speed 15 mph. We cluster the wave components lane by lane since we think the topological features on each lane are different. The cluster visualization clearly shows the simplicity and complexity of traffic waves:  traffic waves with similar travel distances may remain stable without any bifurcation or merging, while others may have bifurcation and merging more than 10 times.

\begin{figure}[H]
    \centering
    \begin{subfigure}{0.3\linewidth}
        \centering
        \includegraphics[width=\linewidth]{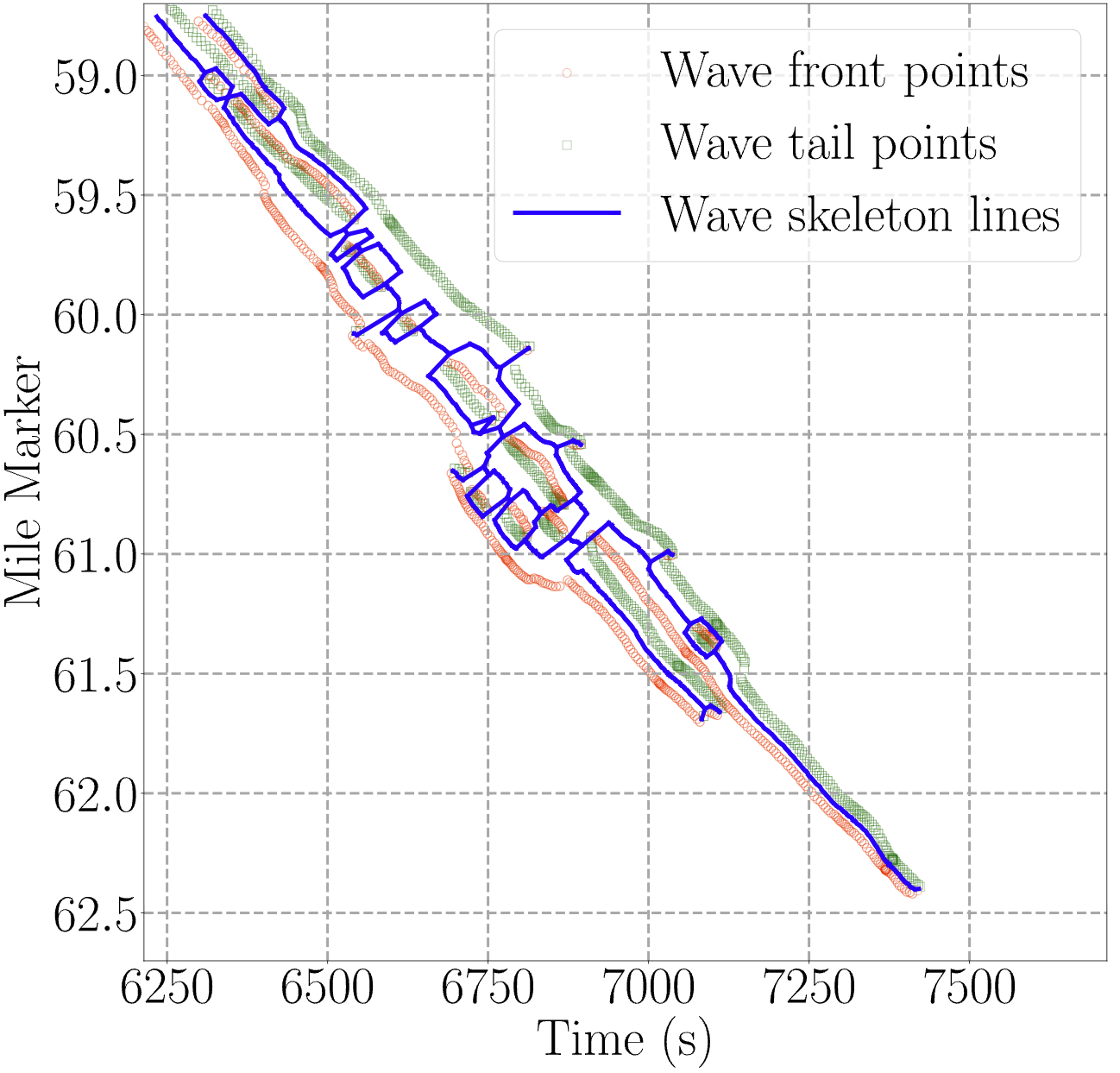}
        \captionsetup{labelformat=empty}
        \caption{(a) Sample topology from cluster 3}
        \label{fig:subfig1}
    \end{subfigure}
    \begin{subfigure}{0.3\linewidth}
        \centering
        \includegraphics[width=\linewidth]{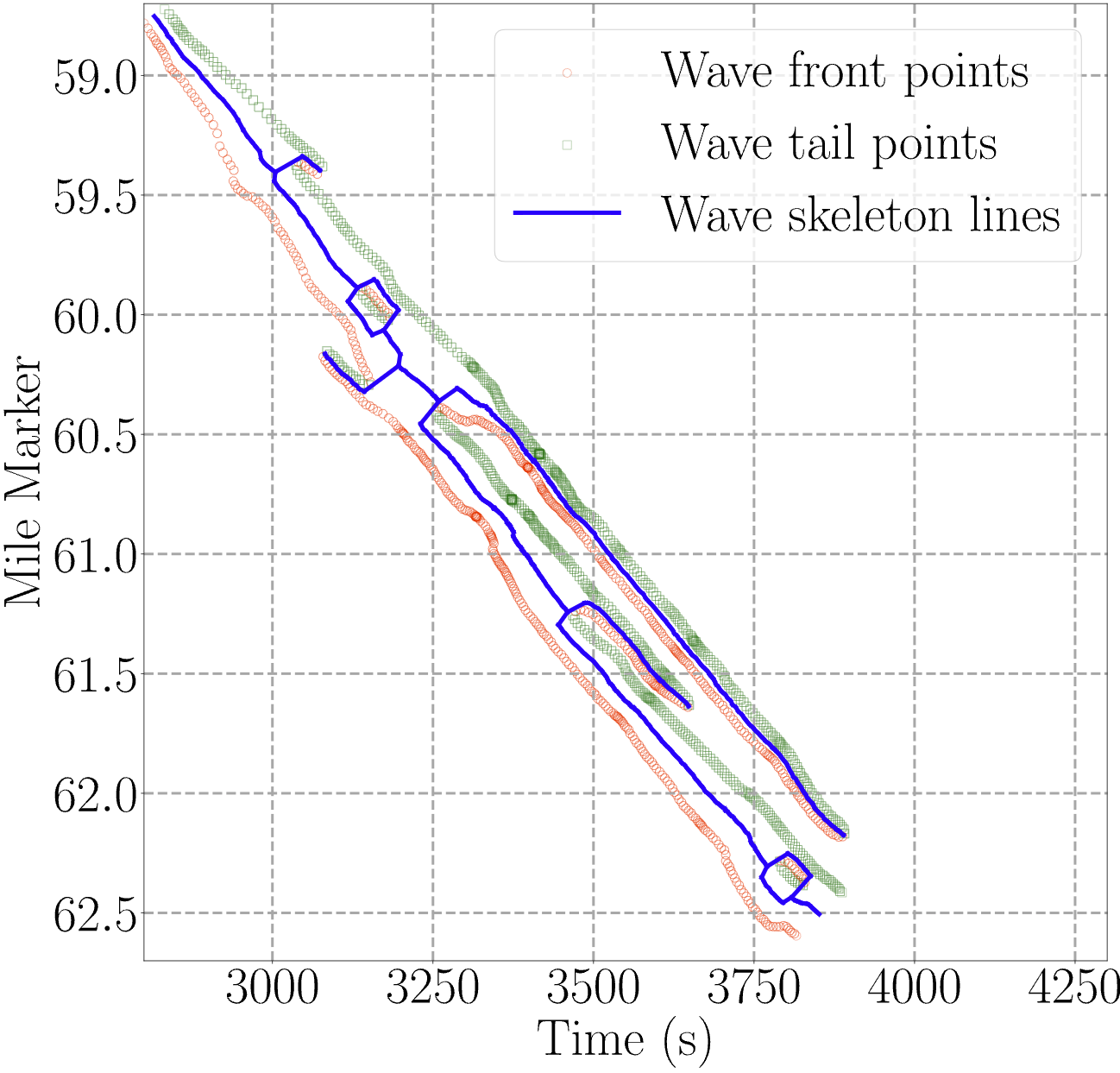}
        \captionsetup{labelformat=empty}
        \caption{(b) Sample topology from cluster 2}
        \label{fig:subfig2}
    \end{subfigure}
    \begin{subfigure}{0.3\linewidth}
        \centering
        \includegraphics[width=\linewidth]{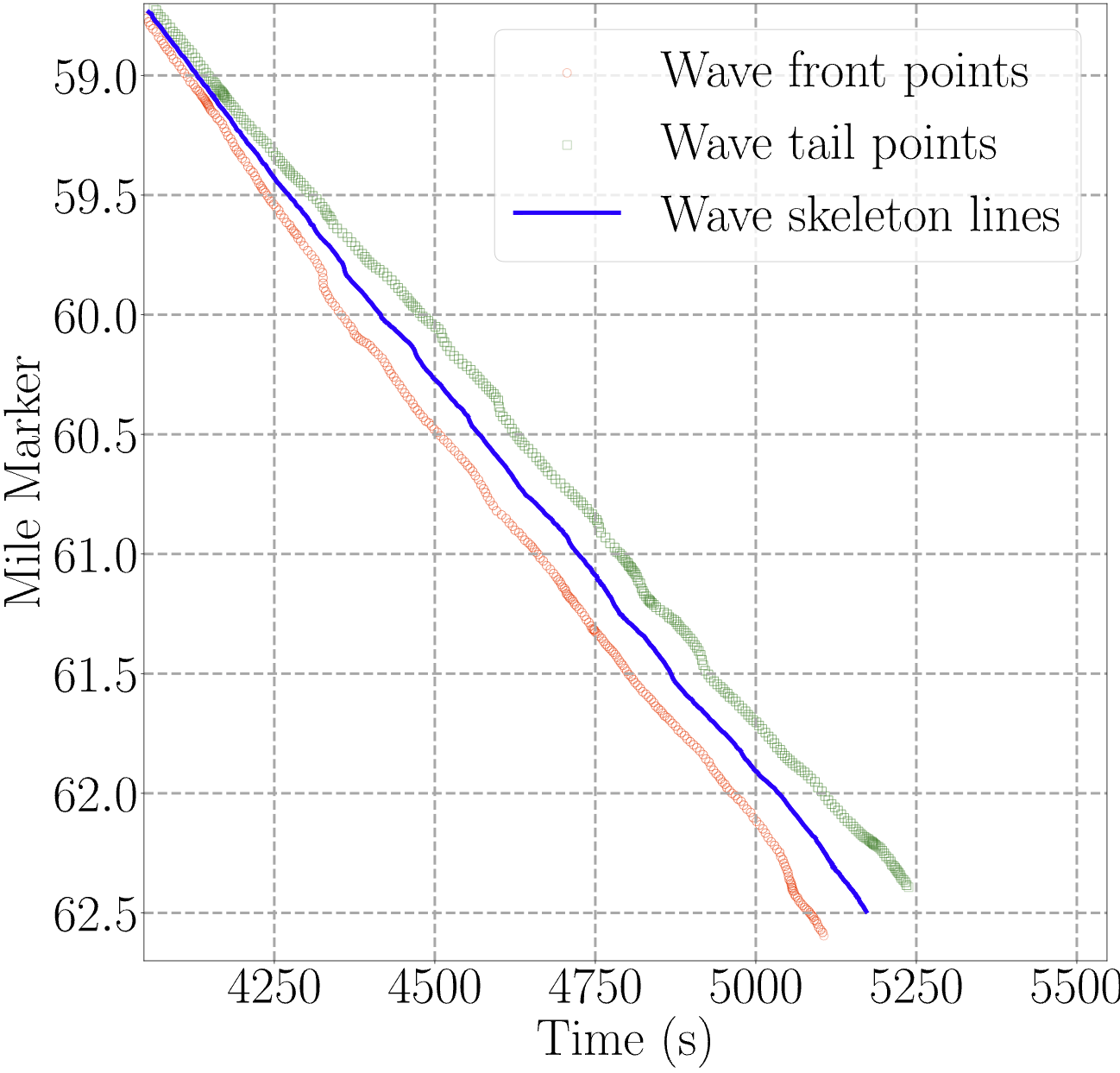}
        \captionsetup{labelformat=empty}
        \caption{(c) Sample topology from cluster 1}
        \label{fig:subfig3}
    \end{subfigure}
    \caption{\textbf{Topology samples from different clusters:} (a) A sample wave topology from the most complex cluster bifurcates 17 times and merges 15 times, traveling 3.62 miles;
    (b) A sample wave topology from the intermediate complexity cluster bifurcates 5 times and merges 2 times, traveling 3.58 miles;
    (c) A sample wave topology from the simplest cluster has no bifurcations or merges, traveling 3.64 miles.}
    \label{fig:complex}
\end{figure}

\begin{itemize}
    \item \textbf{Cluster 1} in Figure~\ref{fig:clustering} are waves with simple topological structures: the wave components in this cluster can travel between 0 and 4 miles, experiencing either no bifurcations and merges or only 1-2 occurrences. The wave topology remains relatively stable. Figure~\ref{fig:subfig3} demonstrate one sample topology from this cluster.
    \item \textbf{Cluster 2} falls in between, being neither highly complex nor overly simple. Figure~\ref{fig:subfig2} demonstrate one sample topology from this cluster. 
    \item \textbf{Cluster 3} consists of complex topologies, where a single wave can bifurcate and merge back more than 10 times over a travel distance of more than 3 miles. Figure~\ref{fig:subfig1} demonstrate one sample topology from this cluster.
\end{itemize}

\subsection{Spatial distribution of wave bifurcation and merge}
\begin{figure}[H]
    \centering
    \begin{subfigure}{0.45\textwidth}
        \centering
        \includegraphics[width=\linewidth]{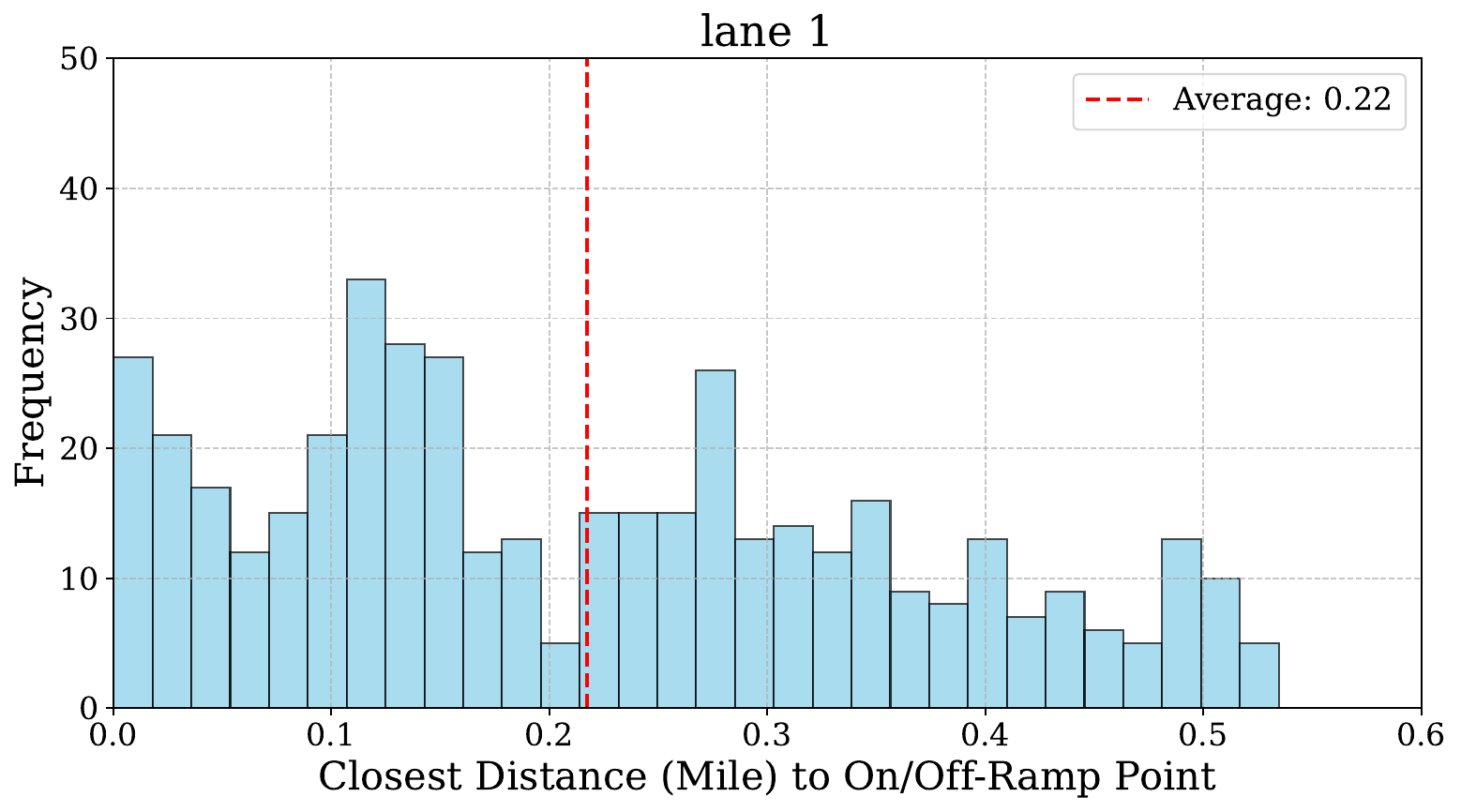}
        \caption{Lane 1 wave bifurcation}
    \end{subfigure}
    \begin{subfigure}{0.45\textwidth}
        \centering
        \includegraphics[width=\linewidth]{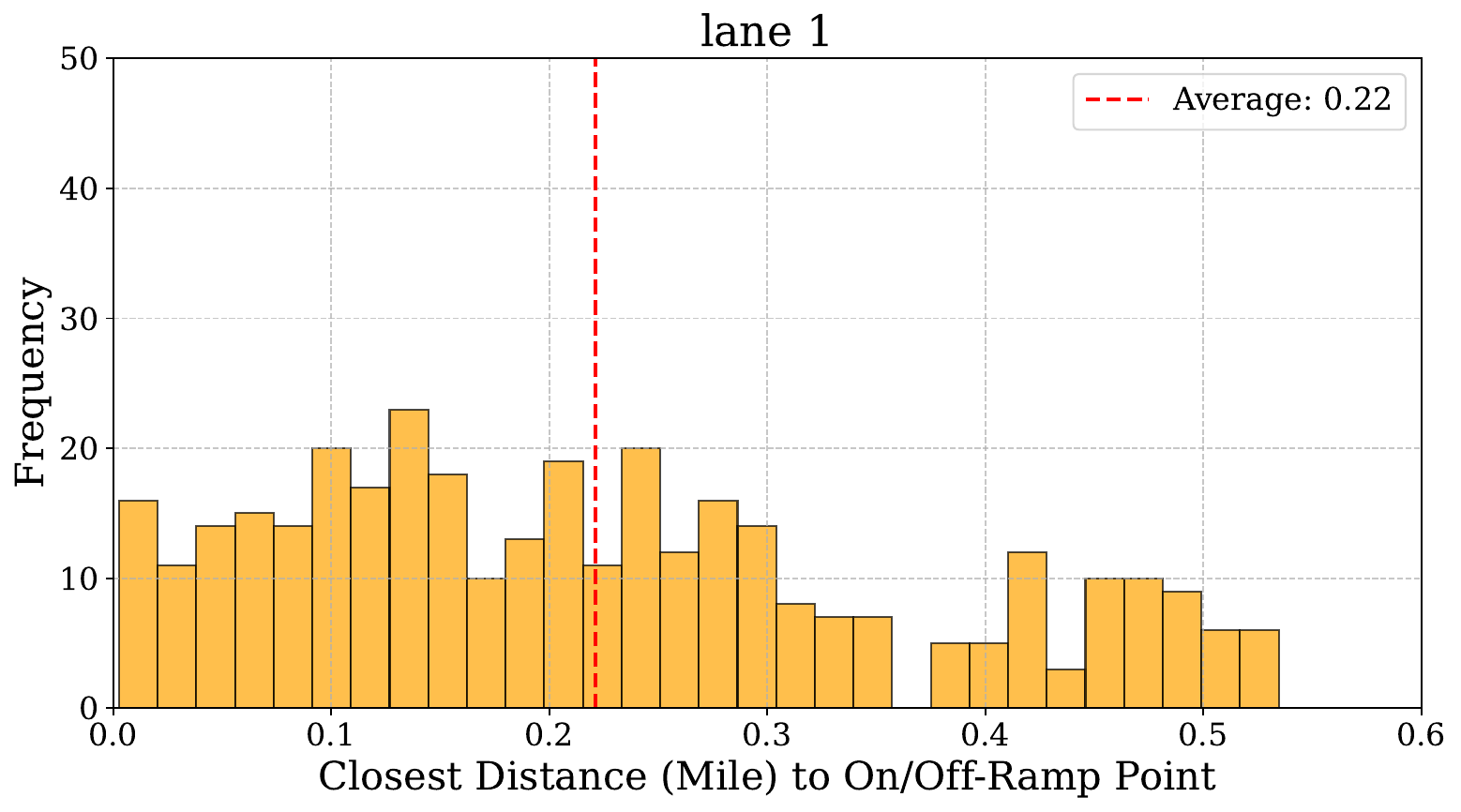}
        \caption{Lane 1 wave merge}
    \end{subfigure}
    \vspace{0.1cm}
        \begin{subfigure}{0.45\textwidth}
        \centering
        \includegraphics[width=\linewidth]{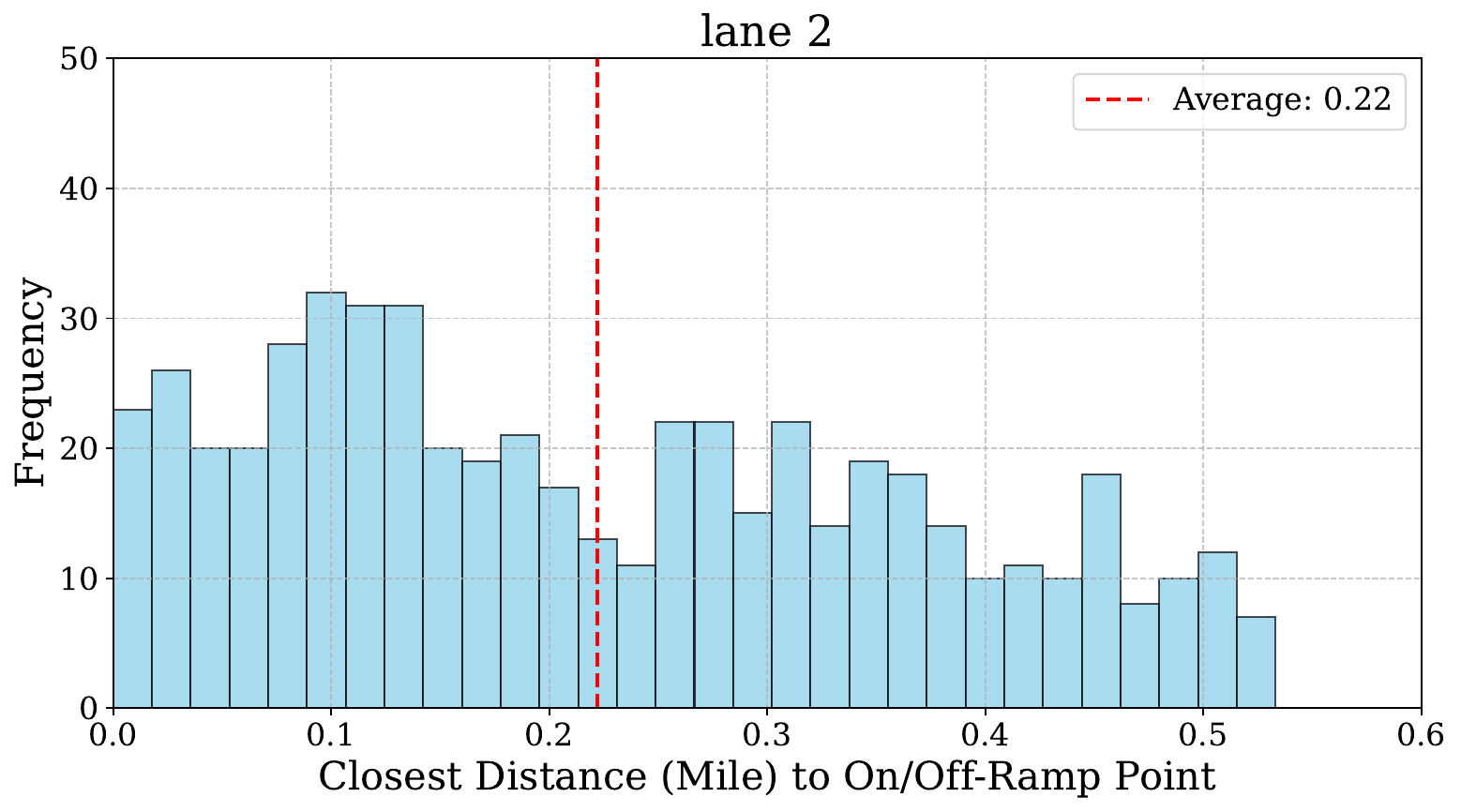}
        \caption{Lane 2 wave bifurcation}
    \end{subfigure}
    \begin{subfigure}{0.45\textwidth}
        \centering
        \includegraphics[width=\linewidth]{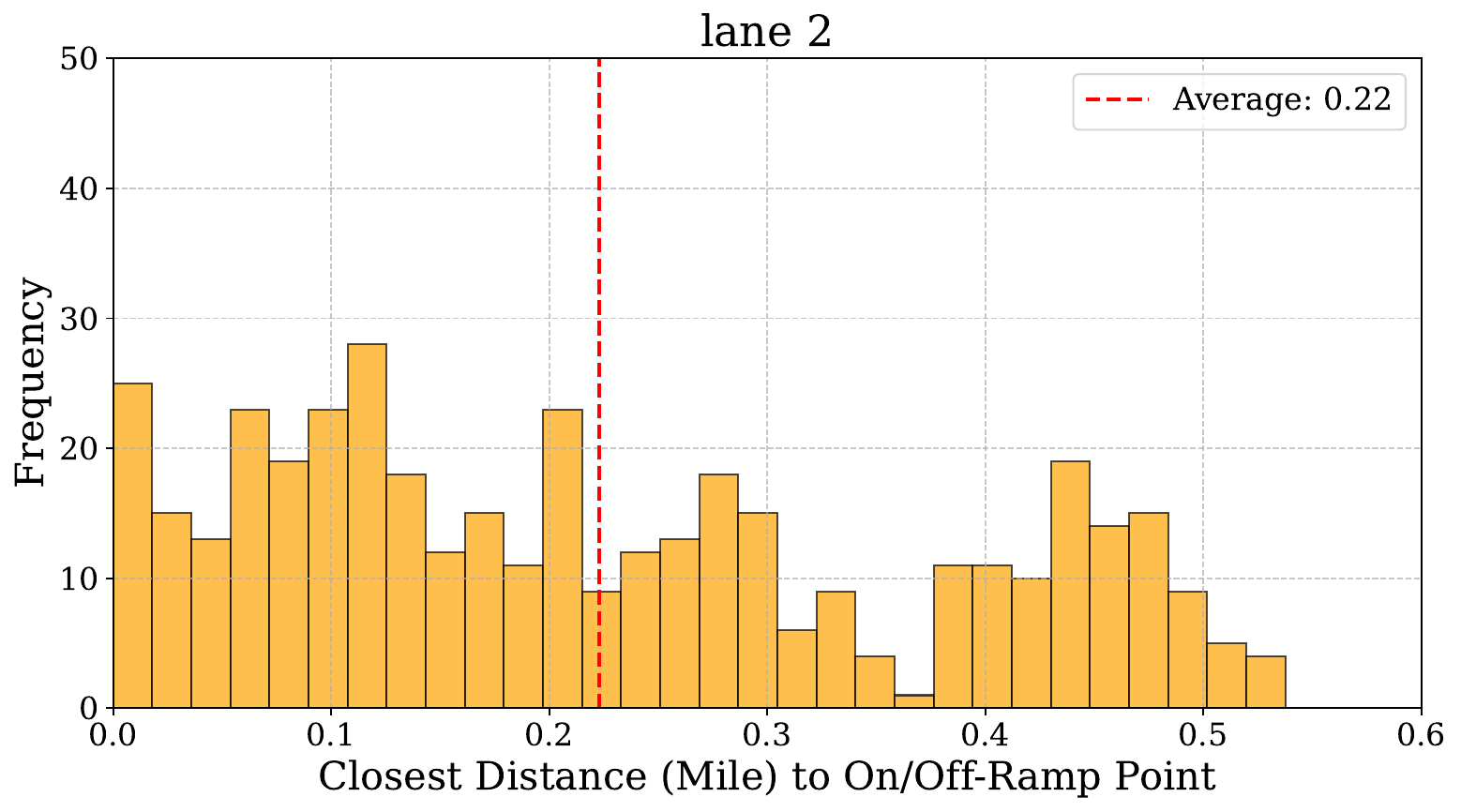}
        \caption{Lane 2 wave merge}
    \end{subfigure}
    \vspace{0.1cm}
    \begin{subfigure}{0.45\textwidth}
        \centering
        \includegraphics[width=\linewidth]{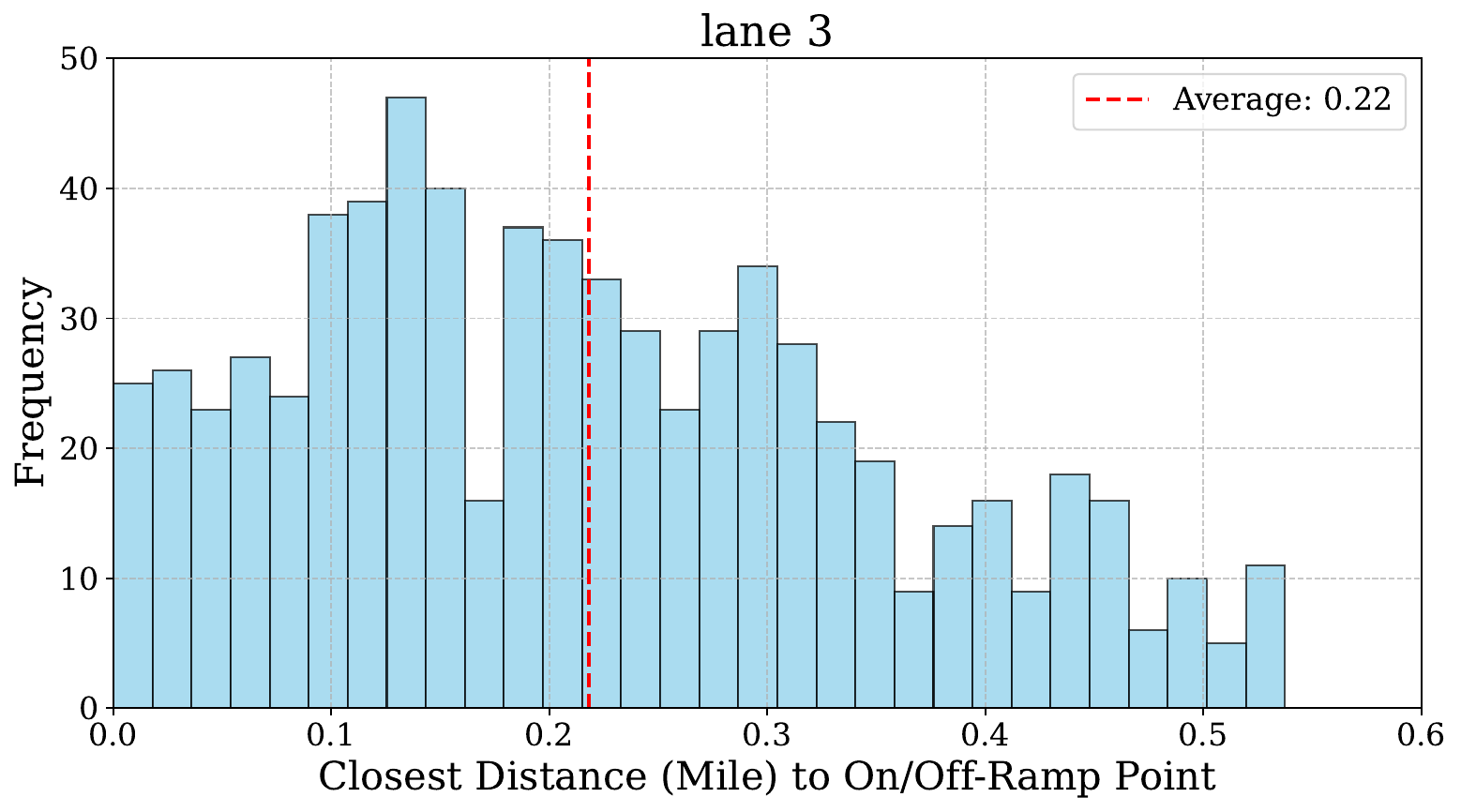}
        \caption{Lane 3 wave bifurcation}

    \end{subfigure}
    \begin{subfigure}{0.45\textwidth}
        \centering
        \includegraphics[width=\linewidth]{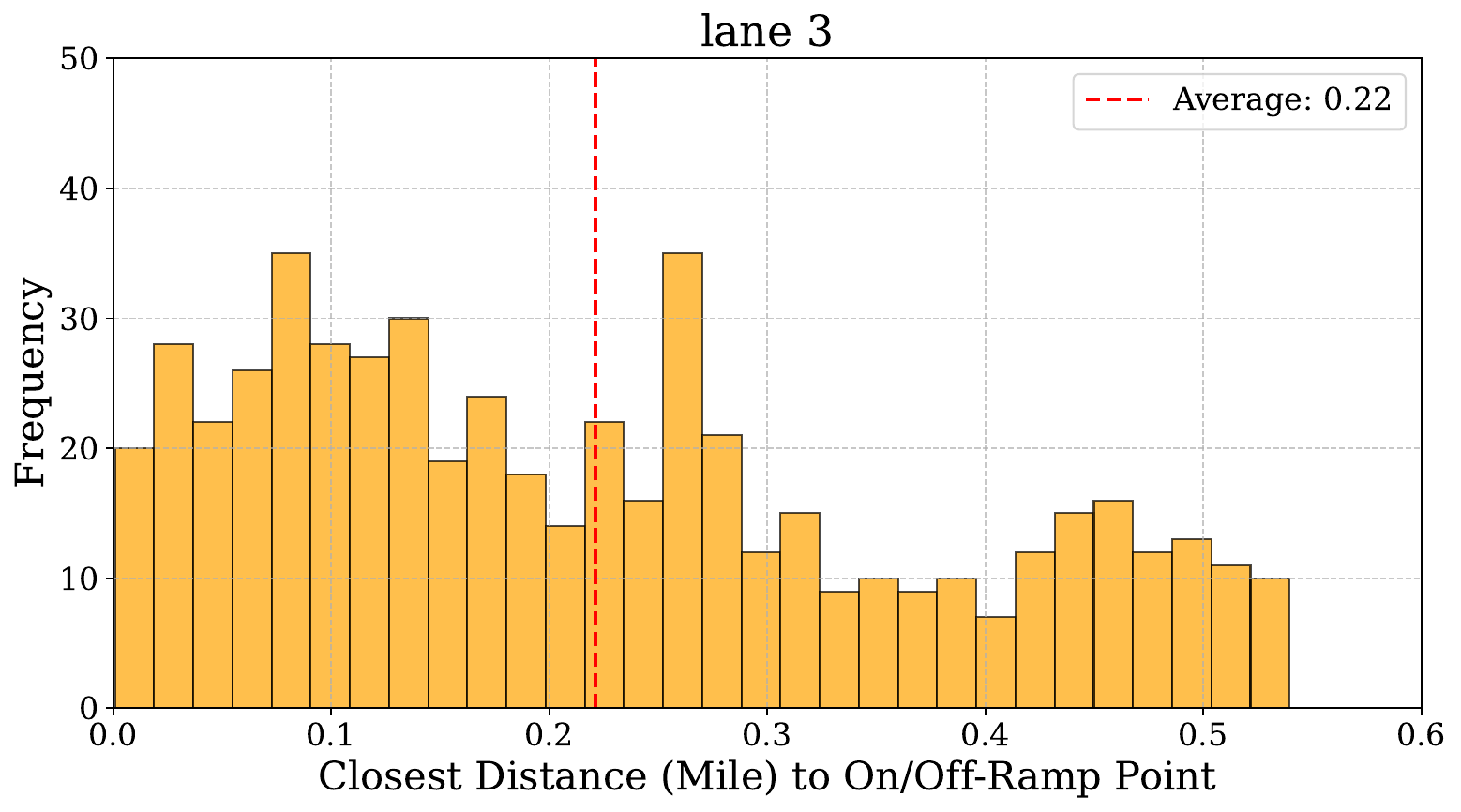}
        \caption{Lane 3 wave merge}

    \end{subfigure}
    \vspace{0.1cm}
    \begin{subfigure}{0.45\textwidth}
        \centering
        \includegraphics[width=\linewidth]{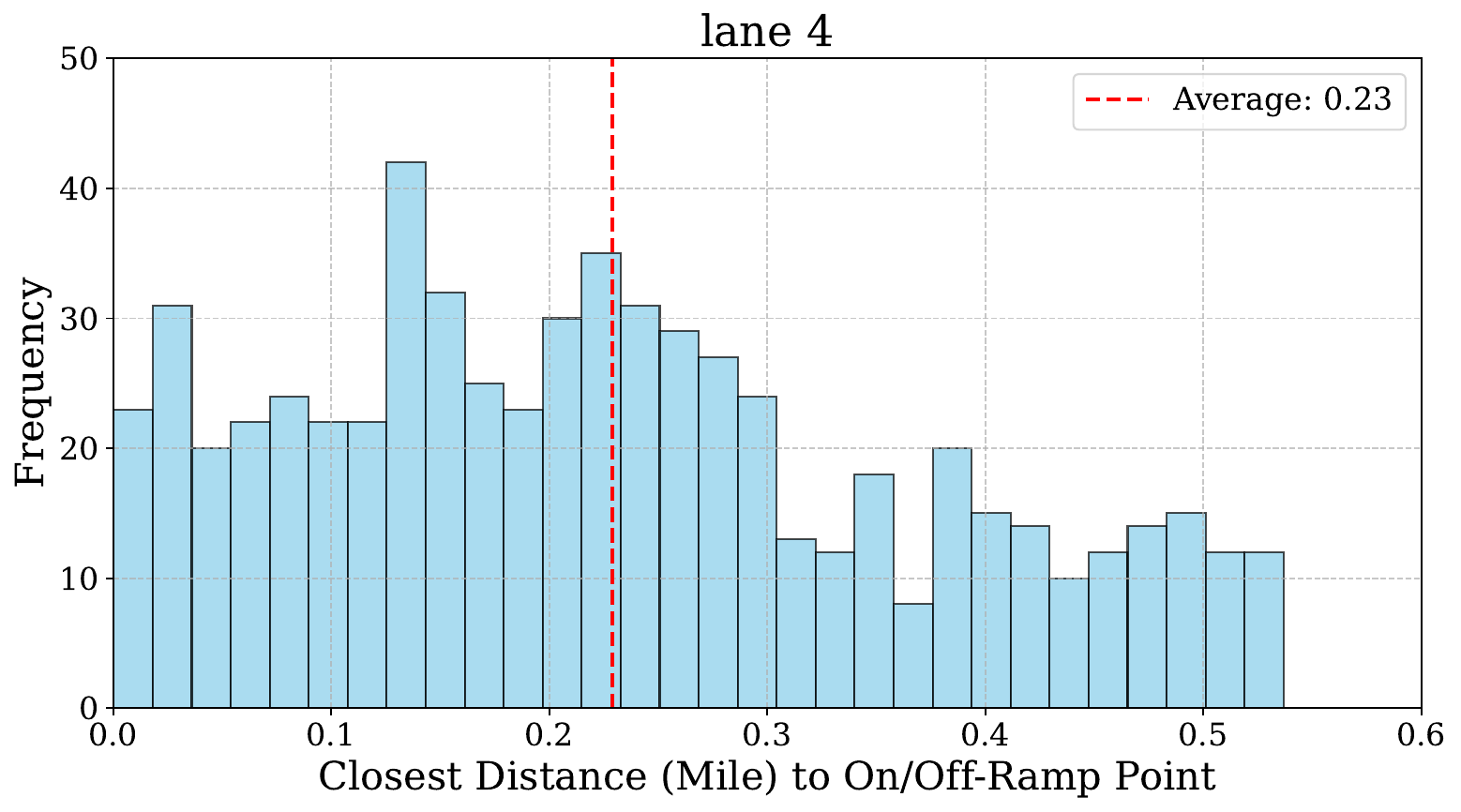}
        \caption{Lane 4 wave bifurcation}

    \end{subfigure}
    \begin{subfigure}{0.45\textwidth}
        \centering
        \includegraphics[width=\linewidth]{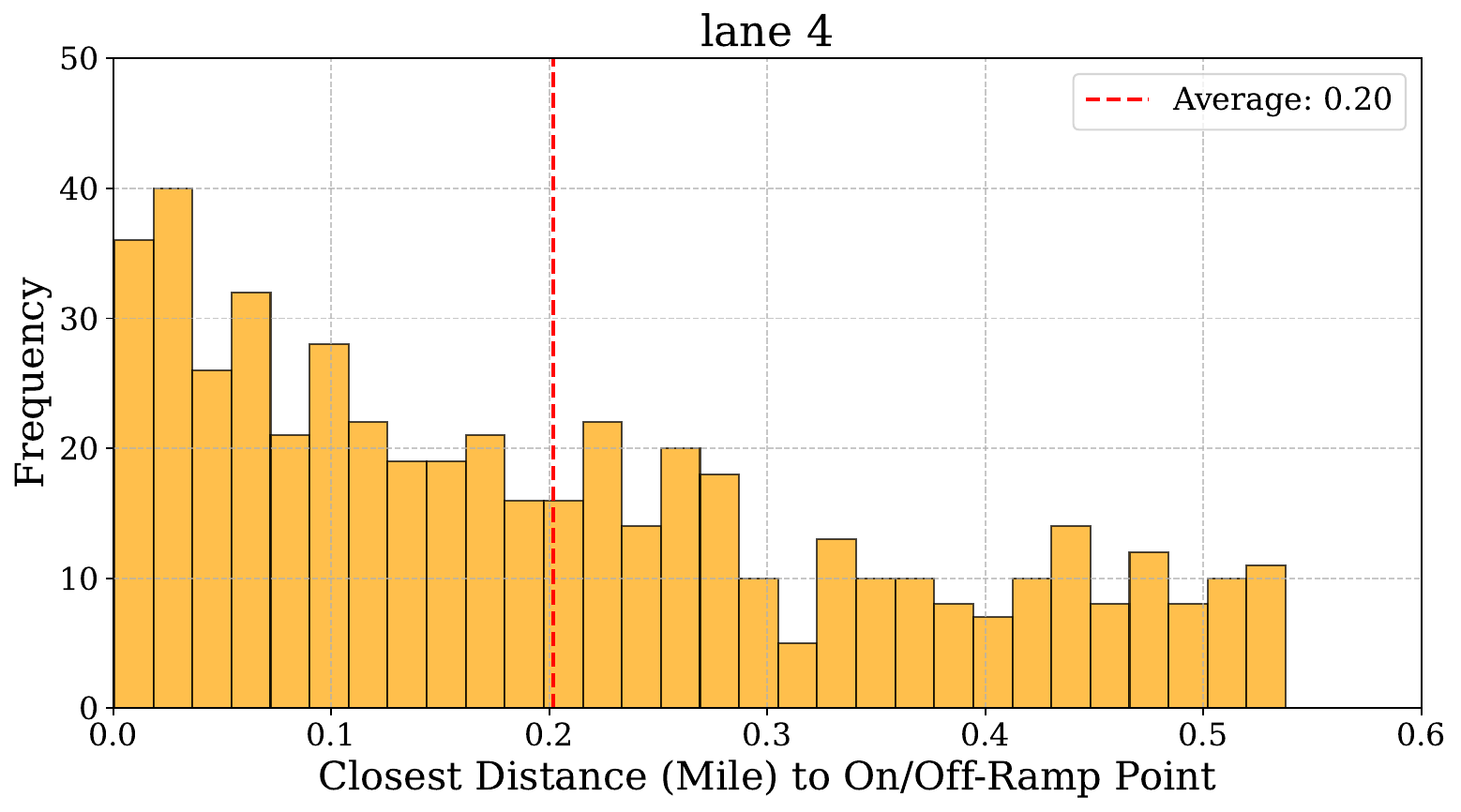}
        \caption{Lane 4 wave merge}

    \end{subfigure}
    \caption{Wave bifurcation and merge points spatial distribution across all lanes}
    \label{fig:birf-merge-dis}
\end{figure}

With all the identified wave topologies, we further plot the spatial locations of wave bifurcation and merging points to examine their relationship with ramps. Figure~\ref{fig:birf-merge-dis} demonstrates the relationship between wave bifurcation (one wave split to multiple waves) and merge points (multiple waves combine to one) across lanes regarding their proximity (absolute distance) to the closest on/off-ramp points \cite{gloudemans202324}.
The histograms show the frequency distribution of wave bifurcations and merges occurrence at different distances from the on/off-ramps. The figures in the left column (shaded light blue) show the distribution of wave bifurcations, while the figures in the right column (shaded orange) show the distribution of wave merges.

From Figure~\ref{fig:birf-merge-dis}, we can observe that wave bifurcation occurs closer to ramps, while wave merge happens slightly further from ramps, with a more balanced distribution. However, it does reveal a pattern indicating that road geometry features can explain the occurrence of wave bifurcation and merging, as they are more concentrated in the proximity distribution. On the other hand, wave bifurcation and merging also occur in areas located 0.5 miles away from any on-ramps or off-ramps, suggesting that driver behavior may also play a role in these complex dynamics.

Although identifying the root cause of bifurcation and merging is challenging, our analysis of road geometry suggests that these phenomena can occur anywhere, although with a concentration within the ramp impact area. We believe the underlying cause is more closely tied to driving behavior, with road geometry acting as a contributing factor by influencing the frequency of cut-ins and lane changes. Factors such as asymmetric driving behavior \cite{yeo2008asymmetric,yeo2009understanding}, anticipation effects \cite{treiber2007influence}, and driver distraction \cite{cooper2009investigation} may play a significant role. A more detailed microscopic analysis of driving behavior, enabled by improvements in data quality \cite{gloudemans2024so}, could provide deeper insights in the future.

\section{Conclusion}
\label{sec:conclusion}
\edit{
This article presents an automatic and scalable approach for identifying stop-and-go wave boundaries, enabling large-scale analysis of such phenomena. The method captures wave generation, propagation, dissipation, as well as bifurcation and merging. It builds on a graph-based representation of the spatio-temporal points associated with stop-and-go waves, specifically wave front (start) points and wave tail (end) points, and frames the solution as a graph component identification problem. This study establishes a foundation for measuring traffic wave properties, a prerequisite for evaluating how microscopic~\cite{shanto2020challenges}, mesoscopic~\cite{leea2024empirical}, and macroscopic~\cite{wang2022macroscopic} models reproduce observed traffic wave phenomena. The approach can also serve as a performance metric for iterative calibration and hyper-parameter optimization. Recent applications, including the use of generative artificial intelligence for traffic wave reconstruction~\cite{ji2025stop}, have applied this method to measure the physical properties of generative outputs.

In addition, this study provides new empirical insights into the complexity of traffic wave dynamics. By identifying and clustering approximately 400 waves per lane through a graph-based representation, we demonstrate  for the first time that traffic waves can bifurcate and merge at scales not previously observed. These findings highlight the need for traffic models to incorporate bifurcation and merging phenomena in order to more realistically capture traffic flow dynamics. While road geometry plays a role in these dynamics, our analysis shows it is not the sole determining factor, suggesting a more combined effects of different factors. 
The scale and the complexity of the stop-and-go waves may encourage researchers in the community to revisit the stop-and-go waves phenomena observed in the NGSIM dataset and other experimental datasets. The rich topologies within the wave boundaries warrant further investigation.
}

The code developed in this article is publicly available at \url{https://github.com/I24-MOTION/wave-analysis} and will be listed on \url{https://i24motion.org}. The tools introduced in this article could open new possibilities for analyzing stop-and-go traffic waves at scale and it could lower the barrier for researchers to deal with large-scale trajectory data. For example, the I-24 MOTION instrument \cite{gloudemans202324} will enable both CAV field experiments \cite{lee2025traffic,mattas2025safety} and ITS deployments \cite{zhang2025real}.
It is worthy of attention that the method and tool presented in this article allow for empirically analyzing how control strategies and vehicle automation influence traffic wave properties, a growing topic within traffic flow theory and characteristics \cite{zhou2025traffic,mattas2025effects,han2025capacity,tgsimwave,ji2025stop,mattas2025safety}. \edit{Further research will focus on investigating the factors influencing wave evolution \cite{suh2016empirical}, accompanied by refinements in data quality.}

\section*{Acknowledgment}
The authors would like to extend their gratitude to Professor Martin Treiber at TU Dresden for his insights regarding the dimensions of Edie's box. Preliminary version of this work is presented at Conference in Emerging Technologies in Transportation Systems (TRC-30). This work is supported by the Tennessee Department of Transportation (TDOT) under Grant No. OTH2023-01F-01 and the National Science Foundation (NSF) under Grant No. 2135579 (Work, Sprinkle), and 2111688 (Sprinkle). The views expressed herein do not necessarily represent the views of the Tennessee Department of Transportation or the United States Government.

\appendix

\section{Virtual Trajectory Generation}
\label{app:virtual}

\subsection{Virtual sensors and virtual vehicles}
Long-range raw vehicle trajectory data is imperfect due to the challenges of tracking and re-identification in dense traffic conditions \cite{gloudemans2024so,wang2024automatic}. The tasks fail when errors occurring at the current frame cannot be corrected using information from subsequent frames. 
These errors may arise from various system sources, including homography estimation \cite{gloudemans2024so} and object occlusion \cite{coifman2024partial}.
However, these challenges do not significantly impact local traffic measurements, such as traffic speeds, which can still be obtained with extremely high fidelity. In this context, generating vehicle trajectories \cite{tsanakas2022generating} from high-fidelity speed field can serve as a surrogate measure for the raw trajectory data, enabling the analysis of stop-and-go waves.We use the virtual trajectory generation method from \cite{ji2024virtual} as a first step in wave identification in this work. We detail this approach below. Table \ref{tab:variables} provides a comprehensive list of the variables and parameters used throughout the appendix.

\begin{table}[ht]
\caption{\textbf{Variables and parameters defined and utilized in the appendix}: the parameter column indicates whether it is a parameter in the paper.}
\centering
{\scriptsize
\begin{tabular}{clcr}
\toprule
\textbf{Notation} & \textbf{Description} & \textbf{Unit}& \textbf{Parameter Setting} \\ \hline
$N$ & total number of trajectories from raw data & -  & -\\
$\rho_E(t,x)$ & Edie's definition for traffic density at time $t$ and space $x$ & veh/mile & - \\
$q_E(t,x)$    & Edie's definition for traffic flow at time $t$ and space $x$ & veh/hr& -\\
$v_E(t,x)$    & Edie's definition for traffic speed at time $t$ and space $x$& mph & -\\
$\Delta t$    & width of the shear box & second& 4\\
$\Delta x$    & height of the shear box & mile& 0.02\\
$w$    & wave propagation speed for the shear box $t$ and space $x$& mph & -12.5\\
\hline
$M_o$    & the startpoint of the empirical data collection site & mile& 62.7\\
$M_d$    & the endpoint of the empirical data collection site & mile& 58.7\\
$T_s$    & the sampling interval for generating virtual trajectories  & second   & 1\\
$T_v$    &  the  interval for each virtual vehicle entering the site  & second   &  5 \\
$N$ &  the number of virtual vehicles & -  & - \\
\bottomrule
\end{tabular}
}
\label{tab:app-variables}
\end{table}

\subsubsection{Virtual sensors: Edie's definition for macroscopic measurements}
\begin{figure}
    \centering
    \includegraphics[width=0.5\textwidth]{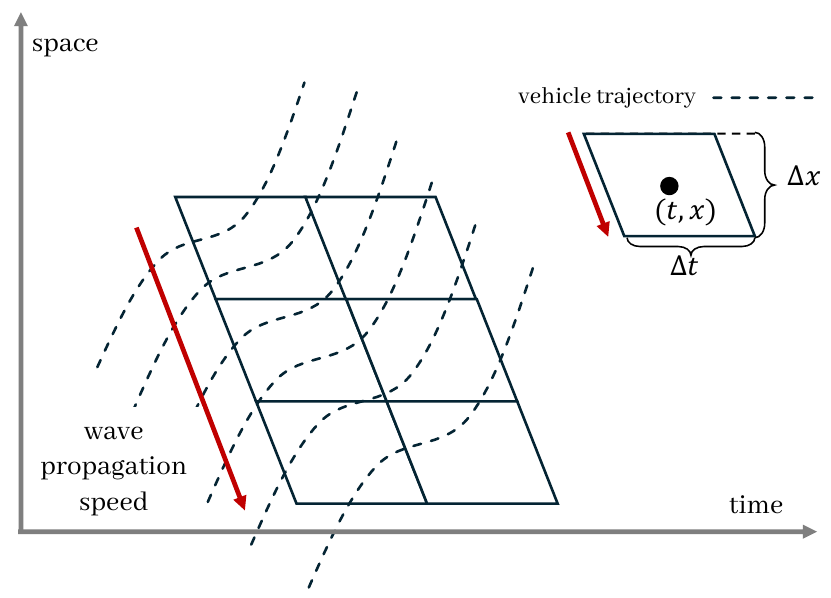}
    \caption{\textbf{Macroscopic measurements field calculation}: Illustration of Edie's definition \cite{edie1963discussion} applied to a shear box of size  $\Delta t$ $\times$ $\Delta x$. Dashed lines shows the vehicle trajectories collected from field, and the shear box marks where we quantify the macroscopic measurements. The red arrow points out the stop-and-go wave propagating against the traffic. The shear box's angle matches this wave direction.}
    \label{fig:edie}
\end{figure}

Edie \cite{edie1963discussion} provides an approach to calculate spatio-temporal mean of density, flow speed, and traffic flow from vehicle trajectories, which is named as Edie's definition (shown in Figure \ref{fig:edie}).  According to the definition, the density and the flow can be computed from the \textit{total travel time} (TTT) and \textit{total travel distance} (TTD) of all the vehicles within an area. Using Edie's definition allows for the parallel calculation of measurements on a trajectory-by-trajectory basis \cite{ji2024virtual}. Consider a shear box \cite{he2017constructing,tsanakas2022generating} centered at a point ($t$, $x$), and let $\Delta t$ and $\Delta x$ denote the height and width of the box, as shown in Figure \ref{fig:edie}. The macroscopic estimates can be computed as:
\begin{align}
    \rho_{E} (t,~x) &= \frac{\text{TTT}(t,~x)}{\Delta x \times \Delta t} = \frac{\sum_i^Nt_i}{\Delta x \times \Delta t}, \label{eq:rhoE} \\
    q_{E} (t,~x) &= \frac{\text{TTD}(t,~x)}{\Delta x \times \Delta t} = \frac{\sum_i^N x_i}{\Delta x \times \Delta t}\label{eq:QE}, \\
    v_{E} (t,~x) &= \frac{q_{E}(t,~x)}{\rho_{E}(t,~x)}, \label{eq:VE}
\end{align}
where $\rho_{E},~q_{E},$ and $v_E$ represent Edie's definition for density, flow and speed respectively, Here, $t_i$ and $x_i$ denote the travel time and travel distance of vehicle $i$, respectively, with a total of $N$ vehicle trajectories from raw data. 

\subsubsection{Adaptive Smoothing}
With the original speed field generated, virtual trajectories can be created. However, the original macroscopic measurements often contain missing values and outliers. As suggested by \cite{tsanakas2022generating}, an adaptive smoothing method should be applied to the raw data to improve accuracy. The \textit{Adaptive Smoothing Method} (ASM), developed by \cite{treiber2003adaptive,treiber2011reconstructing}, is a widely used smoothing and interpolation algorithm for constructing a continuous spatio-temporal mean speed field. While it is especially useful for data collected from fixed infrastructure sensors, such as inductive loops, it is also applicable to the macroscopic speed data generated from trajectories. This method is particularly helpful in interpolating speeds in occluded areas, such as under bridges. We directly apply this method, with parameter settings as listed in Table \ref{tab:ASM}.

\begin{table}[H]
\centering
\caption{Parameters of the adaptive smoothing method}
\scriptsize{
\begin{tabular}{llll}
\toprule
         & Meaning & Value \\ \hline
$\sigma$(mile) &  smoothing width in space coordinate  & 0.12     \\
$\tau$(second) & smoothing width in time coordinate           &  20  \\ 
$c_{\text{free}} $(mph) & wave speed in free traffic        &-12.5     \\ 
$c_{\text{cong}} $(mph)& wave speed in congested traffic        &60.0\\ 
$V_{\text{thr}}$(mph) & crossover from congested to free traffic        &37.29  \\ 
$\Delta V$(mph) & transition width between congested and free traffic        &12.43 \\ 
\bottomrule
\end{tabular}
\label{tab:ASM}
}
\end{table}

\subsubsection{Virtual trajectory generation}
A standard approach to generate trajectories from a macroscopic speed field is to calculate the position $\tau(t)$ of a vehicle assuming the velocity dynamics of the vehicle are computed as follows:
\begin{equation}
\frac{\mathrm{d}\tau(t)}{\mathrm{d}t} = v_E(t,\tau(t)),
\label{eq:euler}
\end{equation}
given an initial condition $\tau(0)=\tau_0$, the process terminates when $\tau(t) = M_d$, where $M_d$ represents the endpoint of the empirical data collection site. The solution to the ordinary differential equation referred to as \eqref{eq:euler} can be approximated using the forward Euler method with a small timestep $T_s$. When the integration timestep employed in solving the ordinary differential equation is small compared to the width $\Delta t$ used to generate the macroscopic speed field, the resulting trajectories may exhibit quantization artifacts. To enhance the smoothness of these trajectories, cubic interpolation \cite{fritsch1980monotone} is recommended by \cite{tsanakas2022generating}, a technique we have also adopted. The virtual vehicles are sent from the start point of the empirical data collection site $M_o$ at a frequency of $T_v$.

\subsubsection{Parameter settings for Virtual Trajectory Generation}
Table \ref{tab:app-variables} outlines the parameters utilized in this study. The Edie's box sizes are set to 4 seconds by 0.02 miles (approximately 32 meters). The wave propagation speed for constructing the shear box is predefined at -12.5 mph \cite{gloudemans202324}. Virtual vehicles are introduced every 5 seconds, with a sampling frequency of 1 second. 

\section{Breadth-First Search Algorithms for Graph Component Identification}
\label{app:bfs}

Algorithm \ref{alg:ccd} details the procedure for identifying components within the wave front $\mathcal{G}_d$ (or wave tail graph $\mathcal{G}_a$).

\begin{small}
\begin{algorithm}[H]
\caption{Identifying components in the wave fronts graph $\mathcal{G}_d$}
\label{alg:ccd}
\KwIn{Graph $\mathcal{G}_d = (\mathcal{D}, \mathcal{E}_{\text{cross}}^\mathcal{D}) $ }
\KwOut{Connected components $\mathcal{C}^d = \{\mathcal{C}_i^d\}_{i=1}^{N_c^d}$ in the graph $\mathcal{G}_d$}
\SetKwComment{Comment}{\# }{}
Initialize a list of components $\mathcal{C}^d$ and $i=0$\;

Initialize a set $\mathcal{V}_v \leftarrow \emptyset$ \Comment{To keep track of visited nodes}
\For{each node $v$ in graph $\mathcal{G}_d$}{
    \If{$v \notin \mathcal{V}_v$}{
        $i \leftarrow i + 1$\;
        Initialize $\mathcal{V}_\mathcal{C} \leftarrow \{v\}$ and $\mathcal{E}_\mathcal{C} \leftarrow \emptyset$\;
        Add $v$ to $\mathcal{V}_v$\;
        Initialize a queue $Q$ and enqueue $v$\;
        \While{$Q$ is not empty}{
            $u \leftarrow$ dequeue $Q$\;
            \For{each neighbor $w$ of $u$ via edge $e = (u, w) \in \mathcal{E}_{\text{cross}}^\mathcal{D})$ }{
                \If{$w \notin \mathcal{V}_v$}{
                    Add $w$ to $\mathcal{V}_{\mathcal{C}}$\;
                    Add $w$ to $\mathcal{V}_v$\;
                    Enqueue $w$ to $Q$\;
                }
                \If{$e \notin \mathcal{E}_\mathcal{C}$}{
                    Add $e$ to $\mathcal{E}_\mathcal{C}$\;
                }
            }
        }
        $\mathcal{C}_i^d \leftarrow (\{\mathcal{V}_\mathcal{C}, \mathcal{E}_\mathcal{C}\})$\;
        Append $\mathcal{C}_i^d$ to $\mathcal{C}^d$\;
    }
}
\Return{$\mathcal{C}^d$}
\end{algorithm}
\end{small}

Algorithm \ref{alg:ccg} details the procedure for identifying the components for $\mathcal{G}$, given the properties defined for the edges:
\begin{equation}
    \mathcal{E} = \mathcal{E}_{\text{inner}} + \mathcal{E}^\mathcal{D}_{\text{cross}} + \mathcal{E}^\mathcal{A}_{\text{cross}}
\end{equation}

The components of $\mathcal{G}$ can be determined by starting from $\mathcal{C}^d$ (the components associated with $\mathcal{E}^\mathcal{D}_{\text{cross}}$) and $\mathcal{C}^a$ (the components associated with $\mathcal{E}^\mathcal{A}_{\text{cross}}$), and using the edges $\mathcal{E}_{\text{inner}}$ to identify any unconnected components within the graph. 

\begin{small}
\begin{algorithm}[H]
\caption{Identifying components in the stop-and-go graph $\mathcal{G}$}
\label{alg:ccg}
\KwIn{Graph $\mathcal{G}$, $\mathcal{C}^d$ and $\mathcal{C}^a$, edges $\mathcal{E}_{\text{inner}}$}
\KwOut{Components $\mathcal{C}= \{C_k\}_{k=1}^{N_c}$ of $\mathcal{G}$}
Initialize a list of components $\mathcal{C}$ and $k=0$\;
Initialize a list of components $\mathcal{C}_v^d$ and $\mathcal{C}_v^a$ \;
\For{each component $\mathcal{C}_i^d \in \mathcal{C}^d$}{
    \If{$\mathcal{C}_i^d \notin \mathcal{C}_v^d$}{
        $k \leftarrow k + 1$\;
        Add $\mathcal{C}_i^d$ to $\mathcal{C}_v^d$\;
        Initialize a new component $\mathcal{C}_k \leftarrow \mathcal{C}_i^d$\;

        \While{there are unvisited components in $\mathcal{C}_k$}{
            \For{each component $c \in \mathcal{C}_k$}{
                \If{$c \in \mathcal{C}^d$}{
                    \For{each component $\mathcal{C}_j^a \in \mathcal{C}^a$}{
                        \If{$\mathcal{C}_j^a \notin \mathcal{C}_v^a$ \textbf{and} $c$ is connected to any $w \in \mathcal{C}_j^a$ via $\mathcal{E}_{\text{inner}}$}{
                            Merge $\mathcal{C}_j^a$ into $\mathcal{C}_k$\;
                            Add $\mathcal{C}_j^a$ to $\mathcal{C}_v^a$\;
                        }
                    }
                } \ElseIf{$c \in \mathcal{C}^a$}{
                    \For{each component $\mathcal{C}_j^d \in \mathcal{C}^d$}{
                        \If{$\mathcal{C}_j^d \notin \mathcal{C}_v^d$ \textbf{and} $c$ is connected to any $w \in \mathcal{C}_j^d$ via $\mathcal{E}_{\text{inner}}$}{
                            Merge $\mathcal{C}_j^d$ into $\mathcal{C}_k$\;
                            Add $\mathcal{C}_j^d$ to $\mathcal{C}_v^d$\;
                        }
                    }
                }
            }
        }
        Add $\mathcal{C}_k$ to $\mathcal{C}$\;
    }
}
\Return $\mathcal{C}$\;
\end{algorithm}
\end{small}

The algorithm takes as input the graph $\mathcal{G}$, initial component sets $\mathcal{C}^d$ and $\mathcal{C}^a$, and the set of inner edges $\mathcal{E}_{\text{inner}}$. Starting by initializing an empty list $\mathcal{C}$ for the final components and a counter $k$ for component tracking, it also sets up visited sets $\mathcal{C}_v^d$ and $\mathcal{C}_v^a$. For each component $\mathcal{C}_i^d$ in $\mathcal{C}^d$, if not visited, the algorithm increments $k$, marks $\mathcal{C}_i^d$ as visited, and initializes a new component $\mathcal{C}_k$ starting with $\mathcal{C}_i^d$. It then expands $\mathcal{C}_k$ by checking connections with $\mathcal{C}^a$ via $\mathcal{E}_{\text{inner}}$ and merges connected, unvisited components into $\mathcal{C}_k$, marking them as visited. This process continues until all connections are searched. The algorithm adds each fully expanded component $\mathcal{C}_k$ to $\mathcal{C}$, repeating for all components in $\mathcal{C}^d$ and $\mathcal{C}^a$ until all possible connections are merged. The resulting list $\mathcal{C}$ represents all connected components in $\mathcal{G}$, providing a comprehensive identification of the components based on the initial sets $\mathcal{C}^d$ and $\mathcal{C}^a$ from the wave fronts and tails graph and inner edges. Each component $\mathcal{C}_k \in \mathcal{C}$ can consist of multiple sub-components from $\mathcal{C}^d$ and $\mathcal{C}^a$, specifically $\mathcal{C}_i^d$ from $\mathcal{C}^d$ and $\mathcal{C}_j^a$ from $\mathcal{C}^a$, as detailed below:
\begin{equation}
    \mathcal{C}_k = \bigcup_{i=1}^{N_k^d} \mathcal{C}_i^d \cup \bigcup_{j=1}^{N_k^a} \mathcal{C}_j^a, \quad \text{where} \ \mathcal{C}_i^d  \in \mathcal{C}^d , \mathcal{C}_j^a \in \mathcal{C}^a, 
\end{equation}
and \( N_k^d \) is the number of wave front components and \( N_k^a \) is the number of wave tail components. The component $\mathcal{C}_k$ enables the detailed analysis of wave dynamics, including how waves merge and bifurcate. By examining the interactions between the different $\mathcal{C}_i^d$ and $\mathcal{C}_j^a$ within each $\mathcal{C}_k$, we can understand the complex behaviors and patterns of wave evolution in the graph $\mathcal{G}$. Detailed analysis can be found in section \ref{sec:wave_topo}.

\section{The size of spatio-temporal search box}
Before diving into the details, the key takeaway is that if the box size is too small, the method fails, whereas a larger box size remains effective. We recommend setting the temporal size to 3 times the sending frequency of virtual vehicles, with the spatial size adjusted accordingly. Our method is computationally efficient and not highly sensitive to box size. Each run for a single day and lane is completed within 60 seconds when tested on a 2022 MacBook Air with an Apple M2 chip and 16GB of memory.

\begin{figure}[H]
    \centering
    \includegraphics[width=0.8\linewidth]{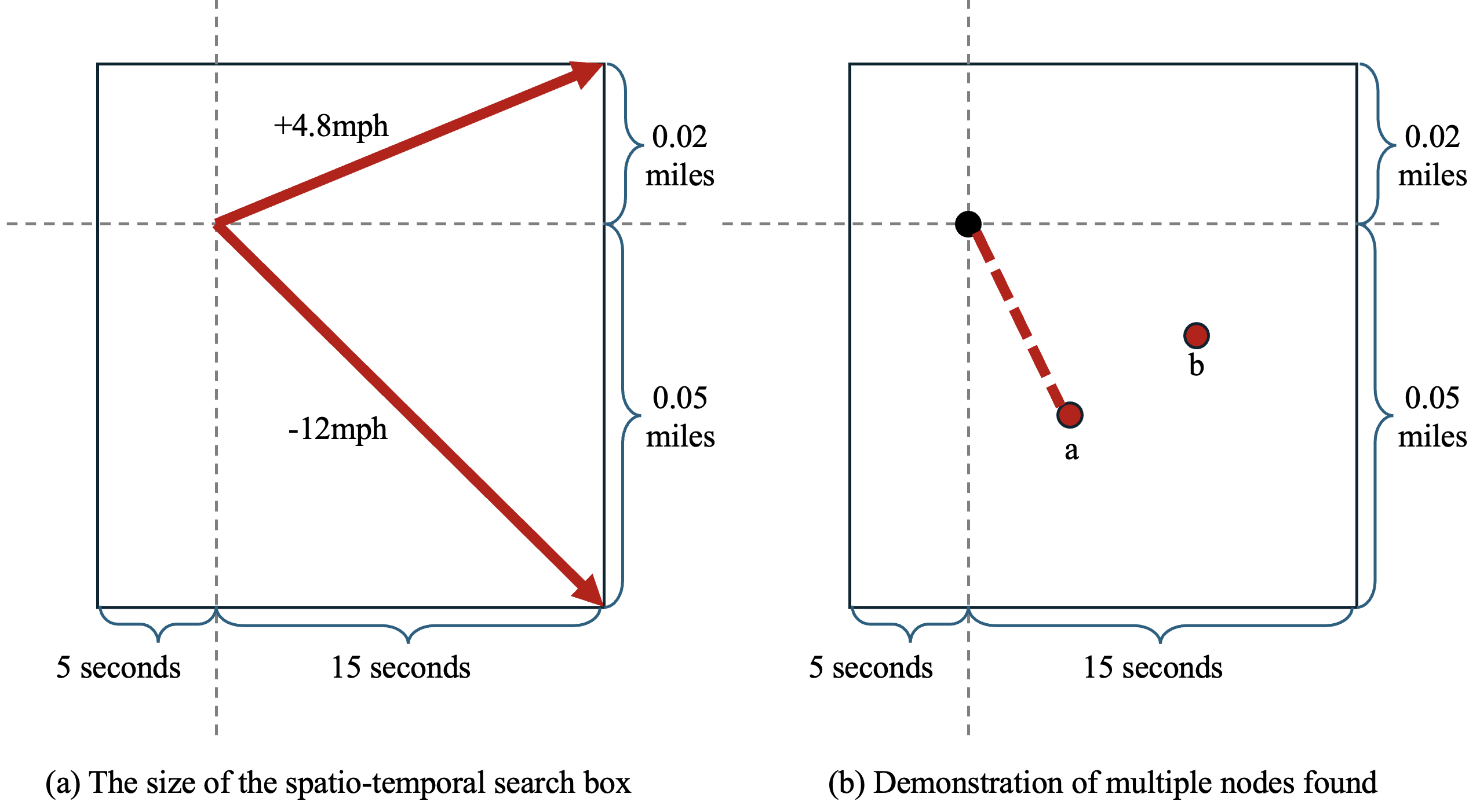}
    \caption{ \textbf{Sensitivity analysis of the spatio-temporal box:} The  effectiveness of our method deteriorates when the box size is too small.}
    \label{fig:box}
\end{figure}

First, we aim to provide details on the choice of the box size in our paper. In this study, the spatio-temporal box is defined as 20 seconds by 0.07 miles, extending spatially from -0.02 miles to 0.05 miles and temporally from -5 seconds to 15 seconds, as illustrated in Figure~\ref{fig:box}(a). The - sign indicates upstream in space and the past in time. The look-ahead box size is determined based on wave travel speed and expected time headway, ensuring it effectively captures the Newell-like car-following characteristics of the next vehicle. 

To achieve this, we use 3 times the sending frequency of virtual vehicles. The backward extension in time and space accounts for scenarios where waves remain stationary \cite{schonhof2007empirical} or when drivers engage in preventive driving, causing braking points to occur even earlier than those of the preceding vehicle. In this paper, virtual vehicles are sent to the mean speed field at 5-second intervals to generate virtual trajectories. A 15-second look-ahead window is then applied to identify the next vehicle. Starting from vehicle index $i$, we check vehicle $i+1$ only. If there are any points from other vehicle indices, such as $i+k$ where $k\geq 2$, we do not attempt to find them. If there are multiple nodes found in the spatio-temporal box, it will connect to the one closest in time. A toy example is shown in Figure~\ref{fig:box}(b), where node $a$ and $b$ are located within the box. The edge will be connected to $a$, as it is closer in time.

\section{Comparison with other traffic wave identification methods}
 In the literature, identifying traffic waves typically involves two steps: (a) detection (e.g., frequency domain method) and (b) grouping (e.g., clustering) \cite{yang2023data, tgsimwave}. The detection method we use is based on the critical speed approach \cite{edie1967generation,li2014stop} and the grouping method we propose involves a graph representation and its connected component identification. Our proposed method contributes the graph topological analysis of traffic waves at the grouping step, offering a novel perspective. In comparison, existing methods are limited in their ability to describe wave bifurcation and merging, as summarized in Table~\ref{tab:methods}.
 \begin{enumerate}
    \item[(a)] Detection: Wavelet transform and critical speed method. 
    
    We would like to clarify that we do not claim our detection method is novel or better than the others. We like the critical speed method due to its straightforward definition, which corresponds to the speed contour line in the space-time diagram \cite{edie1967generation}.
    
    A different approach is the Wavelet transform \cite{zheng2011applications}, which is a frequency-domain technique for detecting stop-and-go waves. We compare the Wavelet transform with the critical speed method to explain the reasons for selecting the critical speed method in Figure~\ref{fig:wavelet}. There are several parameters and hyper-parameters in the Wavelet transform, including the mother wavelet, scale parameters (minimum and maximum of scale), and the width of spread. We use the \verb|pywt| Python package \cite{lee2019pywavelets} to implement the code, and the parameter we choose it the \verb|gaus2| mother wavelet, \verb|min_scale|=1, \verb|max_scale|=32, \verb|width|=2.

\begin{figure}[H]
    \centering
    \includegraphics[width=0.6\linewidth]{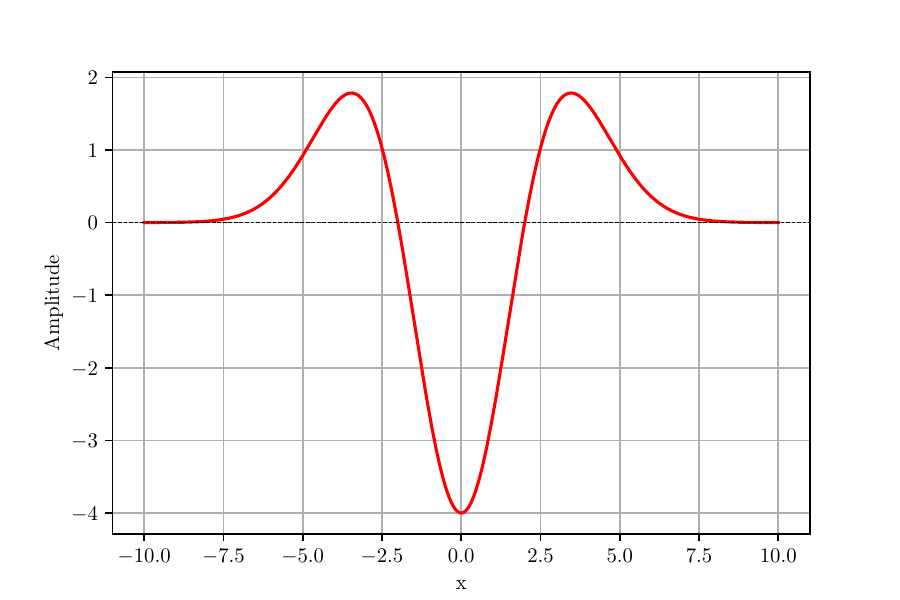}
    \caption{The second-order Gaussian mother wavelet.}
    \label{fig:gaus2}
\end{figure}
The \verb|gaus2| wavelet is visualized in Figure~\ref{fig:gaus2}, with a pattern of slow down then speed up in amplitude. We then determine the maximum coefficient for each scaled signal to identify the best-fitting frequency and distinguish the front and tail based on the sign (+/-) of the coefficient. The detected wave fronts/tails are shown in Figure~\ref{fig:wavelet}.
\begin{figure}[H]
    \centering
    \includegraphics[width=0.6\linewidth]{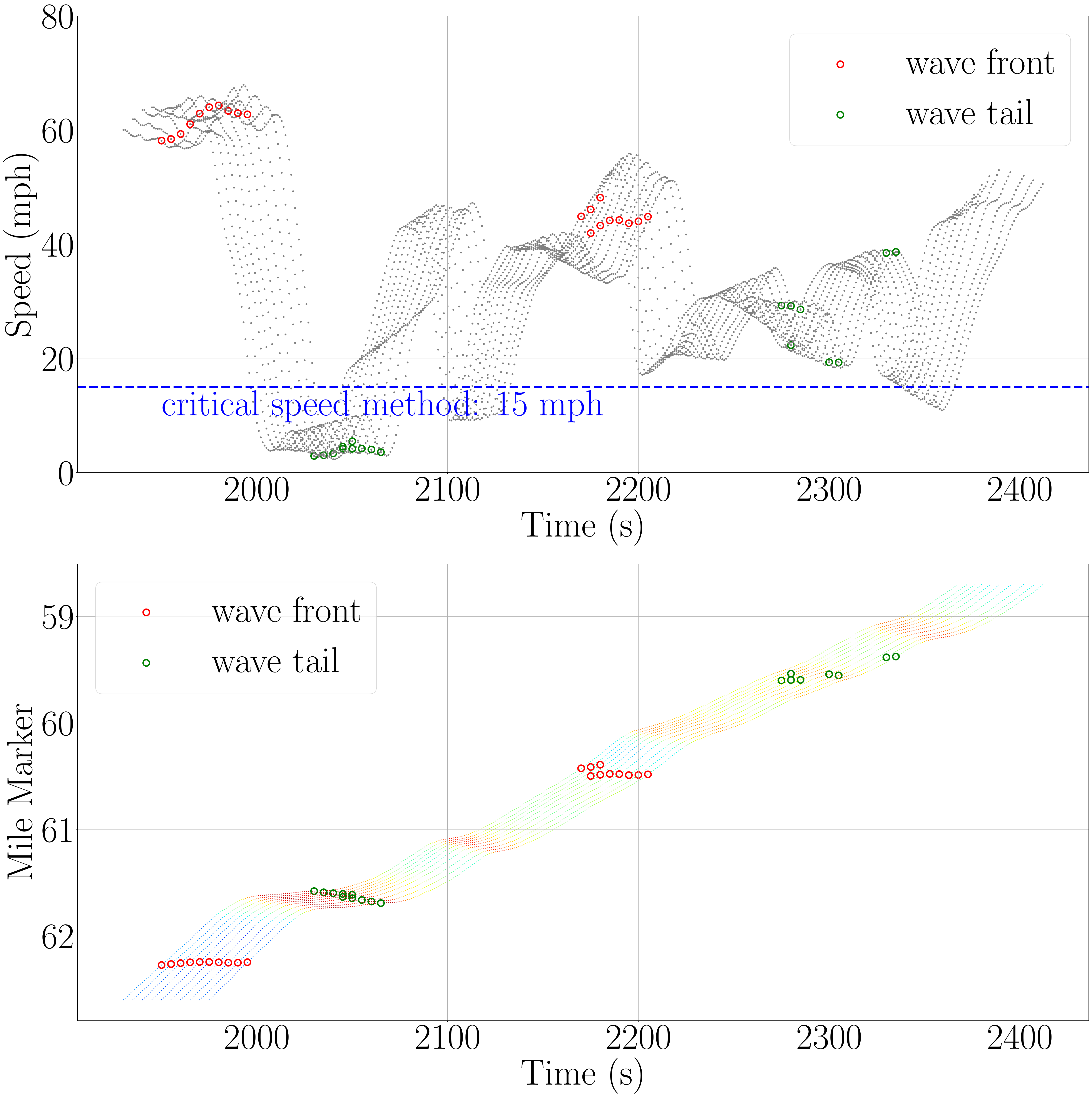}
    \caption{\textbf{Demonstration of Wavelet transform in detecting the wave fronts and tails: } The top diagram represents the speed-time graph with 10 sampled trajectories, where the wave front is marked in red and the tail in green. The bottom diagram illustrates the space-time diagram with all the key spatiotemporal features labeled as well. The data here is from 2022-11-22, covering virtual vehicle indices 380 to 390.}
    \label{fig:wavelet}
\end{figure}
The Wavelet transform offers the advantage of enabling multi-scale signal analysis, but it requires careful selection of the mother wavelet and fine-tuning of its parameters. As can be seen from Figure~\ref{fig:wavelet}, a 10- vehicle platoon experiences multiple traffic oscillations, and each oscillation has different patterns, because the traffic waves are at different stages of development. Tuning the wavelet parameters individually for each wave becomes challenging when applying the technique to large datasets with many types of waves. We thus opt for the critical speed method used in our work to simplify the detection step, as it requires only a single parameter to control the performance. 
    \item[(b)] Grouping: DBSCAN and our proposed graph representation. 
    
    DBSCAN is a widely used clustering method \cite{yang2023data,tgsimwave} for grouping spatio-temporal points in traffic wave analysis. While it performs well, its limitation lies in its inability to identify the topology of the waves.  We replicate the methods in \cite{tgsimwave} and use the \verb|sklearn| Python package to implement.

    We make slight adjustments in the definition to waves to ensure a fair comparison with our method. In \cite{tgsimwave}, waves are defined as instances where the speed has decreased by at least 10\% compared to five seconds earlier. In our case, we define this as the speed falling below the critical speed, using 15 mph as an example. 
\end{enumerate}
    In \cite{tgsimwave}, the distance for clustering is redefined as below. 
\begin{equation}
d((t_1, x_1, y_1), (t_2, x_2, y_2)) =
\begin{cases} 
0, & |x_1 - x_2| \leq x_d \land |t_1 - t_2| \leq t_d \land y_1 = y_2 \\
1, & \text{Otherwise}
\end{cases}
\label{eq:tgsim}
\end{equation}
where the $t$, $x$, $y$ denote time, space, and lane, respectively. As can be seen from the definition, it closely resembles the spatio-temporal search box described in our paper. If the spatial difference $| x_1 - x_2 |$ is within a threshold $x_d$, the temporal difference $|t_1 - t_2| $ is within a threshold $t_d$, and both points are in the same lane $y_1 = y_2 $, the function returns 0, indicating closeness. Otherwise, it returns 1, indicating separation. For a fair comparison, we choose a box size similar to that used in our method, with $x_d = 0.05$ miles and $t_d = 15$ seconds.
 \begin{figure}[H]
     \centering
     \begin{subfigure}{0.4\linewidth}
         \centering
         \includegraphics[width=\linewidth]{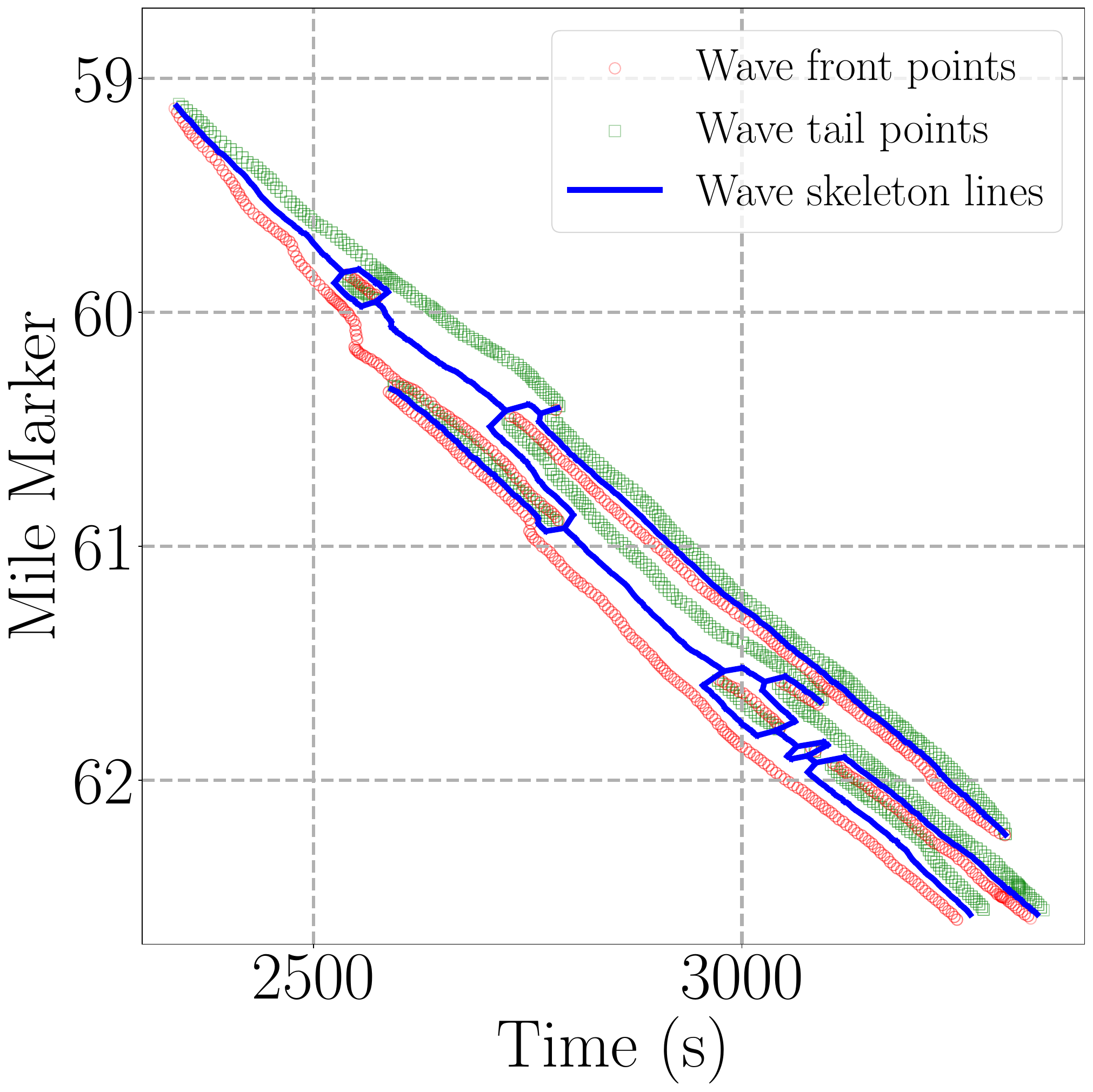}
         \captionsetup{labelformat=empty}
         \caption{(a) Our method}
         \label{fig:ours}
     \end{subfigure}
     \begin{subfigure}{0.4\linewidth}
         \centering
         \includegraphics[width=\linewidth]{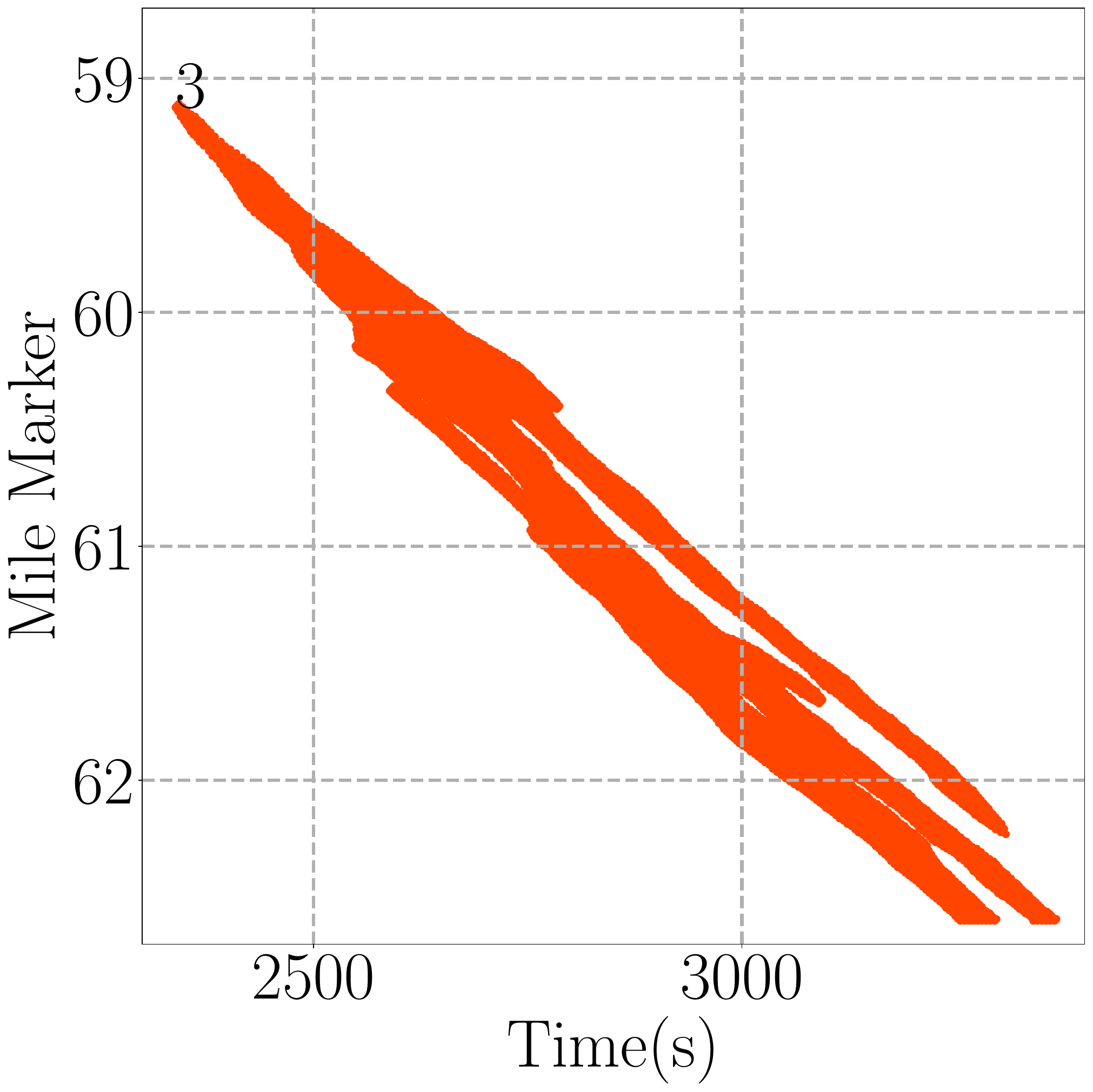}
         \captionsetup{labelformat=empty}
         \caption{(b) DBSCAN}
         \label{fig:dbscan}
     \end{subfigure}
     \caption{\textbf{Comparison of our method and the DBSCAN clustering method:} (a) Our method identifies a set of wave fronts and tails, along with their connections, allowing us to construct wave skeleton lines (blue) for a structured representation. For detailed visualization, click on \url{https://trafficwaves.github.io/2022-11-22/lane_1/4_15.svg}. (b) DBSCAN identifies a set of points under the critical speed.}
     \label{fig:compare}
 \end{figure}
In Figure~\ref{fig:compare}, we selected the same wave identified by both our method and DBSCAN. While DBSCAN provides a set of points within the wave, our method is specifically designed to define the wave boundary and its topology. Our method automatically identifies how many wave fronts and tails are in this wave component, wave bifurcation and merge details, specifying which waves it bifurcates into, as well as the time and location of bifurcation and merging points. While manual inspection from DBSCAN can also determine the time and space of bifurcation and merging from DBSCAN, our method automates this process, making it a scalable solution for analyzing traffic waves from multiple days of data. In summary, the main benefit of our method is to capture these topological elements automatically, which is not done in clustering approaches such as DBSCAN.
\color{black}
\section{Wave fronts and tails travel distance and speed}
\label{sec:appendix3}
More figures on the relationship between the wave travel distance and speed are provided in this section. Mean travel speeds for each lane and critical speed are summarized in Table \ref{tab:speed}.
\begin{figure}[H]
    \centering
    \begin{subfigure}[b]{0.49\linewidth}
        \centering
        \includegraphics[width=\linewidth]{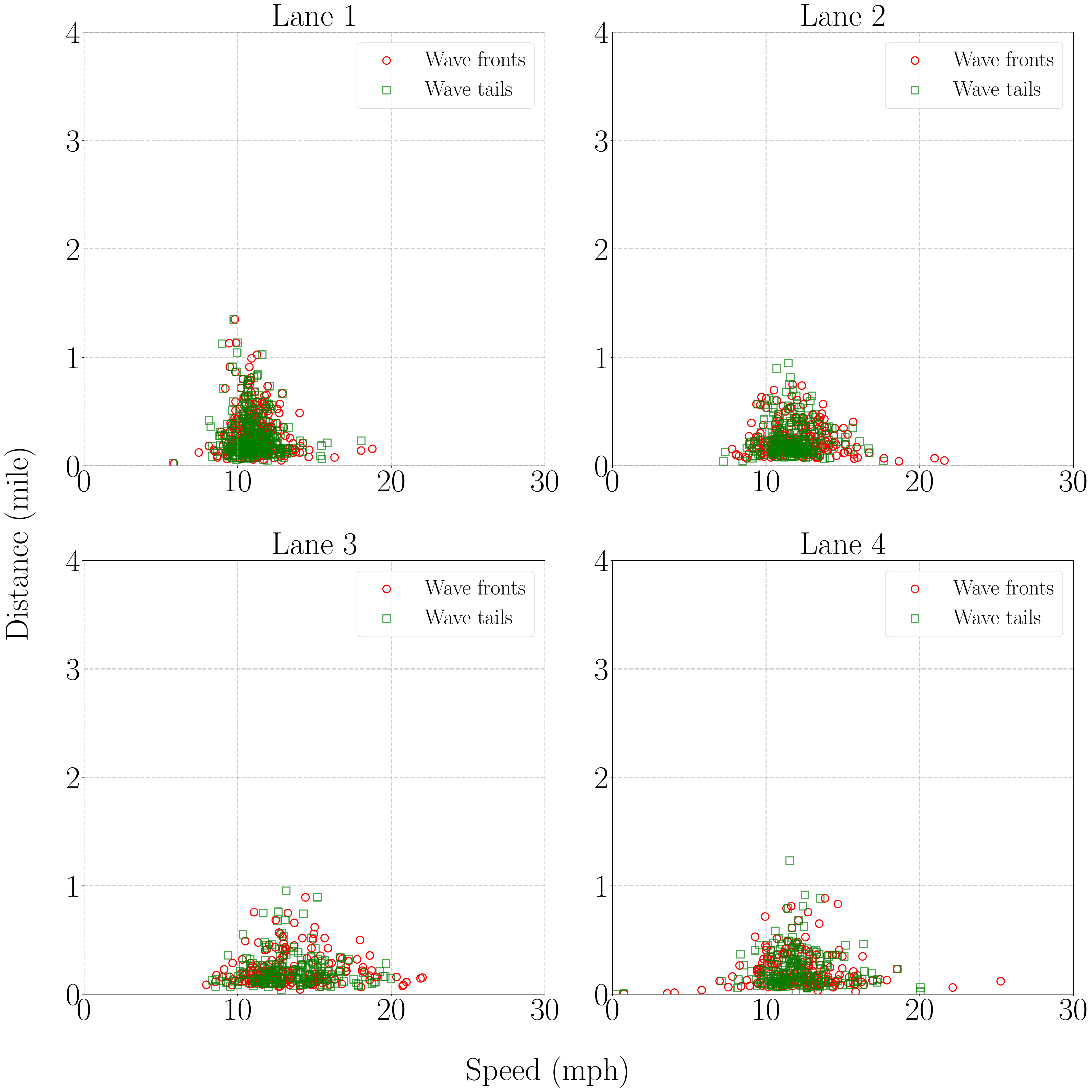}
        \caption{Critical speed at 1 mph}
    \end{subfigure}
    \hfill
    \begin{subfigure}[b]{0.49\linewidth}
        \centering
        \includegraphics[width=\linewidth]{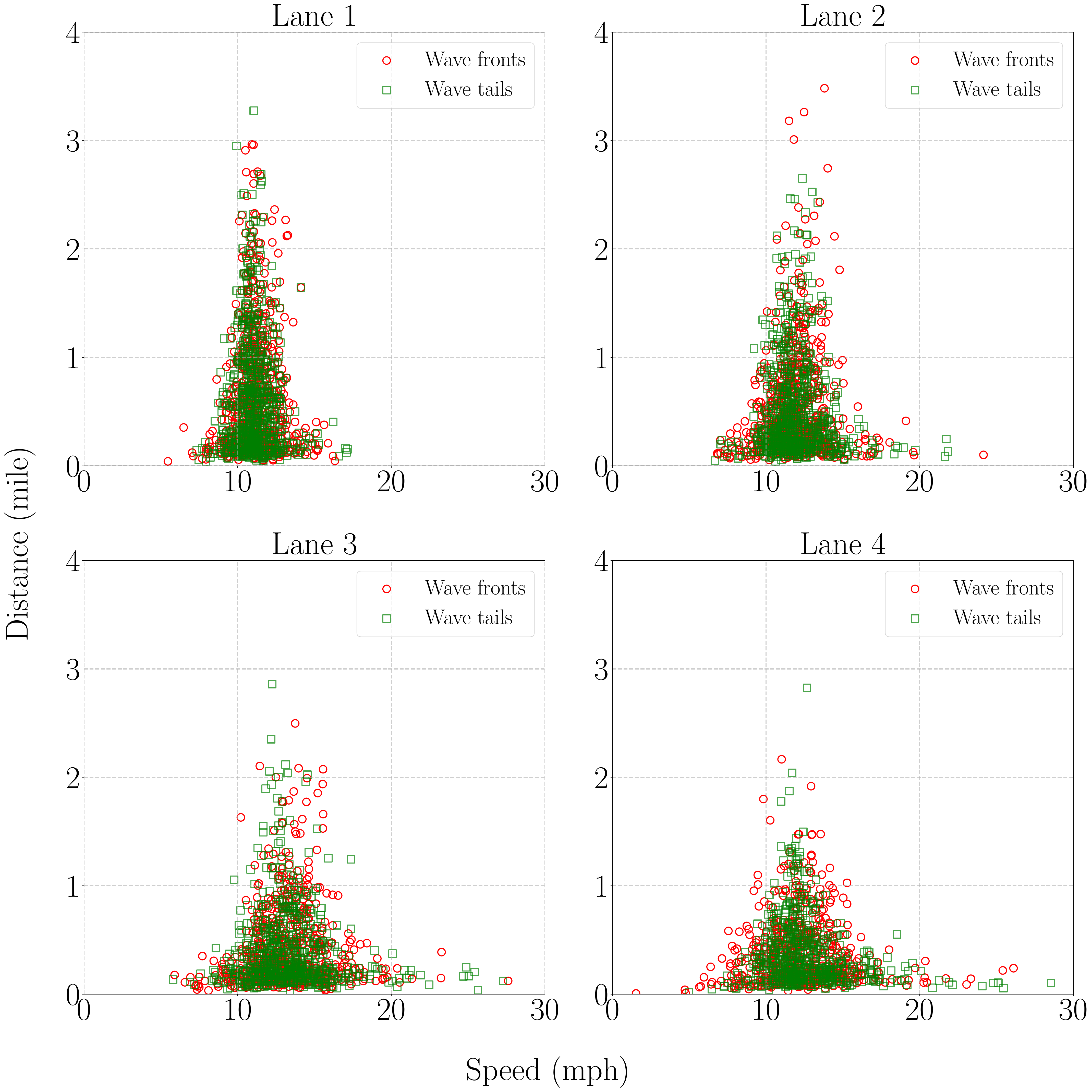}
        \caption{Critical speed at 5 mph}
    \end{subfigure}
    \\
    \begin{subfigure}[b]{0.49\linewidth}
        \centering
        \includegraphics[width=\linewidth]{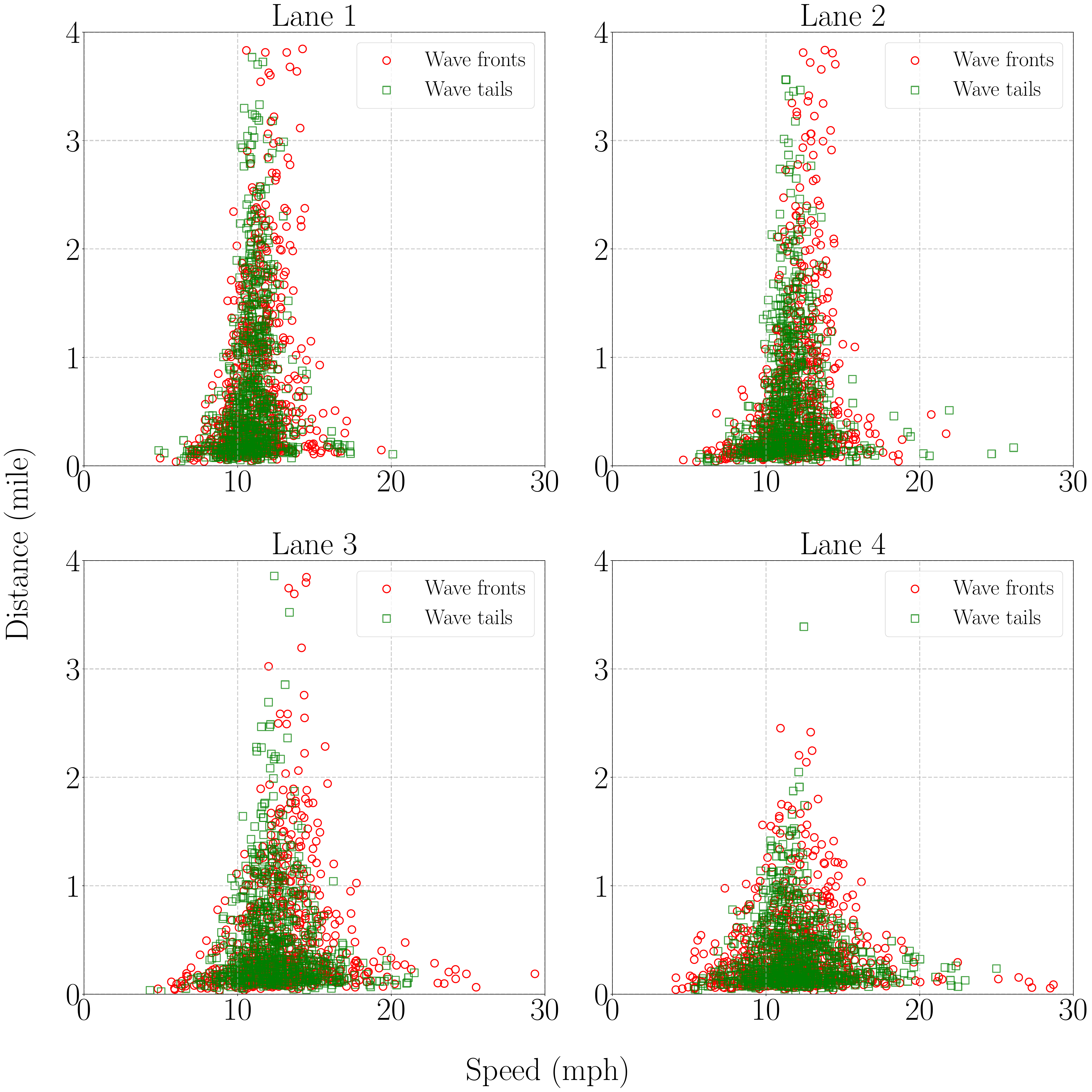}
        \caption{Critical speed at 10 mph}
    \end{subfigure}
    \hfill
    \begin{subfigure}[b]{0.49\linewidth}
        \centering
        \includegraphics[width=\linewidth]{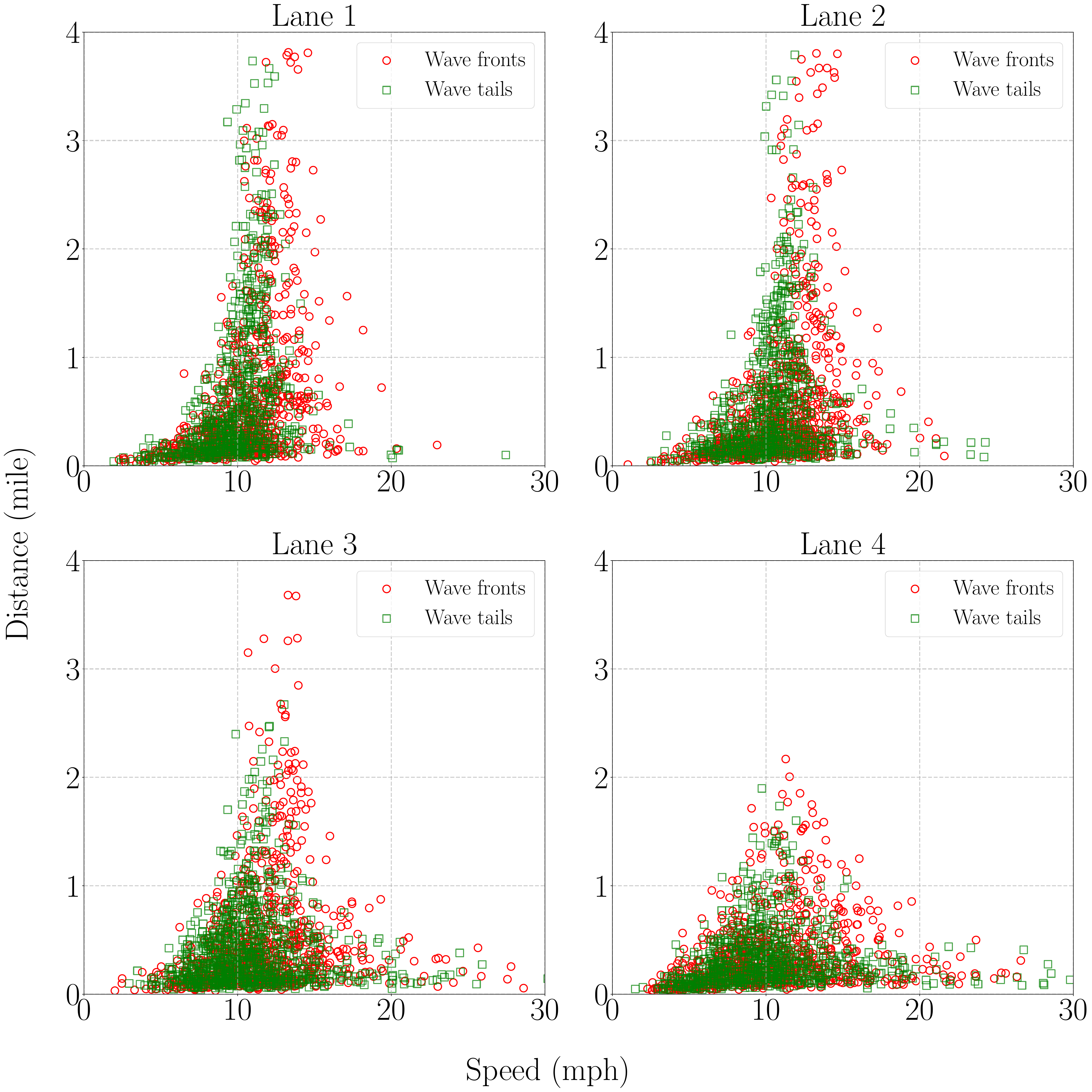}
        \caption{Critical speed at 20 mph}
    \end{subfigure}
    \caption{Wave front travel distance versus travel speed at various critical speeds from 1 to 20 mph.}
\end{figure}

\begin{figure}[H]
    \centering
    \begin{subfigure}[b]{0.49\linewidth}
        \centering
        \includegraphics[width=\linewidth]{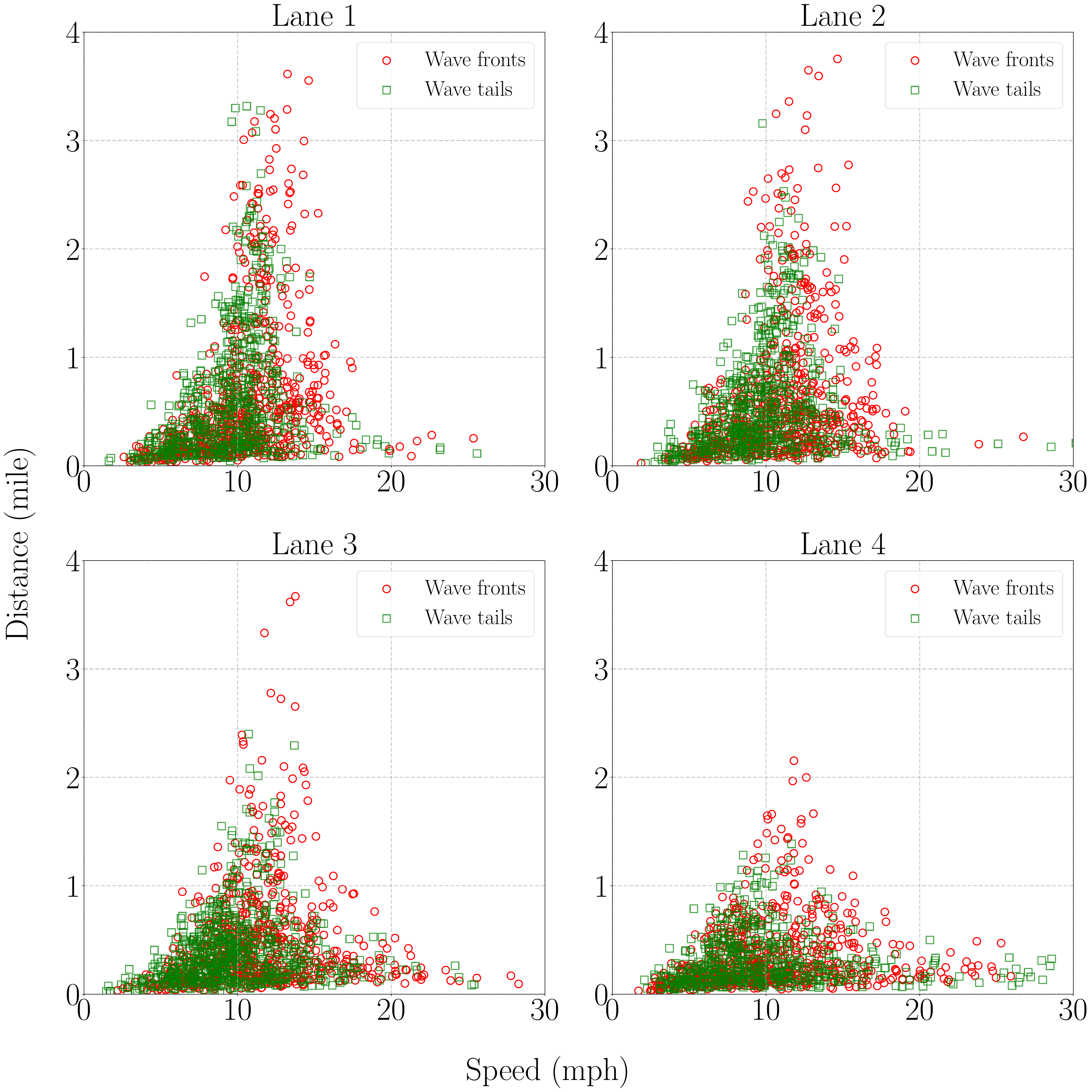}
        \caption{Critical speed at 25 mph}
    \end{subfigure}
    \hfill
    \begin{subfigure}[b]{0.49\linewidth}
        \centering
        \includegraphics[width=\linewidth]{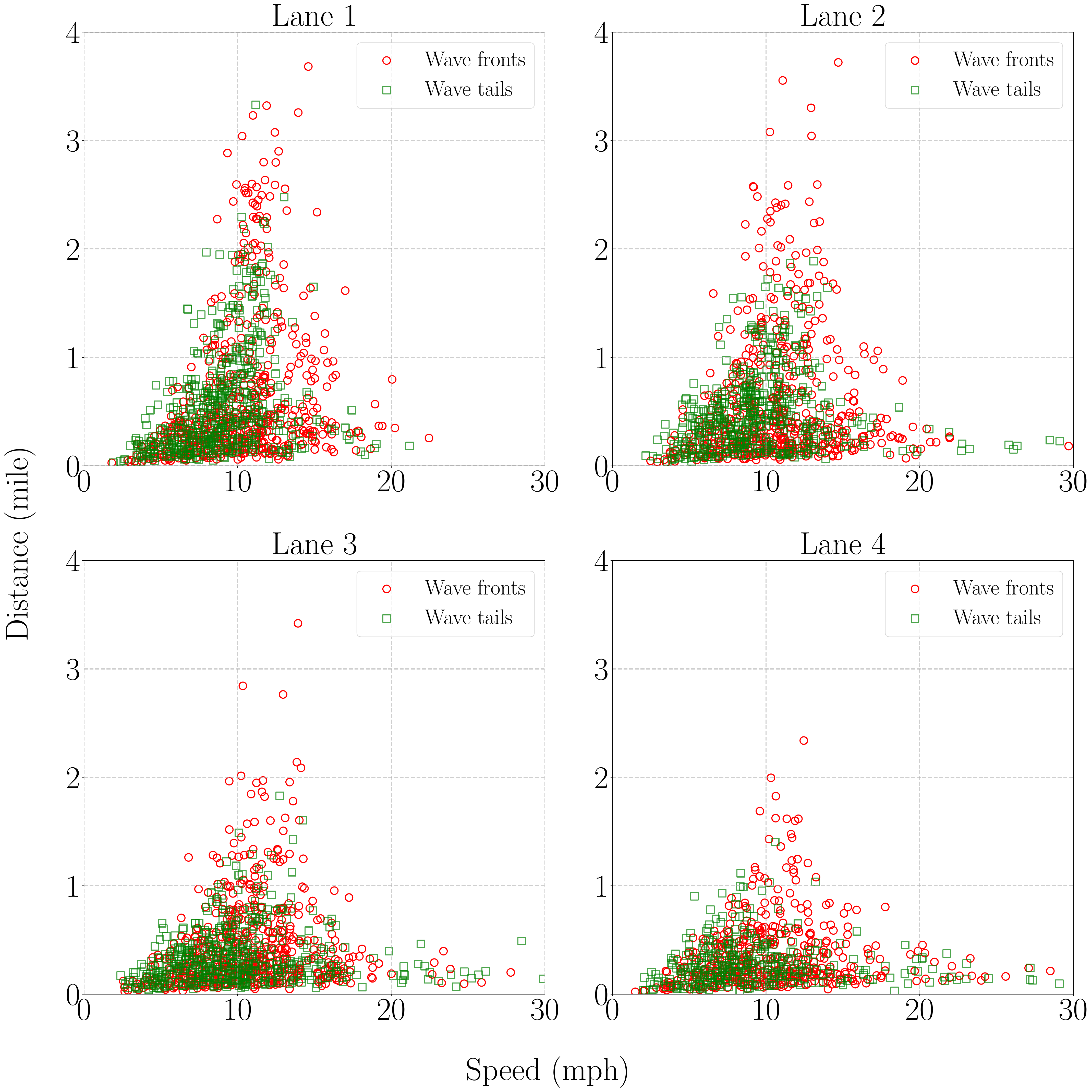}
        \caption{Critical speed at 30 mph}
    \end{subfigure}
    \\
    \begin{subfigure}[b]{0.49\linewidth}
        \centering
        \includegraphics[width=\linewidth]{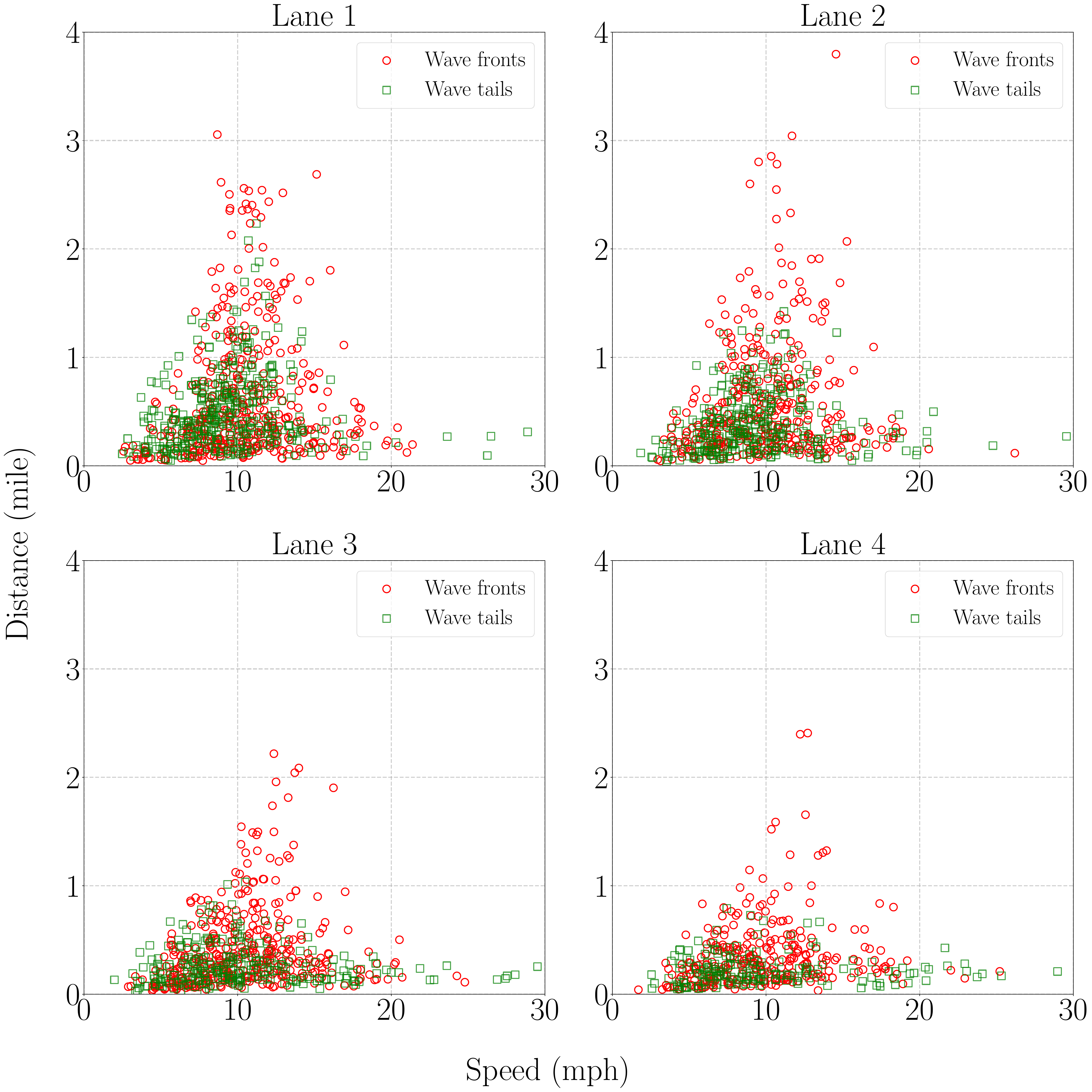}
        \caption{Critical speed at 35 mph}
    \end{subfigure}
    \hfill
    \begin{subfigure}[b]{0.49\linewidth}
        \centering
        \includegraphics[width=\linewidth]{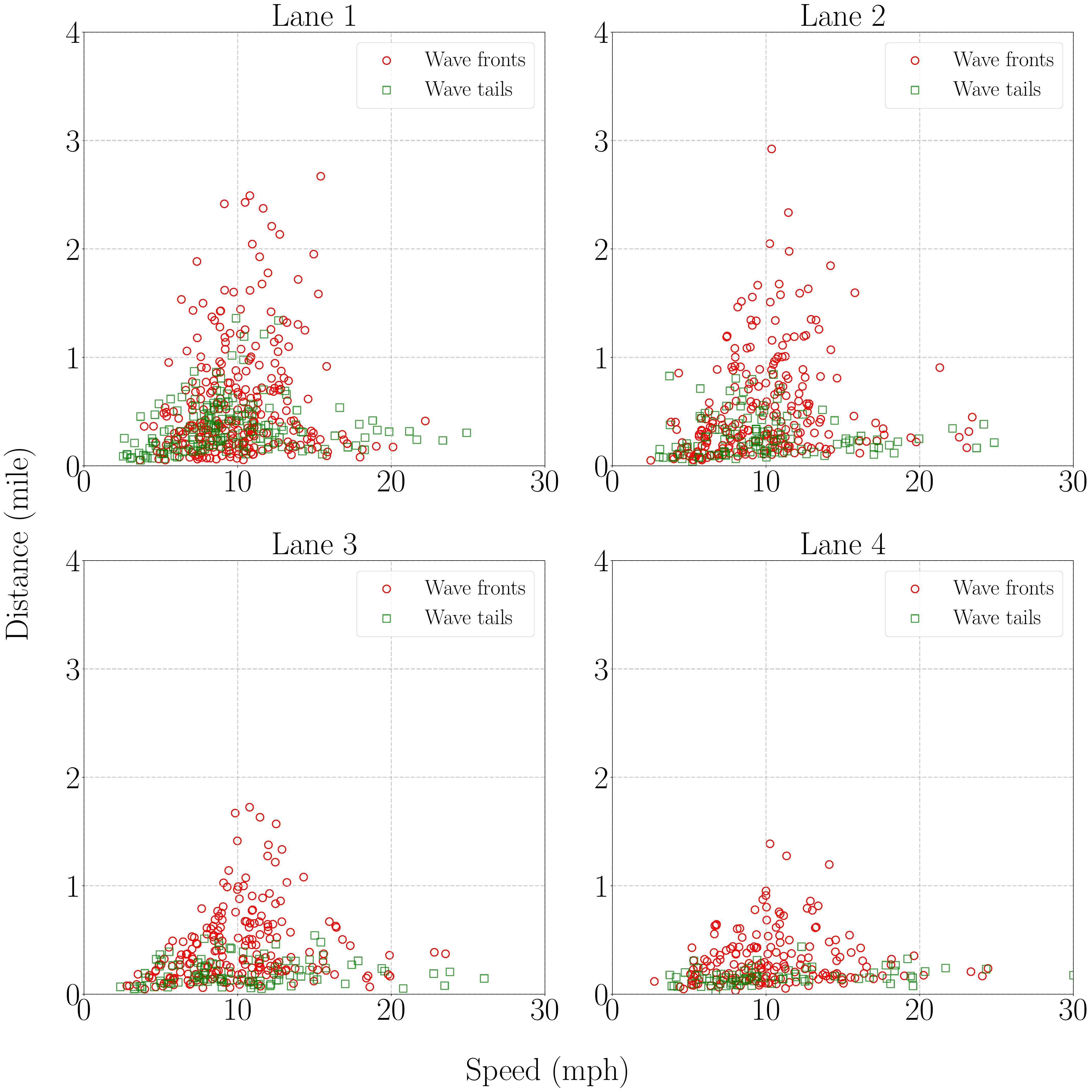}
        \caption{Critical speed at 40 mph}
    \end{subfigure}
    \caption{Wave front travel distance versus travel speed at various critical speeds from 25 to 40 mph.}
\end{figure}

\newpage
\bibliographystyle{ieeetr}
\bibliography{references.bib}
\end{document}